\documentclass[12pt,letterpaper]{article}%
\usepackage{amsmath}
\usepackage{rotating}
\usepackage{version}
\usepackage{amsfonts}
\usepackage{amssymb}
\usepackage{graphicx}
\usepackage{float}
\usepackage{url}
\usepackage[round,authoryear,compress]{natbib}
\usepackage{color}
\usepackage{xr}
\usepackage[normalem]{ulem}
\usepackage[labelfont={bf}]{caption}
\usepackage{wrapfig}
\usepackage{titletoc}
\usepackage[singlespacing,nodisplayskipstretch]{setspace}
\usepackage[pdftex,letterpaper,portrait,twocolumn=false,nomarginpar]{geometry}%
\providecommand{\U}[1]{\protect\rule{.1in}{.1in}}
\begin{document}

\title{The Output Convergence Debate Revisited: \\Lessons from recent developments in the analysis of panel data
models\thanks{This paper was presented as a keynote talk at 19$^{th}$
International Joint Conference on Computational and Financial Econometrics
(CFE) and Computational and Methodological Statistics (CMStatistics), at
Birkbeck, University of London, 13-15 December 2025; seminars at USC 24
February 2026; UC Riverside 27 February 2026; the IAAE Conference Lisbon 24
June 2026; and the 2026 NBER-NSF Time Series Conference, September 11-12,
University of Bologna, Italy. We are grateful for helpful comments from
conference and seminar participants, Daron Acemoglu, Alexander Chudik, Jan
Ditzen, and Oscar Jorda, and to Jingshu Chen, a PhD candidate at USC, for
excellent research assistance.}}
\author{M. Hashem Pesaran\\University of Southern California and Trinity College, University of Cambridge
\and Ron P. Smith\\Birkbeck, University of London}
\maketitle

\begin{abstract}
This paper examines the empirical basis of the output convergence debate in
light of recent developments in dynamic heterogeneous panels with trended
latent factors. Popular tools such as Barro's cross-country regressions and
two-way fixed effects (TWFE) estimators --- which assume parallel trends and
homogeneous dynamics --- substantially underestimate the speed of convergence
and yield misleading inference. Dynamic common correlated effects (DCCE)
estimators (Chudik and Pesaran, 2015) instead provide consistent estimates and
valid inference robust to nonparallel trends and correlated heterogeneity,
even with breaks, trends, or unit roots in the latent technology factor. The
paper also proposes a way to estimate the effects of slowly moving
determinants of growth. Using Penn World Tables data, the empirical results
show that cross-country convergence in per capita output is very slow, while
within-country convergence is reasonably fast. Capital accumulation is an
important determinant of cross-country output differences, and slow-moving
indicators --- potential for conflict and protection of property rights ---
are statistically significant determinants of steady-state output per capita
levels. The paper replicates a positive effect of democratization on output,
but finds its statistical significance declines once nonparallel trends and
dynamic heterogeneity are allowed for.

\end{abstract}


%TCIMACRO{\TeXButton{\thispagestyle{empty}}{\thispagestyle{empty}}}%
%BeginExpansion
\thispagestyle{empty}%
%EndExpansion

\newpage%
%TCIMACRO{\TeXButton{start page numbering}{\pagenumbering{arabic}}}%
%BeginExpansion
\pagenumbering{arabic}%
%EndExpansion%
%TCIMACRO{\TeXButton{\onehalfspacing}{\onehalfspacing}}%
%BeginExpansion
\onehalfspacing
%EndExpansion

\section{Introduction \label{Intro}}

There is an extensive econometric literature on whether log output per capita
($y_{it})$ of different countries $i=1,2,...,n$ converges over time
$t=1,2,...,T$ and the factors that may determine the level and growth of
$y_{it}$. \ This paper provides a critical examination of the empirical basis
of the output convergence literature in the light of recent developments in
the analysis of dynamic heterogeneous panels with interactive effects, which
provides the methodology needed to analyse output determination across countries.

The paper provides a taxonomy of models of the speed of convergence and
potential determinants of growth, time-varying and time-invariant, when the
parallel trend and dynamic homogeneity assumptions that are prevalent in this
literature are relaxed. It revisits the asymptotic biases of the estimators
obtained from Barro (1991) type cross-country regressions of growth on initial
output and from two-way fixed effect (TWFE) panel models. The estimate of the
speed of convergence obtained from Barro type regressions will be biased
towards zero if there are any random differences between countries even if the
parallel trend assumption is maintained. The TWFE estimators are robust to
random or fixed effects, but continue to result in biased estimates if the
parallel trends assumption is relaxed even if dynamic homogeneity is maintained.

By adhering to the parallel trends assumption, cross-country regressions as
well as TWFE panel estimators require that all countries are affected
similarly by the global factors, implying that all countries have the same
steady state growth rate. To allow the average growth rate to differ between
countries requires interactive time effects, where the loadings on the latent
global factor(s) differ across countries and can depend on time-invariant or
slowly moving variables like institutions that influence the degree of
openness to new ideas.\footnote{In this paper, we focus on a single factor
that we label technology for exposition, though models with interactive time
effects can accomodate multiple factors, including climate. We do not pursue
the role of climate change because of the measurement issues involved.}
Furthermore, to generate the observed pattern of global growth the latent
technology factor must be trended. As well as considering time-varying
covariates, we distinguish between time-invariant covariates that influence
the intercepts, which can only have a one-off impact on growth, and those that
influence the loadings on the global factor, which give rise to sustained
growth.\textbf{ }

Nonparallel trends arise naturally if countries are differently placed to
benefit from technological progress, due to historical, political or
institutional factors. In the presence of nonparallel trends, the TWFE
estimator of the speed of convergence is biased downward and this bias does
not vanish even if $T\rightarrow\infty$. More importantly, this paper shows
that the TWFE estimator of the speed of convergence tends to zero,
irrespective of its true value, if the technology factor is trended, which is
required for the world economy to exhibit non-zero output growth. Relaxing the
assumption of dynamic homogeneity is another source of the bias when the
TWFE\ estimator is employed; similar to the ones documented by Pesaran and
Smith (1995) in the context of fixed effects dynamic panel data models. The
empirical importance of dynamic heterogeneity in output convergence
regressions was demonstrated by Lee, Pesaran and Smith (1997, 1998) and more
recently by, for instance, Ong (2024).

The focus of this paper is to provide a systematic methodological
investigation of the implications of relaxing the assumptions of parallel
trends and dynamic homogeneity that are commonly made in the empirical growth
literature. To relax these assumptions we propose the use of correlated common
effect (CCE) type estimators instead of the TWFE estimates. The CCE estimator
was first advanced by Pesaran (2006) for panels with strictly exogenous
regressors, and has since been extended by Kapetanios, Pesaran and Yamagata
(2011) to allow for unit roots in the latent factors, and further by
Westerlund (2018) to account for deterministic trends amongst the latent
factors. The analysis of dynamic panels with latent factors has received less
attention. Chudik and Pesaran (2015) propose dynamic CCE estimators for panels
with interactive effects and heterogeneous dynamics. A related paper by Moon
and Weidner (2017) proposes least squares estimators that allow for
predetermined regressors (including lagged dependent variables), assuming
homogeneous slopes and stationary latent factors. Everaert and De Groote
(2016) have also considered CCE type estimators for panels with homogeneous
dynamics. However, all these papers that allow for dynamics assume the latent
factors are stationary, which is restrictive for the analysis of convergence
in the case of growing economies. Assuming dynamic homogeneity, we show that
the CCE estimator for the speed of convergence is consistent irrespective of
whether the global technological factor is stationary or trended. The trend
could be stochastic, with or without a drift, or could even be subject to a
number of breaks. We refer to this estimator as DCCEP, to highlight the
dynamics and the pooled nature of the underlying panel data model. We also
consider a mean group version of DCCE, which we denote by DCCEMG, that allows
for heterogeneous slopes, which applies generally to dynamics and the effects
of time-varying covariates such as capital stocks across countries.\footnote{A
special issue of \textit{The Journal of Econometrics} on panel data analysis
introduced by Sarafidis and Wansbeek (2021) contains a number of CCE related
papers including Juodis, Karabiyik and Westerlund (2021) on the robustness of
the pooled CCE estimator.}

In comparing Barro, TWFE, and DCCE type estimates, it is important to be clear
that while all these methods aim to estimate the same \textquotedblleft
cross-country average speed of convergence\textquotedblright, they differ in
their treatment of parameter heterogeneity across countries. Barro regressions
provide an estimate of it under the assumption of full parameter homogeneity.
Were it true that there was full parameter homogeneity, convergence would be
to a single level of output for all countries. Were it not true, the estimate
of the average speed of convergence will be biased towards zero. Similarly,
TWFE regressions provide an estimate of it under the assumption that all
countries have the same steady state growth rates. Were this true,
convergence, would be to country specific parallel trends for output. Were
this not true, the estimated average speed of convergence will again be biased
towards zero. The Barro or TWFE biases are asymptotic, not reduced by more
information. In contrast, DCCE type estimates, with or without the assumption
of dynamic homogeneity, focus on the average speed of convergence to
country-specific output paths, and do not impose that countries converge in
either per capita outputs or growth rates. With this in mind, we quantify the
size of the biases of Barro and TWFE estimates of the speed of convergence
using Penn World Table data on log output per capita (and per employee)
covering $157$ countries over seventy years, 1950-2019. Both unconditional and
conditional Barro regressions have speeds of convergence that range between
$0.3$ and $0.6$ percent implying mean lags that fall between $315$ and $171$
years, suggesting that there is little evidence of cross country output
convergence. TWFE estimates of the speed of convergence, that relate to an
assumed \textit{common growth} path across countries, are around $3$ percent
with implied mean lags of around $25$ years. In contrast, the DCCEMG estimates
of speed of convergence that relate to \textit{country-specific} output paths
are much higher, around $11$ percent with mean lags of around $6$ years for
per capita output, corresponding to business cycle frequencies documented in
the time series literature. For example, according to NBER (National Bureau of
Economic Research) the average duration of business cycle in the U.S. over the
period 1945-2020 has been around 75 months or 6.25 years.\footnote{See
https://www.nber.org/research/data/us-business-cycle-expansions-and-contractions}
Similar results are also found for G7 and OECD countries. See, for example,
Chauvet and Yu (2006) who also show that there exists a significant degree of
synchronization of business cycles across countries driven partly by
globalization and common shocks like total factor productivity
shifts.\footnote{Chauvet and Yu (2006, Table 2) report average expansion and
recession durations for OECD and G7 countries over the period 1960Q1-2000Q1.
For OECD their estimate of average business cycle duration is 7 years.} The
rise in the estimated average speed of convergence as we relax parallel trends
and dynamic homogeneity is in line with our theoretical findings, and
highlight the important differences across the models that underlie them.

For determinants of growth in the long run, we focus on capital stock as the
primary time-varying covariate. We argue that the DCCEMG estimators of the
long run effect of capital on output is reliable, despite endogeneity
concerns, since CCE type estimators allow global latent factors to
simultaneously drive both capital and output variables and thus filter out the
common sources of dependence between capital and output. Focussing on DCCEMG,
the long run estimate of capital coefficient is around $0.48$, which is quite
close to the average share of capital at $0.46$, obtained from PWT data.

We also consider the effects of democracy on output using Acemoglu, Naidu,
Restrepo, and Robinson, (2019) replication data and their specification. This
is an important example where the democracy variable is time-varying only for
half of the countries in the panel. In our empirical investigation of the
effects of democracy we also exclude countries with less than 20 time series
data points for estimation, otherwise it would not be possible to relax the
parallel trends assumption. This gives a sample of 148 rather than 175
countries used by Acemoglu et al. (2019). We find the TWFE estimates of the
long run effect of democracy on output to be very similar in size and
significance to the ones they report. However, when the parallel trend
assumption is relaxed, the DCCEP estimate of the long run effect of democracy
is reduced in size and becomes statistically insignificant.

We apply our suggested procedure for time-invariant variables after removing
the time-varying effects using either TWFE or DCCEP estimates. From the large
number of suggested determinants of the level or growth of output we choose:
an indicator of geography, absolute latitude; an indicator of the potential
for conflict, namely ethnolinguistic fractionalization, and an indicator of
property rights, \textit{protection from expropriation}. Whether TWFE or DCCEP
estimates are used to filter out the effects of the time-varying covariates,
we find ethnolinguistic fractionalization and protection from expropriation to
be statistically significant with the expected signs. Latitude was not
statistically significant for both filtering procedures.

\textbf{Related Literature: }In the vast literature on economic growth, there
are many strands. A quantitative literature initiated by Baumol (1986) and
Barro (1991), and more recently Kremer, Willis, and You (2022) is concerned
with modelling convergence in per capita output across countries. Another
quantitative strand associated with Acemoglu and coauthors has emphasised the
role of institutions and methods of identifying causality. A third strand is
concerned with understanding sustained economic growth propelled by
technological progress implemented through a process of learning and
innovation. This strand involved theoretical work by Aghion and Howitt (1992)
on how the new technology surpasses the old through a process of creative destruction.

In terms of the econometric methods, the early contribution by Baumol (1986)
opened the debate. It used a cross-country regression of country growth rates
on initial income and interpreted the coefficient on initial income, beta, as
a measure of "unconditional" or absolute convergence. Barro (1991), Mankiw,
Romer and Weil (1992) and Barro and Sala-i-Martin (1992) added initial
covariates and interpreted the coefficient on initial income as a measure of
"conditional" convergence to country specific steady states. Nerlove (1998,
1999) labelled such cross-country regressions Baumol-Barro regressions, but we
will follow the common usage and call them Barro regressions.

Before allowing for convergence using a Barro type model, Mankiw, Romer and
Weil (1992) estimated a cross-country regression that did not include initial
income, which explained log GDP per working age person in 1985 by
corresponding values of Solow variables like investment and population growth.
Similarly, Acemoglu, Johnson and Robinson (2001) estimate a cross-country
regression explaining log per capita output in 1995 as a function of a measure
of institutions and time-invariant variables like latitude, but not initial
income. Such regressions are not informative about convergence and it is not
clear how one chooses a particular year. Maseland (2021) examines the movement
over time in the coefficients of 43 such "deep" determinants used by Acemoglu
et al. (2001) finding the effect of many of them, including latitude,
increasing from around 1960 until about 2000, around when the paper was
published, and then declining.

Barro regressions were criticised on a number of counts. It was argued that
such cross section snapshots were not be informative about time series
processes and that, so called, "beta convergence", a negative coefficient on
initial income in this regression, does not imply "sigma convergence" a
falling variance of incomes across countries. In Barro regressions any
intercept heterogeneity is inherently correlated because it is inherited by
the regressor, namely initial income. As a result, the estimated speed of
convergence is downward biased unless the country specific intercepts are
identical. This bias is present even if the intercepts are randomly
distributed independently of all errors. To the extent that including
covariates reduces the variance of the intercepts, making them more similar,
the bias is reduced. Thus estimates of the speed of convergence in what are
called conditional convergence regressions are less biased towards zero than
in unconditional ones. Similar issues arise with dynamic heterogeneity in the
coefficients of lagged dependent variables which are inherently correlated
with the regressors.

The title of Nerlove (1998) posed a central question: "growth rate
convergence, fact or artifact?" Like him we will argue that such regressions
are badly biased. However, repeated criticisms of Barro regressions, including
that by Friedman (1992) who asked "Do old fallacies ever die?", did not kill
them and they have been recently resurrected by Kremer et al. (2022). But we
hope the econometric analysis that we provide will put one last nail in the coffin.

Subsequent literature, such as Islam (1995), used panels with country and time
effects. These widely used two-way fixed effects panel data models, also known
as fixed-effect, time-effect, imposed a parallel trend assumption, which we
will relax. After a rapid growth in the literature, interest waned because
there seemed little evidence for "unconditional" convergence and while there
was evidence for "conditional" convergence, specification of the appropriate
covariates was problematic. An early survey is Temple (1999). Durlauf (2009)
documents "The rise and fall of cross-country growth regressions".

Subsequent papers such as Barro (2015) and Acemoglu et al. (2019) have argued
for the importance of institutions, such as democracy, in panel studies of
conditional convergence.

After an extensive survey, Johnson and Papageorgiou (2020) concluded that
"there is a broad consensus of no evidence supporting absolute convergence in
cross-country per capita incomes --- that is poor countries do not seem to be
unconditionally catching up to rich ones." \ However, that consensus was
challenged in papers by Kremer et al. (2022), and Patel, Sandefur and
Subramanian (2021). Kremer et al. (2022) resurrected Barro regressions and
claimed that these provided evidence for unconditional convergence since 2000,
if not from earlier dates, and argued that while conditional convergence held
throughout the period, absolute convergence did not hold initially, but as
human capital and policies improved in poorer countries, the difference in
institutions across countries has shrunk, leading to unconditional
convergence, a conclusion we challenge. Smith (2024) provides a recent review
of some of the econometric methods used in the convergence literature.

\textbf{Outline of the paper:} The rest of the paper is set out as follows:
Section \ref{GEN} provides an overview of the models of output convergence.
Section \ref{BB} onwards considers estimation methods. Section \ref{BB}
investigates the bias of the Barro cross-country regression when there is
intercept heterogeneity with and without conditioning on time-invariant
covariates. Section \ref{HetCon}, examines the bias of the TWFE estimator when
the assumption of parallel trends is relaxed. Section \ref{DCCEP} considers
estimation of models with nonparallel trends and homogeneous dynamics using
the DCCEP estimator. Section \ref{HetConFac} considers estimation of models
with nonparallel trends and heterogeneous dynamics using the DCCEMG estimator.
\textbf{ }Section \ref{TVTI} considers time-varying covariates. Section
\ref{TINV} considers the identification and estimation of the coefficients of
time-invariant variables, like geography, or slowly moving ones like climate
or institutions. Section \ref{empirics} examines the quantitative importance
of the theoretical results using data from the Penn World Tables and the
replication files of Acemoglu et al. (2019) and Kremer et al. (2022). Section
\ref{conc} contains some concluding remarks.

\section{An overview of models of output convergence\label{GEN}}

This section provides an overview of the models that will be examined to
provide context for the subsequent discussion of estimation methods. To
simplify the exposition here we assume balanced panels, but use both balanced
and unbalanced panels in our empirical analysis, depending on the estimation
procedure under consideration. The output data we use in our empirical
application is unbalanced, the length of the time period available differs
quite widely across countries. For instance 15 new countries arrived with the
end of the Soviet Union. Wansbeek and Kapteyn (1989) were the first to show
that the simple within transformations cannot be applied to unbalanced panels
and provided the necessary algorithms for implementation of two way fixed
effects (TWFE) for unbalanced (what they called incomplete) panel data models.
\textbf{ }

Much of the literature can be related to the widely used TWFE autoregressive
(AR) panel data model, which can be viewed as a stochastic version of output
dynamics in context of deterministic Solow-Swan growth models (Solow (1956)
and Swan (1956)). Similar AR specifications also result from stochastic growth
models. See, for example, Brock and Mirman (1972), Merton (1975), Donaldson
and Mehra (1983), and Binder and Pesaran (1999).

Assuming a linear specification with first-order dynamics, the baseline panel
model is:%
\begin{equation}
y_{it}=\alpha_{i}+f_{t}+\rho y_{i,t-1}+u_{it}, \label{FETE}%
\end{equation}
where $y_{it}$ is the logarithm of per capita output in countries
$i=1,2,...,n,$ at time periods $t=1,2,...,T,$ with $\left\vert \rho\right\vert
<1$. The $\alpha_{i}$ are heterogeneous country-specific intercepts. The time
effect, $f_{t}$, can be viewed as a latent global factor, such as technology,
assumed here to have the same effects on $y_{it}$ across all $i$.

What is central in a growth context is that the latent factor charcterizing
technical progress is trended. To highlight this issue, we note that the TWFE
specification (\ref{FETE}) can be written equivalently in a de-factored form
as%
\begin{equation}
\text{ }\mathring{y}_{it}=\alpha_{i}+\rho\mathring{y}_{i,t-1}+u_{it},
\label{gap}%
\end{equation}
where $\mathring{y}_{it}=y_{it}-g_{t}$, and $g_{t}=\sum_{s=0}^{\infty}\rho
^{s}f_{t-s}.$ Since $\left\vert \rho\right\vert <1$, then for each
$i=1,2,...,n$
\[
E\left(  \mathring{y}_{it}\right)  =\frac{\alpha_{i}}{1-\rho},\text{ and
}E\left(  \Delta y_{it}\right)  =E\left(  \Delta g_{t}\right)  .
\]
Thus under the TWFE specification, log output paths will be parallel to one
another, with all countries having the same mean output growth rate, $E\left(
\Delta g_{t}\right)  $. This feature of TWFE is often referred to as the
"parallel trend" assumption.

The parallel trend assumption can be relaxed by allowing heterogeneous
loadings on the global factor giving a more general, interactive time effects,
specification where $f_{t}$ is replaced by $\gamma_{i}f_{t}$, with $\gamma
_{i}$ representing the factor loading for country $i$.\footnote{As noted
above, given the growth context for exposition we treat there as being a
single global factor. But it easily generalises to a multi-factor model:
$\boldsymbol{\gamma}_{i}^{\prime}\mathbf{f}_{t}$, with $\mathbf{f}_{t}$
representing a vector of latent variables, and $\boldsymbol{\gamma}_{i}$ the
associated vector of factor loadings.} With the parallel trend assumption
relaxed, we have $E\left(  \Delta y_{it}\right)  =\gamma_{i}E\left(  \Delta
g_{t}\right)  $, and the mean output growth will no longer be equal across
countries. When $\Delta f_{t}$ is stationary, global growth is given by
$E\left(  \Delta y_{it}\right)  =(1-\rho)^{-1}\gamma_{i}E\left(  \Delta
f_{t}\right)  =\gamma_{i}E\left(  \Delta g_{t}\right)  $. The two
representations of the cross-country mean growth rates are equivalent.
Differences in factor loadings across countries could be due to differences in
trade, capital and immigration policies across countries due to geopolitical
and other considerations. We discuss the implications of differences in factor
loadings in Section \ref{HetCon}. In practice, these loadings could also vary
over time, thus introducing another degree of complexity in the analysis of
output convergence.

Assuming that $\rho$ lies in the range $(-1,+1)$, the parameter of interest is
$\phi=1-\rho>0$, which measures the speed of convergence of de-factored output
$\mathring{y}_{it}=y_{it}-\gamma_{i}g_{t}$ to its steady state value of
$E\left(  \mathring{y}_{it}\right)  =\left(  1-\rho\right)  ^{-1}\alpha_{i}$.
For estimation, or when there are higher order lags, it is more convenient to
work with the error correction representations of (\ref{FETE}) or (\ref{gap})
given by
\begin{equation}
\Delta y_{it}=\alpha_{i}+\gamma_{i}f_{t}-\phi y_{i,t-1}+u_{it},\text{ or
}\Delta\mathring{y}_{it}=\alpha_{i}-\phi\mathring{y}_{i,t-1}+u_{it}.
\label{ECM}%
\end{equation}
In these specifications it is assumed that country-specific error terms,
$u_{it}$, are serially uncorrelated and cross-sectionally independent with
mean zero and variances $\sigma_{iu}^{2}$, for $i=1,2,...,n$ that vary across
countries but are fixed over time. Interactive time effects take account of
cross-sectional dependence due to common factors, it is also important to
allow for possible spatial or spill-over effects that cannot be eliminated by
use of factors. See Pesaran and Xie (2026).

Higher order dynamics might be required to allow for possible serial
correlation in $u_{it}$. In practice, this is achieved by adding lagged values
of the output growth variables to the error correction representations,
(\ref{ECM}). Assuming a $p^{th}$ order panel AR for log output%
\begin{equation}
y_{it}=\alpha_{i}+\gamma_{i}f_{t}+%
%TCIMACRO{\tsum _{\ell=1}^{p}}%
%BeginExpansion
{\textstyle\sum_{\ell=1}^{p}}
%EndExpansion
\rho_{\ell}y_{i,t-\ell}+u_{it} \label{TWFEp}%
\end{equation}
yields the following $p-1$ order panel error correction specification
\begin{equation}
\Delta y_{it}=\alpha_{i}+\gamma_{i}f_{t}-\phi y_{i,t-1}+\sum_{\ell=1}%
^{p-1}\psi_{\ell}\Delta y_{i,t-\ell}+u_{it}, \label{ECMp-1}%
\end{equation}
where $\phi=1-%
%TCIMACRO{\tsum _{\ell=1}^{p}}%
%BeginExpansion
{\textstyle\sum_{\ell=1}^{p}}
%EndExpansion
\rho_{\ell}$. As we shall see, the choice of the lag order, $p$, plays an
important role for the empirical validity of some of the estimation
procedures. One prominent example arises when Barro regressions are used to
estimate $\phi$, which requires that there is only a single lag, in addition
to a number of other assumptions that will be discussed below in some detail.

The TWFE specification, (\ref{FETE}), and its extension in (\ref{ECMp-1})
imposes that all countries converge to their steady state per capita output
levels at the same speed, $\phi$, which is a rather strong assumption.
Countries differ in their political institutions and climatic conditions which
affect the speed with which they adapt to shocks. We discuss the implications
of these assumptions in Section \ref{HetConFac}. Accordingly, we also consider
a fully heterogeneous version of (\ref{ECMp-1}) that allows for $\phi$ and
$\psi_{\ell}$ to differ across $i$, namely%
\begin{equation}
\Delta y_{it}=\alpha_{i}+\gamma_{i}f_{t}-\phi_{i}y_{i,t-1}+\sum_{\ell=1}%
^{p-1}\psi_{i,\ell}\Delta y_{i,t-\ell}+u_{it}. \label{ECMHetro}%
\end{equation}
The parameter of interest is the average speed of convergence across
countries, defined by $E(\phi_{i})$.

Adding additional variables such as labour and capital, to be denoted by the
$k\mathbf{\times}1$ vector $\mathbf{x}_{it}$, to the above output equation is
straightforward. One important issue is the simultaneous determination of
$\mathbf{x}_{it}$ and $y_{it}$. Here we suppose that $y_{it}$ and
$\mathbf{x}_{it}$ are simultaneously determined via the latent factors,
$\mathbf{g}_{t}$, that could be $I(1)$ and trended, with a de-factored
representation:
\begin{equation}
y_{it}=\boldsymbol{\gamma}_{iy}^{\prime}\mathbf{g}_{t}+\mathring{y}%
_{it}\text{, and \ }\mathbf{x}_{it}=\boldsymbol{\Gamma}_{ix}\mathbf{g}%
_{t}+\boldsymbol{\mathring{x}}_{it}, \label{fREP}%
\end{equation}
where $\mathring{y}_{it}$ and $\boldsymbol{\mathring{x}}_{it}$ are the
stationary components assumed to be distributed independently of the latent
factors, as it is standard in the analysis of linear factor models. The factor
loadings, $\boldsymbol{\gamma}_{iy}$ and $\boldsymbol{\Gamma}_{ix}$ are
allowed to vary across countries and need not affect the inputs and the
outputs similarly. As before we refer to $\mathring{y}_{it}$ and
$\boldsymbol{\mathring{x}}_{it}$ as de-factored components and assume that
they are related via the following ARDL specification
\begin{equation}
\mathring{y}_{it}=\phi_{i}(L)\mathring{y}_{i,t-1}+\boldsymbol{\psi}%
_{x,i}^{\prime}(L)\boldsymbol{\mathring{x}}_{it}+u_{it}, \label{DefREP}%
\end{equation}
where $\phi_{i}(L)$ and $\boldsymbol{\psi}_{x,i}(L)$ are finite order
distributed lag functions such that $\mathring{y}_{it}$ and
$\boldsymbol{\mathring{x}}_{it}$ are jointly covariance stationary, and
$\boldsymbol{\mathring{x}}_{it}$ is weakly exogenous (with respect to
$u_{it})$. Although the latent factors and their loadings cannot be separately
identified, the above ARDL specification in de-factored variables does allow
identification of $\phi_{i}(L)$ and $\boldsymbol{\psi}_{x,i}(L)$ for all $i$,
by proxing the latent factors with cross section averages of $y_{it}$ and
$\mathbf{x}_{it}$ so long as the number of latent factors is not larger than
$k+1$. This is a necessary condition. A necessary and sufficient condition for
identification is given by the rank condition in Assumption 6 in Chudik and
Pesaran (2015).

The de-factored specification can be written equivalently as a dynamic panel
with interactive effects:
\begin{equation}
y_{it}=\phi_{i}(L)y_{i,t-1}+\boldsymbol{\psi}_{x,i}^{\prime}(L)\mathbf{x}%
_{it}+\boldsymbol{\gamma}_{i}^{\prime}\mathbf{f}_{t}+\varepsilon_{it},
\label{DPanel}%
\end{equation}
where $\boldsymbol{\gamma}_{i}^{\prime}\mathbf{f}_{t}=\left[
\boldsymbol{\gamma}_{iy}^{\prime}-\phi_{i}(L)L\boldsymbol{\gamma}_{iy}%
^{\prime}-\boldsymbol{\psi}_{x,i}^{\prime}(L)\boldsymbol{\Gamma}_{ix}\right]
\mathbf{g}_{t}$. The short run and long run effects of $\mathbf{x}_{it}$ on
$y_{it}$ can be distinguished using the following error correction
representation%
\begin{align}
\Delta y_{it}  &  =\alpha_{i}+\boldsymbol{\gamma}_{i}^{\prime}\mathbf{f}%
_{t}-\phi_{y,i}y_{i,t-1}+\sum_{\ell=1}^{p-1}\psi_{y,i,\ell}\Delta y_{i,t-\ell
}\label{ECMCov}\\
&  +\boldsymbol{\phi}_{x,i}^{\prime}\mathbf{x}_{i,t-1}+\sum_{\ell=1}%
^{p-1}\boldsymbol{\psi}_{x,i,\ell}^{\prime}\Delta\mathbf{x}_{i,t-\ell}%
+u_{it},\nonumber
\end{align}
where $\mathbf{\psi}_{x,i,}$ and $\mathbf{\psi}_{x,i}/\phi_{y,i}\,$\ refer to
short run and long run effects, respectively.

In addition to $\mathbf{x}_{it}$ that vary over time and across countries, the
literature has emphasised the importance of covariates that do not vary over
time or vary very slowly. These may influence the intercepts $\alpha_{i},$
which under the parallel trends assumption are the main source of differences
in steady state output levels. Once the parallel trend assumption is relaxed,
it is the differences in output growth paths, $\gamma_{i}f_{t},$ that will
matter, since countries that are on different growth paths will diverge and
differences in $\alpha_{i}$ will be of secondary importance. It is the
differences in $\gamma_{i}$ across $i\ $ that determine the changing
dispersion of output per capita across countries. In our empirical analysis we
consider the following models for the determination of $\alpha_{i}$ and
$\gamma_{i}:$
\begin{align}
\alpha_{i}  &  =\alpha_{\alpha}+\boldsymbol{\theta}_{\alpha}^{\prime
}\mathbf{z}_{i\alpha}+\eta_{i\alpha,}\label{aizi}\\
\gamma_{i}  &  =\alpha_{\gamma}+\boldsymbol{\theta}_{\gamma}^{\prime
}\mathbf{z}_{i\gamma}+\eta_{i\gamma}, \label{gizi}%
\end{align}
where $\mathbf{z}_{i\alpha}$ and $\mathbf{z}_{i\gamma}$ are likely to include
climatic, institutional or policy variables; $\eta_{i\alpha}$ and
$\eta_{i\gamma}$ are random errors, representing the unexplained parts of
$\alpha_{i}$ and $\gamma_{i};$ and $\alpha_{a},\alpha_{\gamma},$
$\mathbf{\theta}_{\alpha}$ and $\mathbf{\theta}_{\gamma}$ are fixed
coefficients. There is an important difference between the two types of
variables, $\mathbf{z}_{i\alpha}$ and $\mathbf{z}_{i\gamma}$. Changes in the
latter could induce growth effects that are sustained, whilst the effects of
changing $\mathbf{z}_{i\alpha}$, could at best result in a once and for all
change in the level of output, without generating sustained growth effects.
Thus policies that change the value of $\gamma_{i}$ are likely to have more
fundamental impacts on output.

Abstracting from time-varying effects, and for simplicity assuming that
$\mathbf{z}_{i\alpha}$ and $\mathbf{z}_{i\gamma}$ have no variables in common,
then using (\ref{aizi}) and (\ref{gizi}) in (\ref{ECMHetro}) and averaging
over $t$ we have%
\begin{equation}
a_{iT}=\bar{y}_{i\circ}-%
%TCIMACRO{\tsum _{\ell=1}^{p}}%
%BeginExpansion
{\textstyle\sum_{\ell=1}^{p}}
%EndExpansion
\rho_{i,\ell}\bar{y}_{i,-\ell}=\alpha_{T}+\boldsymbol{\theta}_{T}^{\prime
}\mathbf{z}_{i}+\eta_{iT}+\bar{u}_{i\circ}, \label{CReg}%
\end{equation}
where $\bar{y}_{i,-\ell}=T^{-1}\sum_{t=1}^{T}y_{i,t-\ell}$, $\alpha_{T}%
=\alpha_{\alpha}+\alpha_{\gamma}\bar{f}_{T}$, $\boldsymbol{\theta}%
_{T}=(\boldsymbol{\theta}_{\alpha}^{\prime},\boldsymbol{\theta}_{\gamma
}^{\prime}\bar{f}_{T})^{\prime}$, $\mathbf{z}_{i}=$ $\left(  \mathbf{z}%
_{i\alpha}^{\prime},\mathbf{z}_{i\gamma}^{\prime}\right)  ^{\prime}$,
$\eta_{iT}=\eta_{i\alpha}+\eta_{i\gamma}\bar{f}_{T}$, and $\bar{u}_{i\circ
}=T^{-1}\sum_{t=1}^{T}u_{it}$. Equation (\ref{ECMCov}) could also be used to
filter the dynamics. Therefore, in principle the time-invariant effects,
$\boldsymbol{\theta}_{T}$, can be identified using cross-country regressions
where the dynamics have been filtered out, and $\mathbf{z}_{i}$ are
distributed independently of the composite errors, $\eta_{iT}+\bar{u}_{i\circ
}$.

We shall return to the problem of identification and estimation of
time-invariant effects in Section \ref{TINV}.

\section{Estimation of the speed of convergence by Barro regressions\label{BB}%
}

We begin with the Barro regression. This can be interpreted in terms of the
panel AR(1) model given by (\ref{FETE}) assuming $u_{it}$ are serially
uncorrelated and cross-sectionally independent, and the\ fixed effects,
$\alpha_{i}$, follow a random coefficient specification. The unconditional
Barro regression involves running a cross-country regression of $y_{iT}%
-y_{i0}$ on an intercept and $y_{i0}$. The estimate of $\phi=1-\rho$ is then
recovered from the least squares estimate of the coefficient of $y_{i0}$.
Specifically, the unconditional Barro regression can be written as
\begin{equation}
y_{iT}-y_{i0}=a_{T}-(1-\rho^{T})y_{i0}+v_{iT}, \label{BB0}%
\end{equation}
where it is assumed that the panel is balanced in the sense that $y_{i0}$ and
$y_{iT}$ are available for all $i$. In addition to assuming homogeneous
dynamics, the Barro regression also requires $y_{i0}$ and $v_{iT}$ are
independently distributed.

To investigate the relationship of Barro regressions to the TWFE panel data
model, we assume\footnote{Recall that under parallel trends, assumed when
using Barro regressions, $\gamma_{i}=\gamma$ and only $\alpha_{i}$ is affected
by time-invariant variables, $\mathbf{z}_{i}$}
\begin{equation}
\alpha_{i}=\alpha+\boldsymbol{\theta}^{\prime}\mathbf{z}_{i}+\eta_{i},
\label{ai0}%
\end{equation}
and eliminate the latent factor, $f_{t}$, from (\ref{FETE}) by use of cross
section averages. Averaging $y_{it}$ given by (\ref{FETE}), and augmented by
(\ref{ai0}) over $i$ we have
\begin{equation}
\bar{y}_{\circ t}=\alpha+\boldsymbol{\theta}^{\prime}\mathbf{\bar{z}}%
+\bar{\eta}+f_{t}+\rho\bar{y}_{\circ,t-1}+\bar{u}_{\circ t}, \label{ybart}%
\end{equation}
where $\bar{y}_{\circ t}=n^{-1}\sum_{i=1}^{n}y_{it}$, $\bar{u}_{\circ
t}=n^{-1}\sum_{i=1}^{n}u_{it}$, $\mathbf{\bar{z}}=n^{-1}\sum_{i=1}%
^{n}\mathbf{z}_{i},$ and $\bar{\eta}=n^{-1}\sum_{i=1}^{n}\eta_{i}$. Then
$f_{t}$ can be eliminated by subtracting (\ref{ybart}) from (\ref{FETE}) to
obtain
\begin{equation}
y_{it}-\bar{y}_{\circ t}=\boldsymbol{\theta}^{\prime}(\mathbf{z}%
_{i}-\mathbf{\bar{z}})+\rho\left(  y_{i,t-1}-\bar{y}_{\circ,t-1}\right)
+u_{it}-\bar{u}_{\circ t}+\left(  \eta_{i}-\bar{\eta}\right)  . \label{ARCCE}%
\end{equation}
Now solving forward from the initial values $y_{i0}-\bar{y}_{0}$ , where
$\bar{y}_{0}=n^{-1}\sum_{i=1}^{n}y_{i0}$, we have
\begin{align}
y_{iT}-\bar{y}_{\circ T}  &  =\rho^{T}\left(  y_{i0}-\bar{y}_{0}\right)
+\left(  \frac{1-\rho^{T}}{1-\rho}\right)  \left[  \boldsymbol{\theta}%
^{\prime}(\mathbf{z}_{i}-\mathbf{\bar{z}})+\left(  \eta_{i}-\bar{\eta}\right)
\right] \label{BBder0}\\
&  +\sum_{s=0}^{T-1}\rho^{s}\left(  u_{i,T-s}-\bar{u}_{\circ,T-s}\right)
,\nonumber
\end{align}
\textbf{ }After some rearrangements, (\ref{BBder0}) can be written
equivalently as
\begin{equation}
y_{iT}-y_{i0}=a_{T}+b_{T}y_{i0}+\mathbf{c}_{T}^{\prime}\mathbf{z}_{i}%
+\zeta_{iT}, \label{BB1}%
\end{equation}
where
\begin{equation}
a_{T}=\bar{y}_{\circ T}-\rho^{T}\bar{y}_{0}-\left(  \frac{1-\rho^{T}}{1-\rho
}\right)  \left(  \boldsymbol{\theta}^{\prime}\mathbf{\mathbf{\bar{z}}+}%
\bar{\eta}\right)  -\sum_{s=0}^{T-1}\rho^{s}\bar{u}_{\circ,T-s}, \label{cT}%
\end{equation}%
\begin{equation}
b_{T}=-(1-\rho^{T}),\text{ \ \ }\mathbf{c}_{T}=\left(  \frac{1-\rho^{T}%
}{1-\rho}\right)  \boldsymbol{\theta}\mathbf{,} \label{bT}%
\end{equation}
and
\begin{equation}
\zeta_{iT}=\left(  \frac{1-\rho^{T}}{1-\rho}\right)  \eta_{i}+\sum_{s=0}%
^{T-1}\rho^{s}u_{i,T-s}. \label{BBerror}%
\end{equation}
Comparing this representation with the Barro regression given by (\ref{BB0}),
we note that (\ref{BB1}) coincides with what is known as conditional Barro
regression and reduces to unconditional Barro regression when
$\boldsymbol{\theta}=\mathbf{0}$. The\textbf{ }parameters of Barro regressions
are $b_{T}$ and $\mathbf{c}_{T}$, and $\rho$ and $\boldsymbol{\theta}$ are not
directly estimated when one runs a standard conditional Barro regression.
However, $\phi$ and $\boldsymbol{\theta}$ can be estimated using (recalling
that $\left\vert \rho\right\vert <1$, by assumption)
\begin{align*}
\phi &  =1-\rho=1-\left(  1+b_{T}\right)  ^{1/T}\\
\boldsymbol{\theta}  &  =\left(  \frac{1-\rho}{1-\rho^{T}}\right)
\mathbf{c}_{T}=-\left[  \frac{1-\left(  1+b_{T}\right)  ^{1/T}}{b_{T}}\right]
\mathbf{c}_{T}.
\end{align*}
Clearly, $a_{T}$ and $\mathbf{c}_{T}$ varies with $T$ but not $\phi$ and
$\mathbf{\theta}$. For comparability with pooled panel estimates of
$\mathbf{\theta}$ when reporting Barro's estimates one needs to report
estimates of $\boldsymbol{\theta}$, in addition to the estimates of $b_{T}$
and $\mathbf{c}_{T}$ that are routinely reported in the literature. It is also
clear that estimates of $\phi$ and $\boldsymbol{\theta}$ based on Barro's
regression estimates of $b_{T}$ and $\mathbf{c}_{T}$ will be biased, due to
the non-zero correlations between $y_{i0}$ and $\zeta_{iT}$, and between
$y_{i0}$ and $\mathbf{z}_{i},$ as established below.

Considering (\ref{BBder0}), there are a number of distinct possible sources of
bias which come from the correlation between either of the regressors $y_{i0}$
and $\mathbf{z}_{i}$ and any of the three elements of the composite error term
of the Barro regression, namely $\left(  \frac{1-\rho^{T}}{1-\rho}\right)
\eta_{i},$ $\sum_{s=0}^{T-1}\rho^{s}u_{i,T-s}$ and $\sum_{s=0}^{T-1}\rho
^{s}\bar{u}_{T-s}.$ Below we show that as long as $\sigma_{\eta}^{2}\neq0,$
then any heterogeneity in the $\alpha_{i}$, even if it is random, will cause
bias because of the inevitable correlation between initial output $y_{i0}$ and
$\eta_{i},$ the random component of $\alpha_{i}$. In addition, serial
correlation in the $u_{it}$ will cause $y_{i0}$ to become correlated with
cross section averages of $u_{iT},u_{i,T-1},....$ In the case of
$\mathbf{z}_{i},$ while the shocks - pandemics, revolutions and the like - may
be exogenous, many of the elements of $\mathbf{z}_{i}$ are likely to be
adaptive to past shocks, causing them to be correlated with them.

\subsection{Asymptotic bias of the Barro estimator\label{BBbias}}

For the OLS estimator of $b_{T}$ from (\ref{BB1}) to be consistent it is
required that $E\left(  y_{i0}\zeta_{iT}\right)  =0$, for $i=1,2,...,n;$
irrespective of whether we consider unconditional or conditional Barro
regressions. When conditioning variables are added to the regression it is
also required that $E(\mathbf{z}_{i}\zeta_{iT})=0$. Both of these conditions
involve restrictions on the way initial values, $y_{i0},$ are generated. Since
the random component of $\alpha_{i}$, namely $\eta_{i}$, is specific to
country $i$ it is likely that $y_{i0}$ will also depend on $\eta_{i}$. Such a
dependence arises irrespective of whether the $\left\{  y_{it}\right\}  $
processes have started from a finite past prior to the initial date $t=0$, or
from a very distant past. It also does not matter if the countries differ in
their start dates, say at time $t=-M_{i}$, where $M_{i}$ is a positive
integer. In this general set up, supposing country $i$ was established
(started) with $y_{i,-M_{i}}$ at time $t=-M_{i}$, then iterating
(\ref{FETE})\ forward from $y_{i,-M_{i}}$, (similar to the way equation
(\ref{BBder0}) was derived) we obtain
\begin{align}
y_{i0}-\bar{y}_{0}  &  =\rho^{M_{i}}(y_{i,-M_{i}}-\bar{y}_{\circ,-M_{i}%
})+\left(  \frac{1-\rho^{M_{i}}}{1-\rho}\right)  \left[  \boldsymbol{\theta
}^{\prime}(\mathbf{z}_{i}-\mathbf{\bar{z}})+\left(  \eta_{i}-\bar{\eta
}\right)  \right] \label{yi0}\\
&  +\sum_{s=0}^{M_{i}-1}\rho^{s}(u_{i,-s}-u_{\circ,-s}).\nonumber
\end{align}
Under the panel AR(1) model (\ref{FETE}) $u_{it}$ are serially uncorrelated.
As a first order approximation it is also plausible to assume that $u_{it}$ is
distributed independently of $f_{t}$, $\eta_{i}$ and $\mathbf{z}_{i}$. Even
under these assumptions, using (\ref{BBerror}), we have a non-zero correlation
between $y_{i0}$ and $\zeta_{iT}$. Specifically,
\begin{equation}
E\left(  y_{i0}\zeta_{iT}\right)  =\left(  \frac{1-\rho^{T}}{1-\rho}\right)
\left(  \frac{1-\rho^{M_{i}}}{1-\rho}\right)  \sigma_{\eta}^{2},
\label{Covyi0egi}%
\end{equation}
where $\sigma_{\eta}^{2}=var(\eta_{i})$, and renders Barro regression
estimators of $\phi$ and $\boldsymbol{\theta}$ inconsistent if $\sigma_{\eta
}^{2}>0$, irrespective of whether the regression is estimated unconditionally
or conditional on $\mathbf{z}_{i}$. The magnitude of the bias could depend on
the conditioning variables used and whether they are uncorrelated with
$u_{it}$.\footnote{The endogeneity of the conditioning variables, which has
been widely discussed in the literature, is not essential for the bias which
arises if $\sigma_{\eta}^{2}>0,$ which can be allowed for by using TWFE in a
model with parallel trends.}

We now derive the asymptotic bias of the unconditional Barro estimator of
$\rho_{0}$ the true value of $\rho,$ which is obtained as the least squares
estimator of $b_{T}=-(1-\rho_{0}^{T})$ from the OLS regression of
$y_{iT}-y_{i0}$ on $y_{i0}-\bar{y}_{0}$, namely
\[
\hat{b}_{T}=\frac{\sum_{i=1}^{n}\left(  y_{i0}-\bar{y}_{0}\right)  \left(
y_{iT}-y_{i0}\right)  }{\sum_{i=1}^{n}\left(  y_{i0}-\bar{y}_{0}\right)  ^{2}%
},
\]
recalling that $\bar{y}_{\circ0}=$ $n^{-1}\sum_{i=1}^{n}y_{i0}$. Then using
(\ref{BB1})
\begin{align*}
\hat{b}_{T}-b_{T}  &  =\frac{n^{-1}\sum_{i=1}^{n}\left(  y_{i0}-\bar{y}%
_{0}\right)  \left(  \mathbf{c}_{T}^{\prime}\mathbf{z}_{i}+\zeta_{iT}\right)
}{n^{-1}\sum_{i=1}^{n}\left(  y_{i0}-\bar{y}_{0}\right)  ^{2}}\\
&  \rightarrow_{p}\frac{\lim_{n\rightarrow\infty}n^{-1}\sum_{i=1}^{n}E\left[
\left(  y_{i0}-\bar{y}_{0}\right)  \left(  \mathbf{c}_{T}^{\prime}%
\mathbf{z}_{i}+\zeta_{iT}\right)  \right]  }{\lim_{n\rightarrow\infty}%
n^{-1}\sum_{i=1}^{n}E\left(  y_{i0}-\bar{y}_{0}\right)  ^{2}}.
\end{align*}
To simplify the derivations we assume that $M_{i}\rightarrow\infty$ for all
$i$ (all processes have started in a distant past with $\left\vert
\rho\right\vert <1$) and note that in this case, using (\ref{yi0}), we have
\[
y_{i0}-\bar{y}_{0}=\left(  \frac{1}{1-\rho_{0}}\right)  \left[
\boldsymbol{\theta}^{\prime}\left(  \mathbf{z}_{i}-\mathbf{\bar{z}}\right)
+\eta_{i}-\bar{\eta}\right]  +\sum_{s=0}^{\infty}\rho_{0}^{s}\left(
u_{i,-s}-\bar{u}_{\circ,-s}\right)  .
\]
Using this result together with (\ref{BBerror}) and noting that $u_{it}$ and
$\eta_{i}$ are assumed to be cross-sectionally independent we obtain
\begin{equation}
E\left[  \left(  y_{i0}-\bar{y}_{0}\right)  \left(  \mathbf{c}_{T}^{\prime
}\mathbf{z}_{i}+\zeta_{iT}\right)  \right]  =\frac{\left(  1-\rho_{0}%
^{T}\right)  }{\left(  1-\rho_{0}\right)  ^{2}}\left[  \boldsymbol{\theta
}^{\prime}Var\left(  \mathbf{z}_{i}\right)  \boldsymbol{\theta}\mathbf{+}%
\sigma_{\eta}^{2}-E\left(  \eta_{i}\bar{\eta}\right)  \right]  .
\label{covyi0rgzi}%
\end{equation}
But since $\eta_{i}$ are independently distributed then $E\left(  \eta_{i}%
\bar{\eta}\right)  =n^{-1}\sigma_{\eta}^{2}$. Similarly
\[
E\left(  y_{i0}-\bar{y}_{0}\right)  ^{2}=\frac{1-\rho_{0}^{T}}{\left(
1-\rho_{0}\right)  ^{2}}\left[  \boldsymbol{\theta}^{\prime}Var\left(
\mathbf{z}_{i}\right)  \boldsymbol{\theta}\mathbf{+}\sigma_{\eta}^{2}\right]
+\frac{\sigma_{iu}^{2}}{1-\rho_{0}^{2}}+O(n^{-1}).
\]
Hence, as $n\rightarrow\infty$
\begin{equation}
\hat{b}_{T}-b_{T}=\hat{\rho}^{T}-\rho_{0}^{T}\rightarrow_{p}\frac
{\frac{\left(  1-\rho_{0}^{T}\right)  \left(  1+\rho_{0}\right)  }{1-\rho
}\left(  \frac{\boldsymbol{\theta}^{\prime}\boldsymbol{\Omega}_{z}%
\boldsymbol{\theta}\mathbf{+}\sigma_{\eta}^{2}}{\bar{\sigma}_{u}^{2}}\right)
}{1+\left(  \frac{1+\rho_{0}}{1-\rho_{0}}\right)  \left(  \frac
{\boldsymbol{\theta}^{\prime}\boldsymbol{\Omega}_{z}\boldsymbol{\theta
}\mathbf{+}\sigma_{\eta}^{2}}{\bar{\sigma}_{u}^{2}}\right)  }>0, \label{BIASB}%
\end{equation}
where
\[
\mathbf{\Omega}_{z}=\lim_{n\rightarrow\infty}n^{-1}\sum_{i=1}^{n}Var\left(
\mathbf{z}_{i}\right)  \text{, and }\bar{\sigma}_{u}^{2}=\lim_{n\rightarrow
\infty}n^{-1}\sum_{i=1}^{n}\sigma_{iu}^{2}%
\]

The magnitude of the bias, $\hat{\rho}^{T}-\rho_{0}^{T},$ depends on $\rho
_{0},$ $T$ and the ratio $\left(  \boldsymbol{\theta}^{\prime}%
\boldsymbol{\Omega}_{z}\boldsymbol{\theta}\mathbf{+}\sigma_{\eta}^{2}\right)
/\bar{\sigma}_{u}^{2}$, which represent the size of cross section dispersions
relative to the average of the time series dispersions. If we set this ratio
to unity, the bias of the Barro estimator is $(1-\rho_{0}^{T})(1+\rho_{0})/2$,
which could be substantial considering that $\rho_{0}$ is likely to be close
to $1$. For instance, if $\rho_{0}=0.8,$ and $T=10$, then the asymptotic bias
of estimating $\rho_{0}^{T}$ is given by $(1-\rho_{0}^{T})(1+\rho
_{0})/2=0.803,$which implies $\hat{\rho}\rightarrow_{p}0.9907,$ namely a
convergence rate of $1$ percent instead of the actual value of $20$ percent.
Also the bias in $\hat{b}_{T}$ is upward for any $T$, and rises with $T$ since
$\left\vert \rho\right\vert <1$. The longer the time series data the more
pronounced the bias of the Barro\ estimator will become.\footnote{Kremer et
al. (2022, p338) using $T=10$ found divergence in income per capita at a rate
of $0.5$ percent annually for 1985 to 1995, a convergence at rate of $0.7$
percent over the more recent period 2005-2015.}

By conditioning on the observed characteristics, it is possible to reduce the
heterogeneity bias of the Barro estimator if the conditioning variables,
$\mathbf{z}_{i}$, are independently distributed of the shocks, $u_{it}$. The
larger the size of $\boldsymbol{\theta}^{\prime}\boldsymbol{\Omega}%
_{z}\boldsymbol{\theta}/\bar{\sigma}_{u}^{2}$ the greater the scope for bias
reduction by means of conditioning.

In short, the Barro regression can only deliver consistent estimates of the
speed of convergence of output across countries, if $\alpha_{i}$ is a
\textit{known deterministic }function of $\mathbf{z}_{i}$, the errors of the
underlying panel AR(1) model are serially uncorrelated, and the idiosyncratic
errors are unrelated to all time-invariant regressors such as climate,
latitude, demographic and institutional factors, for example. These are very
strong assumptions and are unlikely to hold.

Even if it is argued that $\alpha_{i}$ is a deterministic function of a known
vector of time-invariant variables $\mathbf{z}_{i}$ and consequently
$\sigma_{\eta}^{2}=0,$ the Barro estimator could still be inconsistent due
possible correlations between $\mathbf{z}_{i}$ and current and past values of
the idiosyncratic errors, $\left\{  u_{i,T+1},u_{i,T},....,u_{i0},...\right\}
$, as well as error serial correlation. Derivation of the Barro cross-country
regression requires the underlying panel data model to follow an AR(1)
specification and cannot accommodate higher order dynamics. As a result if
such higher order dynamics are ignored, the errors in (\ref{FETE}), $u_{it}$,
will become serially correlated and the assumption that $y_{i0}$ and
$\zeta_{iT}$ in (\ref{BB1}) are uncorrelated will not hold any longer. The
non-zero correlation between $y_{i0}$ and $\zeta_{iT}$ also renders the least
squares estimates of the time-invariant effects, $\boldsymbol{\theta}$
inconsistent if $\sigma_{\eta}^{2}>0$ and/or higher order dynamics are required.

Whilst random coefficient models are good at dealing with parameter
heterogeneity in the context of static panels, assuming random effects
(intercepts and/or slopes) in dynamic panels invariably lead to inconsistent
estimates when heterogeneity is ignored. The Barro regression basically fails
because it cannot allow for random differences across countries. All
randomness must be taken into account by assuming an exact relationship
between $\alpha_{i}$ and $\mathbf{z}_{i}$.

\section{Asymptotic bias of TWFE estimator in presence of nonparallel
trends\label{HetCon}}

To avoid the bias of Barro regressions we need to turn to panel data
techniques, the most popular of which is TWFE specification, given by
(\ref{FETE}). But, as is widely acknowledged, an important limitation of the
TWFE estimator is the parallel trends assumption that requires the time
effects, $f_{t}$, have the same impact on all countries.\footnote{The parallel
trends is also a core assumption in causal inference, particularly in the
Difference-in-Difference (DiD) analysis where it is assumed that the trend
components of the outcomes for treated and control groups are the same. In the
case of current application, the analysis is further complicated due to the
dynamics, often abstracted from in DiD analysis.} This imposes identical
steady state growth rates across all countries, irrespective of their
differences in geography, endowments, and climatic conditions. Countries also
differ in degree of openness to flows of goods, capital, people and ideas and
these factors will influence the extent to which technology diffuses across
countries. The obstacles to such flows may come from domestic institutions or
from external constraints. Sanctions for instance are designed to cut
countries off. Any similarities in the effects of latent factors across
countries should be empirically investigated rather than assumed, \textit{a
priori}.

Starting from (\ref{FETE}), we relax the parallel trends assumption but for
now retain the slope homogeneity assumption, $\rho_{i}=\rho_{0},$ where
$\rho_{0}$ is the true value of $\rho$. Specifically, we suppose that
\begin{equation}
y_{it}=\alpha_{i}+\gamma_{i}f_{t}+\rho_{0}y_{i,t-1}+u_{it}, \label{IFE}%
\end{equation}
where $\left\vert \rho_{0}\right\vert <1$. We first consider the implications
of assuming parallel trends when in reality the effects of the global factor,
$f_{t}$, on country outputs are heterogeneous, namely $Var\left(  \gamma
_{i}\right)  =\sigma_{\gamma}^{2}>0$. We also assume that $f_{t}$ is strong in
the sense that its effects is pervasive across countries and $E\left(
\gamma_{i}\right)  =\lim_{n\rightarrow\infty}n^{-1}\sum_{i=1}^{n}\gamma_{i}%
>0$. This condition is necessary for the global output to have a non-zero mean
growth rate defined by $\bar{g}_{t}=n^{-1}\sum_{i=1}^{n}\Delta y_{it}$.

The TWFE estimator of $\rho_{0}$ is given by
\begin{equation}
\hat{\rho}_{_{TWFE}}=\frac{(\mathbf{y}_{i,-1}-\mathbf{\bar{y}}_{-1})^{\prime
}\mathbf{M}_{T}(\mathbf{y}_{i\circ}-\mathbf{\bar{y}})}{\sum_{i=1}%
^{n}(\mathbf{y}_{i,-1}-\mathbf{\bar{y}}_{-1})^{\prime}\mathbf{M}%
_{T}(\mathbf{y}_{i,-1}-\mathbf{\bar{y}}_{-1})}, \label{rhoFETE}%
\end{equation}
where $\mathbf{M}_{T}=\mathbf{I}_{T}-T^{-1}\mathbf{\tau}_{T}\mathbf{\tau}%
_{T}^{\prime},$ $\mathbf{\tau}_{T}$ is a $T\times1$ vector of ones,
$\mathbf{y}_{i\circ}=(y_{i1},y_{i2},...,y_{iT})^{\prime}$, $\mathbf{y}%
_{i,-1}=(y_{i0},y_{i1},...,y_{i,T-1})^{\prime}$, $\mathbf{\bar{y}=}(\bar
{y}_{\circ1},\bar{y}_{\circ2},...,\bar{y}_{\circ T})^{\prime}$, $\mathbf{\bar
{y}}_{-1}\mathbf{=}(\bar{y}_{\circ0},\bar{y}_{\circ1},...,\bar{y}_{\circ
,T-1})^{\prime}$, and $\bar{y}_{\circ t}=n^{-1}\sum_{i=1}^{n}y_{it}$. To
derive an expression for the bias of $\hat{\rho}_{TWFE}$ we assume that
$f_{t}$ that $u_{it}$ are independently distributed, $u_{it}$ are
cross-sectionally independent. Under (\ref{IFE})
\begin{equation}
\mathbf{y}_{i\circ}-\mathbf{\bar{y}=}\left(  \alpha_{i}-\bar{\alpha}\right)
\mathbf{\tau}_{T}+\left(  \gamma_{i}-\bar{\gamma}\right)  \mathbf{f+}\rho
_{0}(\mathbf{y}_{i,-1}-\mathbf{\bar{y}}_{-1})+\mathbf{u}_{i\circ}%
-\mathbf{\bar{u}} \label{gap1}%
\end{equation}
where $\mathbf{f=(}f_{1},f_{2},...,f_{T})^{\prime}$, $\mathbf{\bar{u}=(}%
\bar{u}_{\circ1},\bar{u}_{\circ2},...,\bar{u}_{\circ T})^{\prime}$, with
$\bar{u}_{\circ t}=n^{-1}\sum_{i=1}^{n}u_{it}$, Using it in (\ref{rhoFETE})
\begin{equation}
\hat{\rho}_{TWFE}-\rho_{0}=\frac{T^{-1}n^{-1}\sum_{i=1}^{n}(\mathbf{y}%
_{i,-1}-\mathbf{\bar{y}}_{-1})^{\prime}\mathbf{M}_{T}\left[  \left(
\gamma_{i}-\bar{\gamma}\right)  \mathbf{f}+\mathbf{u}_{i\circ}-\mathbf{\bar
{u}}\right]  }{T^{-1}n^{-1}\sum_{i=1}^{n}(\mathbf{y}_{i,-1}-\mathbf{\bar{y}%
}_{-1})^{\prime}\mathbf{M}_{T}(\mathbf{y}_{i,-1}-\mathbf{\bar{y}}_{-1})}.
\label{FETEbias}%
\end{equation}
Also solving (\ref{IFE}) forward, from an initial value in a distant past, we
have
\begin{equation}
y_{it}=\alpha_{i}/(1-\rho)+\gamma_{i}g_{t}+v_{it}, \label{yig}%
\end{equation}
where
\begin{equation}
g_{t}=%
%TCIMACRO{\dsum \limits_{j=0}^{\infty}}%
%BeginExpansion
{\displaystyle\sum\limits_{j=0}^{\infty}}
%EndExpansion
\rho^{j}f_{t-j},\text{ and }v_{it}=%
%TCIMACRO{\dsum \limits_{j=0}^{\infty}}%
%BeginExpansion
{\displaystyle\sum\limits_{j=0}^{\infty}}
%EndExpansion
\rho^{j}u_{i,t-j}, \label{gtvit}%
\end{equation}
which in turn gives
\begin{equation}
\mathbf{y}_{i,-1}-\mathbf{\bar{y}}_{-1}=\left(  \gamma_{i}-\bar{\gamma}%
_{n}\right)  \mathbf{g}_{-1}+\mathbf{v}_{i,-1}-\mathbf{\bar{v}}_{-1},
\label{gapy}%
\end{equation}
where $\bar{\gamma}_{n}=n^{-1}\sum_{i=1}^{n}\gamma_{i}$, $\mathbf{g}%
_{-1}\mathbf{=(}g_{0},g_{1},...,g_{T-1})^{\prime}$, and $\mathbf{v}_{i,-1}$
and $\mathbf{\bar{v}}_{-1}$ are defined analogously to $\mathbf{y}_{i,-1}$ and
$\mathbf{\bar{y}}_{-1}$. Using this result, the numerator of (\ref{FETEbias})
becomes%
\begin{align*}
Num_{TWFE}=\left(  n^{-1}\sum_{i=1}^{n}\left(  \gamma_{i}-\bar{\gamma}%
_{n}\right)  ^{2}\right)  \left(  T^{-1}\mathbf{g}_{-1}^{\prime}\mathbf{M}%
_{T}\mathbf{f}\right)   &  \mathbf{+}T^{-1}n^{-1}\sum_{i=1}^{n}\mathbf{\tilde
{v}}_{i,-1}^{\prime}\mathbf{M}_{T}\left(  \mathbf{u}_{i\circ}-\mathbf{\bar{u}%
}\right) \\
&  +T^{-1}n^{-1}\sum_{i=1}^{n}\left(  \gamma_{i}-\bar{\gamma}\right)
\mathbf{g}_{-1}^{\prime}\mathbf{M}_{T}\left(  \mathbf{u}_{i\circ}%
-\mathbf{\bar{u}}\right)  \mathbf{.}%
\end{align*}
where $\mathbf{\tilde{v}}_{i,-1}=\mathbf{v}_{i,-1}-\mathbf{\bar{v}}_{-1}$.
Since by assumption $u_{it}$ and $f_{t}$ (and hence $g_{t})$ are independently
distributed, and $u_{it}$ are cross-sectionally independent, then for a fixed
$T$ we have%
\begin{equation}
Num_{TWFE}=\sigma_{\gamma}^{2}\left(  T^{-1}\mathbf{g}_{-1}^{\prime}%
\mathbf{M}_{T}\mathbf{f}\right)  +T^{-1}\lim_{n\rightarrow\infty}n^{-1}%
\sum_{i=1}^{n}E\left[  \mathbf{\tilde{v}}_{i,-1}^{\prime}\mathbf{M}_{T}\left(
\mathbf{u}_{i\circ}-\mathbf{\bar{u}}\right)  \right]  +O_{p}\left(
n^{-1/2}\right)  , \label{Num2}%
\end{equation}
where $\sigma_{\gamma}^{2}=\lim_{n\rightarrow\infty}n^{-1}\sum_{i=1}%
^{n}E\left(  \gamma_{i}-\bar{\gamma}_{n}\right)  ^{2}$.

Consider now the denominator of (\ref{FETEbias}) and using (\ref{gapy}) note
that%
\begin{align}
Den_{TWFE}  &  =T^{-1}n^{-1}\sum_{i=1}^{n}(\mathbf{y}_{i,-1}-\mathbf{\bar{y}%
}_{-1})^{\prime}\mathbf{M}_{T}(\mathbf{y}_{i,-1}-\mathbf{\bar{y}}%
_{-1})\label{Den1}\\
&  =\sigma_{\gamma}^{2}\left(  T^{-1}\mathbf{g}_{-1}^{\prime}\mathbf{M}%
_{T}\mathbf{g}_{-1}\right)  +T^{-1}\lim_{n\rightarrow\infty}n^{-1}\sum
_{i=1}^{n}E\left(  \mathbf{\tilde{v}}_{i,-1}^{\prime}\mathbf{M}_{T}%
\mathbf{\tilde{v}}_{i,-1}^{\prime}\right)  +O_{p}\left(  n^{-1/2}\right)
\nonumber
\end{align}
Using (\ref{Num2}) and (\ref{Den1}) in (\ref{FETEbias}) now yields (for a
fixed $T$)%
\begin{equation}
p\lim_{n\rightarrow\infty}\left(  \hat{\rho}_{TWFE}-\rho_{0}\right)
=\frac{\sigma_{\gamma}^{2}\left(  T^{-1}\mathbf{g}_{-1}^{\prime}\mathbf{M}%
_{T}\mathbf{f}\right)  +\lim_{n\rightarrow\infty}\left(  nT\right)  ^{-1}%
\sum_{i=1}^{n}E\left[  \mathbf{\tilde{v}}_{i,-1}^{\prime}\mathbf{M}_{T}\left(
\mathbf{u}_{i\circ}-\mathbf{\bar{u}}\right)  \right]  }{\sigma_{\gamma}%
^{2}\left(  T^{-1}\mathbf{g}_{-1}^{\prime}\mathbf{M}_{T}\mathbf{g}%
_{-1}\right)  +\lim_{n\rightarrow\infty}\left(  nT\right)  ^{-1}\sum_{i=1}%
^{n}E\left(  \mathbf{\tilde{v}}_{i,-1}^{\prime}\mathbf{M}_{T}\mathbf{\tilde
{v}}_{i,-1}^{\prime}\right)  } \label{plimrho}%
\end{equation}
When the parallel trend assumption does not hold $\sigma_{\gamma}^{2}>0$ and
in general the TWFE estimator of $\rho_{0}$ will be biased.

The second term in the numerator of (\ref{plimrho}) is similar to the Nickell
(1981) bias and is of order $T^{-1}$, and vanishes as $T\rightarrow\infty$. It
is present whether there is a factor or not. The first term arises from the
presence of the factor and the magnitude of the asymptotic bias of the TWFE
estimator critically depends on the time series properties of $f_{t}$. To
highlight this point we consider two cases: $(a)$ when $f_{t}$ is covariance
stationary, (b) when $f_{t}$ is trended or follows a random walk with a
non-zero drift, namely $f_{t}=\mu t+s_{t}$ ($\mu\neq0$). The latter
specification is particularly relevant to the growth convergence literature
where it is important to allow for the possibility of a non-zero steady state
output growth for the global economy.

\subsection{The bias of the TWFE estimator under nonparallel trends with a
stationary latent factor}

Under the assumption of a stationary factor, without loss of generality, we
assume that $f_{t}$ has mean zero with the autocovariance function,
$\gamma_{f}(\left\vert s-s^{\prime}\right\vert )=E(f_{t-s}f_{t-s^{\prime}})$.
In this case, using (\ref{gtvit}) we have
\begin{align*}
T^{-1}\mathbf{g}_{-1}^{\prime}\mathbf{M}_{T}\mathbf{f}  &  =T^{-1}\sum
_{t=1}^{T}f_{t}g_{t-1}-\bar{f}\bar{g}_{-1}\rightarrow_{p}\sum_{s=0}^{\infty
}\rho^{s}\gamma_{f}(s+1),\\
T^{-1}\mathbf{g}_{-1}^{\prime}\mathbf{M}_{T}\mathbf{g}_{-1}  &  =T^{-1}%
\sum_{t=1}^{T}g_{t-1}^{2}-\bar{g}_{-1}^{2}\rightarrow_{p}\sum_{s=0}^{\infty
}\sum_{s^{\prime}=0}^{\infty}\rho^{s+s^{\prime}}\gamma_{f}(\left\vert
s-s^{\prime}\right\vert )>0.
\end{align*}
Therefore, when the parallel trends assumption does not hold ($\bar{\sigma
}_{\gamma}^{2}>0)$ and $f_{t}$ is covariance stationary, the asymptotic bias
of $\hat{\rho}_{TWFE}$ for \textit{both} $n$ and $T$ large is given
\begin{equation}
p\lim_{n,T\rightarrow\infty}\left(  \hat{\rho}_{TWFE}-\rho_{0}\right)
=\frac{\left(  \frac{\sigma_{\gamma}^{2}}{\omega^{2}}\right)  \sum
_{s=0}^{\infty}\rho_{0}^{s}\gamma_{f}(s+1)}{1+\left(  \frac{\sigma_{\gamma
}^{2}}{\omega^{2}}\right)  \sum_{s=0}^{\infty}\sum_{s^{\prime}=0}^{\infty}%
\rho_{0}^{s+s^{\prime}}\gamma_{f}(\left\vert s-s^{\prime}\right\vert )},
\label{BiasStat}%
\end{equation}
where
\[
\omega^{2}=\lim_{n,T\rightarrow\infty}T^{-1}n^{-1}\sum_{i=1}^{n}E\left[
\left(  \mathbf{v}_{i,-1}-\mathbf{\bar{v}}_{-1}\right)  ^{\prime}%
\mathbf{M}_{T}\left(  \mathbf{v}_{i,-1}-\mathbf{\bar{v}}_{-1}\right)  \right]
>0.
\]
The sign of this bias depends on the sign of $\sum_{s=0}^{\infty}\rho_{0}%
^{s}\gamma_{f}(s+1)$, which is likely to be positive since $\rho_{0}>0,$ and
one would expect $f_{t}$ to be positively autocorrelated. The magnitude of the
bias depends on the relative importance of the heterogeneity of $\gamma_{i}$,
given by $\kappa^{2}=\sigma_{\gamma}^{2}/\omega^{2}$, and the degree of
persistence of the time effects, given by $\sum_{s=0}^{\infty}\rho_{0}%
^{s}\gamma_{f}(s+1)$. The TWFE estimator is asymptotically unbiased (as $n$
and $T\rightarrow\infty$) only if $f_{t}$ is serially independent. Everaert
and De Groote (2016) derive expressions for the inconsistency of TWFE
estimators under (\ref{IFE}) assuming that $f_{t}$ follows a stationary AR(1) process.

\subsection{The bias of the TWFE estimator under non parallel trends with a
trended latent factor}

Consider now the case where $f_{t}$ is trended, and suppose that $f_{t}=\mu
t+s_{t}$, which is decomposed into a linear trend, $\mu t$, and a stochastic
component $s_{t}$. The latter could be stationary or a unit root process. In
either case $s_{t}$ will be dominated by the trend component, so long as the
drift term $\mu\neq0$. Then it is easily seen that
\[
g_{t-1}-\bar{g}_{-1}=\frac{\mu}{1-\rho_{0}}\left(  t-\frac{T+1}{2}\right)
+\left(  q_{t-1}-\bar{q}_{-1}\right)  ,
\]
where $q_{t}=\rho_{0}q_{t-1}+s_{t}$. Since $\left\vert \rho_{0}\right\vert
<1,$ then $q_{t}$ will have the same order of integration as $s_{t},$ namely
$q_{t}$ could be stationary or a unit root process. In either case the linear
trend in $g_{t-1}-\bar{g}_{-1}$ will dominate $\left(  q_{t-1}-\bar{q}%
_{-1}\right)  $ and we have
\[
T^{-3}\mathbf{g}_{-1}^{\prime}\mathbf{M}_{T}\mathbf{f}=\frac{\mu^{2}%
}{12(1-\rho_{0})}+O_{p}(T^{-1}),\text{ and }T^{-3}\mathbf{g}_{-1}^{\prime
}\mathbf{M}_{T}\mathbf{g}_{-1}=\frac{\mu^{2}}{12(1-\rho_{0})^{2}}+O_{p}%
(T^{-1}).
\]
Hence, scaling down the terms in (\ref{plimrho}) by $T^{-3}$, and using the
above results it follows that $p\lim_{n,T\rightarrow\infty}\left(  \hat{\rho
}_{TWFE}-\rho_{0}\right)  =1-\rho_{0}$, or%
\begin{equation}
p\lim_{n,T\rightarrow\infty}\left(  \hat{\rho}_{TWFE}\right)  =1\text{. }
\label{RWD}%
\end{equation}
This is a striking result that follows from the deterministic trend component
of $f_{t}$\thinspace, and holds irrespective of whether the stochastic
component, $s_{t}$, is stationary or a unit root process.

In short, the TWFE estimator of $\rho_{0}$ is biased upward under nonparallel
trends and the bias becomes highly pronounced when $f_{t}$ is trended. In the
latter case even a small degree of heterogeneity in $\gamma_{i}$ can lead to
substantial upward bias in estimation of $\rho_{0}$. In the context of growth
regressions, the random walk model with drift is likely to be the most
relevant case, and largely explains the very small estimates obtained in the
literature for the speed of convergence, $\phi_{0}=1-\rho_{0},$ when TWFE
estimators are used.

\subsection{Panel data models with group-specific time effects\label{GStrend}}

The restrictive nature of the parallel trend assumption is becoming increasing
recognized in the empirical growth literature. For example in addition to
using TWFE, Acemoglu et al. (2019) let the time effect vary over groups. For
instance, in column 8 of Table 4, they interact the time effects with the
product of 7 regional dummies and the initial regime (democratic or non
democratic). Having 14 groups in each year leaves very small numbers in some groups.

In general, the panel data model with group-specific interactive effects can
be written as
\[
y_{it}=\alpha_{i}+\gamma_{g}f_{t}\left(  \sum_{g=1}^{G}d_{ig}\right)
+\rho_{0}y_{i,t-1}+u_{it},
\]
where $d_{ig}=1$ if country $i$ belongs to group (region) $g=1,2,...,G$, and
$zero$ otherwise. The number of groups, $G$, and the group membership is
typically assumed known.\footnote{For large $n$ and $T$ panels, machine
learning techniques, such as cluster analysis, can be used to estimate the
number of groups and the group membership. Such grouping is particularly
relevant if the aim is to identify convergence clubs that are based on
geographical proximity and political and historical similarities.} As before,
$f_{t}$ is the common latent factor. This model can be written more compactly
as
\begin{equation}
y_{ig,t}=\alpha_{i,g}+\gamma_{g}f_{t}+\rho_{0}y_{i,g,t-1}+u_{i,g,t}
\label{group_time}%
\end{equation}
for country $\left(  i,g\right)  $, that denotes country $i$ in group $g,$
with $i=1,2,...,n_{g},$ $g=1,2,...,G$, and $n=%
%TCIMACRO{\tsum _{g=1}^{G}}%
%BeginExpansion
{\textstyle\sum_{g=1}^{G}}
%EndExpansion
n_{g}$. The fixed effects, $\alpha_{i,g}$, and the group-specific time effects
can be eliminated using the group-specific filtering, with $\rho_{0}$
estimated by
\begin{equation}
\hat{\rho}_{FE-GTE}=\frac{\sum_{t=1}^{T}\sum_{g=1}^{G}\sum_{i=1}^{n_{g}}%
\tilde{y}_{ig,t-1}\tilde{y}_{ig,t}}{\sum_{t=1}^{T}\sum_{g=1}^{G}\sum
_{i=1}^{n_{g}}\tilde{y}_{ig,t-1}^{2}} \label{rhoFE-GTE}%
\end{equation}
where $\tilde{y}_{ig,t}=\left(  y_{ig,t}-\bar{y}_{ig,\circ}\right)  -(\bar
{y}_{\circ,gt}-\bar{y}_{\circ g\circ})$ and $\tilde{y}_{ig,t-1}=\left(
y_{ig,t-1}-\bar{y}_{ig,-1}\right)  -(\bar{y}_{\circ,gt}-\bar{y}_{\circ
g,-1}),$ where $\bar{y}_{ig,-s}=T^{-1}\sum_{t=1}^{T}y_{ig,t-s}$, $\bar
{y}_{\circ,gt}=n_{g}^{-1}\sum_{i=1}^{n_{g}}y_{ig,t}$, and $\bar{y}_{\circ
g,-s}=n_{g}^{-1}\sum_{i=1}^{n_{g}}y_{ig,t-s}$ for $s=0$ and $1$.

As an estimator of $\rho_{0}$, $\hat{\rho}_{FE-GTE}$ is subject to the Nickell
bias as well as the group-size bias. For a fixed $G$, $\hat{\rho}_{FE-GTE}$ is
a consistent estimator of $\rho_{0}$ if $\left(  T,n_{1},n_{2},...,n_{G}%
\right)  \rightarrow\infty$. Under this condition group-specific time effects,
$\gamma_{g},$ can be consistently estimated and hence eliminated form the
analysis. The key condition on $n_{g}$, namely $n_{g}\rightarrow\infty$, for
$g=1,2,...,G$ is likely to be restrictive in the case of cross country
analysis and arises due to the dynamic nature of the panel data model and will
not be required in the case of static panel data models with strictly
exogenous regressors.

\section{The DCCEP estimator: nonparallel trends with homogeneous
dynamics\label{DCCEP}}

The use of group-specific time effects can be avoided by following the common
correlated effects (CCE) approach introduced in Pesaran (2006) and later
implemented for dynamic panels by Chudik and Pesaran (2015), referred to as
dynamic CCE (DCCE). The basic idea behind DCCE is to proxy the latent factor,
$f_{t}$, by current and lagged cross country averages of log output, namely
$\bar{y}_{\circ t}$ and $\bar{y}_{\circ,t-1}$. The case with a homogeneous,
$\rho$ is referred to as DCCEP. Averaging (\ref{IFE}) over $i$, we have
$\bar{y}_{\circ t}=\bar{\alpha}+\bar{\gamma}f_{t}+\rho\bar{y}_{\circ,t-1}%
+\bar{u}_{\circ t}$, and assuming $\bar{\gamma}\neq0$, then $f_{t}=\bar
{\gamma}^{-1}\left(  \bar{y}_{\circ t}-\rho\bar{y}_{\circ,t-1}-\bar{\alpha
}-\bar{u}_{\circ t}\right)  $. Using this result back in (\ref{IFE}) now
yields%
\begin{equation}
y_{it}=\left(  \alpha_{i}-\delta_{i0}\bar{\alpha}\right)  +\delta_{i0}\bar
{y}_{\circ t}+\delta_{i1}\bar{y}_{\circ,t-1}+\rho y_{i,t-1}+u_{it}-\delta
_{i0}\bar{u}_{\circ t}, \label{IFEAUG}%
\end{equation}
where $\delta_{i0}=\bar{\gamma}^{-1}\gamma_{i}$, $\delta_{i1}=-\delta_{i0}%
\rho$, and $\bar{u}_{\circ t}=n^{-1}\sum_{i=1}^{n}u_{it}$. Running the above
panel AR(1) regression in $y_{it}$, augmented with the cross section averages,
$\bar{y}_{\circ t}$ and $\bar{y}_{\circ,t-1}$, we obtain the following pooled
DCCE estimator%
\begin{equation}
\hat{\rho}_{DCCEP}=\frac{n^{-1}T^{-1}\sum_{i=1}^{n}\mathbf{y}_{i,-1}^{\prime
}\mathbf{M}_{\bar{h}}\mathbf{y}_{i\circ}}{n^{-1}T^{-1}\sum_{i=1}^{n}%
\mathbf{y}_{i,-1}^{\prime}\mathbf{M}_{\bar{h}}\mathbf{y}_{i,-1}},
\label{rhoDCCEP}%
\end{equation}
where $\mathbf{M}_{\bar{h}}=\mathbf{I}_{T}-\mathbf{\bar{H}}\left(
\mathbf{\bar{H}}^{\prime}\mathbf{\bar{H}}\right)  ^{-}\mathbf{\bar{H}}%
^{\prime}$, $\mathbf{\bar{H}=}\left(  \mathbf{\tau}_{T},\mathbf{\bar{y}%
,\bar{y}}_{-1}\right)  $, and $\mathbf{A}^{-}$ denotes the generalized inverse
of the square matrix $\mathbf{A}$. Note that $\mathbf{M}_{\bar{h}}%
\mathbf{y}_{i\circ}$ and $\mathbf{M}_{\bar{h}}\mathbf{y}_{i,-1}$ are residuals
from the least squares regressions of $\mathbf{y}_{i\circ}$ and $\mathbf{y}%
_{i,-1}$ on $\mathbf{\bar{H}}$ and are invariant to the choice of the
generalized inverse. The use of generalized inverse allows for possible
perfect collinearity that arise between $\bar{y}_{\circ t}$ and $\bar
{y}_{\circ,t-1}$ when $f_{t}$ is trended, as discussed below.

To establish the consistency of $\hat{\rho}_{DCCEP}$ for $\rho_{0}$, writing
(\ref{IFEAUG}) in vector notations gives%
\[
\mathbf{y}_{i\circ}=\mathbf{\bar{H}}\boldsymbol{\pi}_{i}+\rho_{0}%
\mathbf{y}_{i,-1}+\mathbf{u}_{i\circ}-\delta_{i0}\mathbf{\bar{u},}%
\]
where $\boldsymbol{\pi}_{i}=(\alpha_{i}-\delta_{i0}\bar{\alpha},\delta
_{i0},\delta_{i1})^{\prime}$. Then
\[
\hat{\rho}_{DCCEP}-\rho_{0}=\frac{n^{-1}\sum_{i=1}^{n}\mathbf{y}%
_{i,-1}^{\prime}\mathbf{M}_{\bar{h}}\left(  \mathbf{u}_{i\circ}-\delta
_{i0}\mathbf{\bar{u}}\right)  }{n^{-1}\sum_{i=1}^{n}\mathbf{y}_{i,-1}^{\prime
}\mathbf{M}_{\bar{h}}\mathbf{y}_{i,-1}}.
\]
Also, from (\ref{yig}) we obtain$\ \mathbf{y}_{i,-1}=\alpha_{i}/(1-\rho
)\mathbf{\tau}_{T}+\gamma_{i}\mathbf{g}_{-1}+\boldsymbol{v}_{i,-1},$ and by
cross section averaging, $\mathbf{\bar{y}}_{-1}=\bar{\alpha}/(1-\rho
)\mathbf{\tau}_{T}+\bar{\gamma}\mathbf{g}_{-1}+\mathbf{\bar{v}}_{-1}$, which
in turn yields%
\[
\mathbf{y}_{i,-1}=\left(  \frac{\alpha_{i}-\delta_{i0}\bar{\alpha}}{1-\rho
_{0}}\right)  \mathbf{\tau}_{T}+\delta_{i0}\mathbf{\bar{y}}_{-1}%
+\boldsymbol{v}_{i,-1}-\delta_{i0}\mathbf{\bar{v}}_{-1}.
\]
Recall that $\boldsymbol{v}_{i,-1}=(v_{i0},v_{i1},...,v_{i,T-1}),$
$\mathbf{\bar{v}}_{-1}=n^{-1}\sum_{i=1}^{n}\boldsymbol{v}_{i,-1}$, where
$v_{it}=%
%TCIMACRO{\dsum \limits_{j=0}^{\infty}}%
%BeginExpansion
{\displaystyle\sum\limits_{j=0}^{\infty}}
%EndExpansion
\rho^{j}u_{i,t-j}$. Hence%
\[
\hat{\rho}_{DCCEP}-\rho_{0}=\frac{\sum_{i=1}^{n}\left(  \boldsymbol{v}%
_{i,-1}-\delta_{i0}\mathbf{\bar{v}}_{-1}\right)  ^{\prime}\mathbf{M}_{\bar{h}%
}\left(  \mathbf{u}_{i\circ}-\delta_{i0}\mathbf{\bar{u}}\right)  }{\sum
_{i=1}^{n}\left(  \boldsymbol{v}_{i,-1}-\delta_{i0}\mathbf{\bar{v}}%
_{-1}\right)  ^{\prime}\mathbf{M}_{\bar{h}}\left(  \boldsymbol{v}%
_{i,-1}-\delta_{i0}\mathbf{\bar{v}}_{-1}\right)  }.
\]
Let $\mathbf{M}_{g}=\mathbf{I}_{T}-\mathbf{P}_{g},$ $\mathbf{P}_{g}%
=\mathbf{G}\left(  \mathbf{G}^{\prime}\mathbf{G}\right)  ^{-}\mathbf{G}%
^{\prime}$, and $\mathbf{G=}\left(
\begin{array}
[c]{ccc}%
\mathbf{\tau}_{T} & \mathbf{g} & \mathbf{g}_{-1}%
\end{array}
\right)  $. Also
\[
\mathbf{\bar{H}}=\left(  \mathbf{\tau}_{T},\mathbf{\bar{y},\bar{y}}%
_{-1}\right)  =\left(  \mathbf{\tau}_{T},\text{ }\bar{\alpha}/(1-\rho
)\mathbf{\tau}_{T}+\bar{\gamma}\mathbf{g+\bar{v}}\text{, }\bar{\alpha}%
/(1-\rho)\mathbf{\tau}_{T}+\bar{\gamma}\mathbf{g}_{-1}+\mathbf{\bar{v}}%
_{-1}\right)  .
\]
Then $\mathbf{\bar{H}}=\mathbf{G\bar{Q}}+\mathbf{\bar{V}}$, where
$\mathbf{\bar{Q}}$ is a non-singular matrix (by assumption $\bar{\gamma}\neq
0$) given by%
\[
\mathbf{\bar{Q}=}\left(
\begin{array}
[c]{ccc}%
1 & \bar{\alpha}/(1-\rho_{0}) & \bar{\alpha}/(1-\rho_{0})\\
0 & \bar{\gamma} & 0\\
0 & 0 & \bar{\gamma}%
\end{array}
\right)  ,
\]
and $\mathbf{\bar{V}=}\left(  \mathbf{0},\mathbf{\bar{v},\bar{v}}_{-1}\right)
$. Since $u_{it}$ (and hence $v_{it}$) are cross-sectionally independent, then
$\mathbf{\bar{u}=}O_{p}(n^{-1/2})$, $\mathbf{\bar{v}}=O_{p}(n^{-1/2})$, and
$\mathbf{\bar{v}}_{-1}=O_{p}(n^{-1/2})$. It is then easily established that
for any fixed $T>3$, and as $n\rightarrow\infty,$ we have
\[
\mathbf{M}_{\bar{h}}-\mathbf{M}_{g}=\mathbf{G}\left(  \mathbf{G}^{\prime
}\mathbf{G}\right)  ^{-}\mathbf{G}^{\prime}-\left(  \mathbf{G\bar{Q}%
}+\mathbf{\bar{V}}\right)  \left[  \left(  \mathbf{G\bar{Q}}+\mathbf{\bar{V}%
}\right)  ^{\prime}\left(  \mathbf{G\bar{Q}}+\mathbf{\bar{V}}\right)  \right]
^{-}\left(  \mathbf{G\bar{Q}}+\mathbf{\bar{V}}\right)  ^{\prime}%
=O_{p}(n^{-1/2})
\]
This follows since $\mathbf{\bar{V}=}O_{p}\left(  n^{-1/2}\right)  $ and
$\mathbf{G}\left(  \mathbf{G}^{\prime}\mathbf{G}\right)  ^{-}\mathbf{G}%
^{\prime}=\mathbf{G\bar{Q}}\left[  \left(  \mathbf{G\bar{Q}}\right)  ^{\prime
}\left(  \mathbf{G\bar{Q}}\right)  \right]  ^{-}\mathbf{\bar{Q}}^{\prime
}\mathbf{G}^{\prime}$. Using this result, for a fixed\textbf{ }$T$ and
conditional on $\mathbf{f}$, or $\mathbf{G}$, we now have
\begin{equation}
\hat{\rho}_{DCCEP}-\rho_{0}\rightarrow_{p}\frac{lim_{n\rightarrow\infty}%
n^{-1}\sum_{i=1}^{n}E\left(  T^{-1}\boldsymbol{v}_{i,-1}^{\prime}%
\mathbf{M}_{g}\mathbf{u}_{i\circ}\left\vert \mathbf{f}\right.  \right)
}{lim_{n\rightarrow\infty}\text{ }n^{-1}\sum_{i=1}^{n}E\left(  T^{-1}%
\boldsymbol{v}_{i,-1}^{\prime}\mathbf{M}_{g}\boldsymbol{v}_{i,-1}\left\vert
\mathbf{f}\right.  \right)  }. \label{gapDCCEP}%
\end{equation}
For the denominator of (\ref{gapDCCEP}) we have $E\left(  T^{-1}%
\boldsymbol{v}_{i,-1}^{\prime}\mathbf{M}_{g}\boldsymbol{v}_{i,-1}\left\vert
\mathbf{f}\right.  \right)  =Tr\left[  \mathbf{M}_{g}E\left(  \boldsymbol{v}%
_{i,-1}\boldsymbol{v}_{i,-1}^{\prime}\right)  \right]  $, and
\[
E\left(  \boldsymbol{v}_{i,-1}\boldsymbol{v}_{i,-1}^{\prime}\right)
=\frac{\sigma_{i}^{2}}{1-\rho_{0}^{2}}\left(
\begin{array}
[c]{cccc}%
1 & \rho_{0} & \cdots & \rho_{0}^{T-1}\\
\rho_{0} & 1 & \cdots & \rho_{0}^{T-2}\\
\vdots & \vdots & \ddots & \vdots\\
\rho_{0}^{T-1} & \rho_{0}^{T-2} & \cdots & 1
\end{array}
\right)  =\sigma_{i}^{2}\mathbf{V}_{0},
\]
where $\lambda_{\max}\left(  \mathbf{V}_{_{0}}\right)  $ is bounded noting
that $\left\vert \rho_{0}\right\vert <1$. Hence, $lim_{n\rightarrow\infty}$
$n^{-1}\sum_{i=1}^{n}E\left(  T^{-1}\boldsymbol{v}_{i,-1}^{\prime}%
\mathbf{M}_{g}\boldsymbol{v}_{i,-1}\left\vert \mathbf{f}\right.  \right)
=\bar{\sigma}^{2}Tr\left(  T^{-1}\mathbf{M}_{g}\mathbf{V}_{0}\right)  $, where
$\bar{\sigma}^{2}=\lim_{n\rightarrow\infty}n^{-1}\sum_{i=1}^{n}\sigma_{i}^{2}%
$. Also, $Tr\left(  T^{-1}\mathbf{M}_{g}\mathbf{V}_{0}\right)  =Tr\left(
T^{-1}\mathbf{M}_{g}\mathbf{V}_{0}\mathbf{M}_{g}\right)  \leq T^{-1}%
\lambda_{\max}\left(  \mathbf{V}_{0}\right)  Tr\left(  \mathbf{M}_{g}\right)
=\left[  (T-3)/T\right]  \lambda_{\max}\left(  \mathbf{V}_{0}\right)  $ which
is bounded in $T$, irrespective of whether $f_{t}$ is trended or not.

Now consider the numerator of (\ref{gapDCCEP}) and note that
\begin{equation}
E\left(  T^{-1}\boldsymbol{v}_{i,-1}^{\prime}\mathbf{M}_{g}\mathbf{u}_{i\circ
}\left\vert \mathbf{f}\right.  \right)  =T^{-1}\sum_{t=1}^{T}E\left(
v_{i,t-1}u_{it}\right)  -T^{-1}E\left(  \boldsymbol{v}_{i,-1}^{\prime
}\mathbf{P}_{g}\mathbf{u}_{i\circ}\left\vert \mathbf{f}\right.  \right)  .
\label{Num3}%
\end{equation}
Since $v_{it}=%
%TCIMACRO{\dsum \limits_{s=0}^{\infty}}%
%BeginExpansion
{\displaystyle\sum\limits_{s=0}^{\infty}}
%EndExpansion
\rho^{s}u_{i,t-s}$ and $u_{it}$ is serially uncorrelated, then $E\left(
v_{i,t-1}u_{it}\right)  =0$, and $E\left(  T^{-1}\boldsymbol{v}_{i,-1}%
^{\prime}\mathbf{M}_{g}\mathbf{u}_{i\circ}\left\vert \mathbf{f}\right.
\right)  =-T^{-1}E\left(  \boldsymbol{v}_{i,-1}^{\prime}\mathbf{P}%
_{g}\mathbf{u}_{i\circ}\left\vert \mathbf{f}\right.  \right)  $. Further,
since $f_{t}$ (and $g_{t}$) is independently distributed of $u_{it}$ (and
$v_{it}$) then conditional on $\mathbf{f}$, we have $T^{-1}E\left(
\boldsymbol{v}_{i,-1}^{\prime}\mathbf{P}_{g}\mathbf{u}_{i\circ}\right)
=Tr\left[  \mathbf{P}_{g}E\left(  T^{-1}\mathbf{u}_{i\circ}\boldsymbol{v}%
_{i,-1}^{\prime}\right)  \right]  $. It is also easily seen that $E\left(
\mathbf{u}_{i\circ}\boldsymbol{v}_{i,-1}^{\prime}\right)  =\sigma_{i}%
^{2}\mathbf{W}_{\rho}$, where%
\[
\mathbf{W}_{\rho}=\left(
\begin{array}
[c]{cccccc}%
0 & 1 & \rho_{0} & \rho_{0}^{2} & \cdots & \rho_{0}^{T-2}\\
0 & 0 & 1 & \rho_{0} & \cdots & \rho_{0}^{T-3}\\
\vdots & \vdots & \vdots &  & \cdots & \vdots\\
0 & 0 & \ddots & 0 & \cdots & 1\\
0 & 0 & \cdots & 0 & 0 & 0
\end{array}
\right)  =(w_{tt^{\prime}}).
\]
Then
\begin{align*}
E\left(  T^{-1}\boldsymbol{v}_{i,-1}^{\prime}\mathbf{M}_{g}\mathbf{u}_{i\circ
}\left\vert \mathbf{f}\right.  \right)   &  =-T^{-1}E\left(  \boldsymbol{v}%
_{i,-1}^{\prime}\mathbf{P}_{g}\mathbf{u}_{i\circ}\left\vert \mathbf{f}\right.
\right) \\
&  =T^{-2}\sigma_{i}^{2}Tr\left(  \mathbf{AW}_{\rho}\right)  =T^{-2}\sigma
_{i}^{2}\sum_{t=1}^{T}\sum_{t=1}^{T}a_{tt^{\prime}}w_{tt^{\prime}}%
\end{align*}
where $a_{tt^{\prime}}$ is the $(t,t^{\prime})$ element of $\mathbf{A=}%
T\mathbf{G}\left(  \mathbf{G}^{\prime}\mathbf{G}\right)  ^{-1}\mathbf{G}%
^{\prime}$. Hence%
\begin{align*}
\left\vert E\left(  T^{-1}\boldsymbol{v}_{i,-1}^{\prime}\mathbf{M}%
_{g}\mathbf{u}_{i\circ}\left\vert \mathbf{f}\right.  \right)  \right\vert  &
\leq T^{-2}\sigma_{i}^{2}\sup_{t,t^{\prime}}\left\vert a_{tt^{\prime}%
}\right\vert \sum_{t=1}^{T}\sum_{t=1}^{T}\left\vert w_{tt^{\prime}}\right\vert
\\
&  \leq T^{-1}\sup_{t,t^{\prime}}\left\vert a_{tt^{\prime}}\right\vert
\sigma_{i}^{2}\sum_{s=1}^{T-1}\left(  1-\frac{s}{T}\right)  \left\vert
\rho\right\vert ^{s-1},
\end{align*}
and%
\[
lim_{n\rightarrow\infty}n^{-1}\sum_{i=1}^{n}E\left(  T^{-1}\boldsymbol{v}%
_{i,-1}^{\prime}\mathbf{M}_{g}\mathbf{u}_{i\circ}\left\vert \mathbf{f}\right.
\right)  \leq T^{-1}\sup_{t,t^{\prime}}\left\vert a_{tt^{\prime}}\right\vert
\bar{\sigma}^{2}\sum_{s=1}^{T-1}\left(  1-\frac{s}{T}\right)  \left\vert
\rho\right\vert ^{s-1}.
\]
Since $\bar{\sigma}^{2}\sum_{s=1}^{T-1}\left(  1-\frac{s}{T}\right)
\left\vert \rho_{0}\right\vert ^{s-1}$ is bounded in $T$, then conditional on
$\mathbf{G}$, the order of the numerator of (\ref{gapDCCEP}) is determined by
the order of $T^{-1}\sup_{t,t^{\prime}}\left\vert a_{tt^{\prime}}\right\vert
$. It is interesting that the probability order of this term does not depend
on whether $f_{t}$ is stationary or trended, which is in contrast to the order
of the bias of the TWFE estimator that crucially depended on whether $f_{t}$
is stationary or trended.

When $f_{t}$ is stationary then $g_{t}=O_{p}(1)$ and $T^{-1}\mathbf{G}%
^{\prime}\mathbf{G=}O_{p}(1)$ and since $\mathbf{A=G}\left(  T^{-1}%
\mathbf{G}^{\prime}\mathbf{G}\right)  ^{-1}\mathbf{G}$, then $\sup
_{t,t^{\prime}}\left\vert a_{tt^{\prime}}\right\vert =O_{p}(1)$. In the case
where $f_{t}$ is trended we first write $\mathbf{A}$ as
\[
\mathbf{A}=\left(  \sqrt{T}\mathbf{GD}_{T}\right)  \left(  \mathbf{D}%
_{T}\mathbf{G}^{\prime}\mathbf{GD}_{T}\right)  ^{-1}\left(  \sqrt{T}%
\mathbf{D}_{T}\mathbf{G}^{\prime}\right)  ,
\]
where $\mathbf{D}_{T}=diag(T^{-1/2},T^{-3/2},T^{-3/2})$. Then it is easily
established that $\mathbf{D}_{T}\mathbf{G}^{\prime}\mathbf{GD}_{T}=O_{p}(1)$
and $\sqrt{T}\mathbf{GD}_{T}=O_{p}(1)$. Hence, irrespective of whether $f_{t}$
is stationary or trended, $\sup_{t,t^{\prime}}\left\vert a_{tt^{\prime}%
}\right\vert =O_{p}(1)$ and numerator of (\ref{gapDCCEP}) is $O_{p}(T^{-1})$,
and overall we have
\begin{equation}
\hat{\rho}_{DCCEP}-\rho_{0}\rightarrow_{p}O_{p}\left(  T^{-1}\right)  \text{,
as }n\rightarrow\infty. \label{rhoDCCEP2}%
\end{equation}
Namely, the DCCEP estimator of $\rho_{0}$\ is consistent when $n\,$and $T$ are
both large, and applies irrespective of whether $f_{t}$ is stationary or
trended. The trend in $f_{t}$ could be stochastic and subject to breaks.
However, not surprisingly, DCCEP suffers from the small $T$ bias analogous to
the Nickell bias of the TWFE estimator. De Vos and Everaert (2021) derive the
fixed $T$ bias under the assumption of a stationary factor. This small $T$
bias will be present irrespective of whether the panel data model includes a
latent factor or not. It is simply due to dynamic nature of the panel
regression under consideration.

To avoid the small $T$ bias of DCCEP estimator, Hayakawa, Pesaran and Smith
(2023) allow for interactive time effects through a multi-factor error
structure $\boldsymbol{\gamma}_{i}^{\prime}\mathbf{f}$ in dynamic fixed
effects panel data models in addition to the standard fixed and time effects,
but assume that $\boldsymbol{\gamma}_{i}$ are cross-sectionally independent.
They apply maximum likelihood estimation after first-differencing the data to
remove the fixed effects. This transformed quasi maximum likelihood estimator
is shown to be robust to the heterogeneity of the initial values and common
unobserved effects, and is applicable to both stationary and unit root cases.
They also propose a procedure for selection of the number of latent factors.
They apply the estimator to the Acemoglu et al. (2019) growth data using
averages taken over five yearly intervals.

\section{The DCCEMG estimator: nonparallel trends with heterogeneous
dynamics\label{HetConFac}}

In this case the output equations with heterogeneous dynamics as well as
nonparallel trends are given by
\begin{equation}
y_{it}=\alpha_{i}+\gamma_{i}f_{t}+\rho_{i}y_{i,t-1}+u_{it},\,\ \text{for
}i=1,2,...,n. \label{HetroInter}%
\end{equation}
This model is investigated in detail by Chudik and Pesaran (2015), who show
that once we allow for $\rho_{i}$ to differ across $i$, the identification and
estimation of $f_{t}$ become much more complicated due to the heterogeneity in
the dependence of $y_{it}$ on $f_{t}$, and its lagged values. To identify
$f_{t}$ it is assumed that $\rho_{i}$ lies in the range $(-1,1)$, such that
$E\left\vert \rho_{i}\right\vert ^{s}\leq\bar{\rho}^{s}$, where $\bar{\rho}%
<1$. This ensures that $E\left(  \frac{1}{1-\rho_{i}^{2}}\right)  $ exists
which is required for identification of $f_{t}$ from the macro outcomes,
$\bar{y}_{\circ t}$. This assumption rules out the possibility of long memory
processes.\footnote{On this see Robinson (1978) and Granger (1980). A more
general treatment is provided in Pesaran and Chudik (2014).}

\subsection{DCCEMG with stationary latent factors}

Chudik and Pesaran (2015) further assume that $f_{t}$ is covariance stationary
which is rather restrictive as noted above. Here we provide a sketch of the
proof to highlight the way nonparallel trends and dynamic heterogeneity
interact. Solving (\ref{HetroInter}) from a distant past we have
\[
y_{it}=\mu_{i}+\gamma_{i}%
%TCIMACRO{\dsum \limits_{s=0}^{\infty}}%
%BeginExpansion
{\displaystyle\sum\limits_{s=0}^{\infty}}
%EndExpansion
\rho_{i}^{s}f_{t-s}+%
%TCIMACRO{\dsum \limits_{s=0}^{\infty}}%
%BeginExpansion
{\displaystyle\sum\limits_{s=0}^{\infty}}
%EndExpansion
\rho_{i}^{s}u_{i,t-s},
\]
with $\mu_{i}=\alpha_{i}/(1-\rho_{i})$, $\theta_{is}=\gamma_{i}\rho_{i}^{s}$
and $v_{it}=$ $\sum_{s=0}^{\infty}\rho_{i}^{s}u_{i,t-s}$, which if averaged
over $i$ yields
\begin{equation}
\bar{y}_{\circ t}=\bar{\mu}_{n}+%
%TCIMACRO{\dsum \limits_{s=0}^{\infty}}%
%BeginExpansion
{\displaystyle\sum\limits_{s=0}^{\infty}}
%EndExpansion
\bar{\theta}_{ns}f_{t-s}+\bar{v}_{\circ t}, \label{ybardot}%
\end{equation}
where $\bar{\mu}_{n}=n^{-1}\sum_{i=1}^{n}\mu_{i},$ $\bar{\theta}_{ns}%
=n^{-1}\sum_{i=1}^{n}\theta_{is},$ and $\bar{v}_{\circ t}=n^{-1}\sum_{i=1}%
^{n}v_{it}.$It is clear that conditional on $\rho_{i}$, $v_{it}\,^{\prime}s$
continue to be cross-sectionally independent, and $Var\left(  \bar{v}_{\circ
t}\right)  =n^{-2}\sum_{i=1}^{n}Var\left(  v_{it}\right)  $. Also
\[
Var\left(  v_{it}\right)  =Var\left[  E\left(  v_{it}\left\vert \rho
_{i}\right.  \right)  \right]  +E\left[  Var\left(  v_{it}\left\vert \rho
_{i}\right.  \right)  \right]  =E\left(  \frac{\sigma_{i}^{2}}{1-\rho_{i}^{2}%
}\right)  ,
\]
and it is easily seen that $E\left(  \frac{\sigma_{i}^{2}}{1-\rho_{i}^{2}%
}\right)  <C$. Hence, $Var\left(  \bar{v}_{\circ t}\right)  =O(n^{-1})$. Using
these results in (\ref{ybardot}) we have%
\begin{equation}
\bar{\theta}_{n}(L)f_{t}=\bar{y}_{\circ t}-\bar{\mu}_{n}+O_{p}(n^{-1/2}).
\label{thetabar}%
\end{equation}
where $\bar{\theta}_{n}(L)=\sum_{s=0}^{\infty}\bar{\theta}_{ns}L^{s},$ and $L$
is a lag operator. In the case of homogeneous dynamics, $\bar{\theta}_{n}%
(L)$\thinspace$=\bar{\gamma}_{n}(1-\rho L)^{-1}$ and $f_{t}$ can be identified
as an affine function of $\bar{y}_{\circ t}$ and $\bar{y}_{\circ,t-1}$. When
$\rho_{i}$ is heterogeneous such a simple inversion is not possible and
additional restrictions are required. Chudik and Pesaran (2015), building on
an earlier paper by Pesaran and Chudik (2014), show that when $f_{t}$ is
covariance stationary then $f_{t}$ can be approximated by a distributed lag
function of $y_{\circ t},y_{\circ,t-1},....,y_{\circ.t-p_{T}}$, where the lag
order, $p_{T},$ rises with $T$ but at the slower rate of $T^{1/3}$.
Specifically, they show that
\begin{equation}
y_{it}=\alpha_{in}-\gamma_{i}\psi(1)\bar{\mu}_{n}+\gamma_{i}\sum
_{s=0}^{p_{_{T}}}\psi_{s}\bar{y}_{\circ,t-s}+\rho_{i}y_{i,t-1}+\zeta
_{it,n},\,\ \text{for }i=1,2,...,n. \label{yithetint}%
\end{equation}
where%
\[
\tilde{\alpha}_{in}=\alpha_{i}-\gamma_{i}\psi(1)\bar{\mu}_{n}\text{, and
}\zeta_{it,n}=u_{it}+\gamma_{i}\left(  \sum_{s=q_{T}+1}^{\infty}\psi_{s}%
\bar{y}_{\circ,t-s}\right)  +O_{p}(n^{-1/2}).
\]

\subsection{DCCEMG with trended latent factors}

The above approach can be readily extended to panel data models with trended
latent factors and higher order dynamics:
\begin{equation}
\Delta y_{it}=\alpha_{i}+\gamma_{i}f_{t}-\phi_{i}y_{i,t-1}+\sum_{s=1}%
^{p-1}\delta_{is}\Delta y_{i,t-s}+u_{it}. \label{parwite}%
\end{equation}
Despite the higher order dynamics, $f_{t}$ can still be well approximated by
$\bar{y}_{\circ,t-s}$, for $s=0,1,...,q_{T}$, with $\phi_{i}$ estimated as
\begin{equation}
\hat{\phi}_{i,DCCE}=\left(  \mathbf{y}_{i,-1}^{\prime}\mathbf{\bar{M}}%
_{i}\mathbf{y}_{i,-1}\right)  ^{-1}\mathbf{y}_{i,-1}^{\prime}\mathbf{\bar{M}%
}_{i}\Delta\mathbf{y}_{i}, \label{phiHetP}%
\end{equation}
where $\mathbf{\bar{M}}_{i}=\mathbf{I}_{T}-\mathbf{\bar{W}}_{i}\left(
\mathbf{\bar{W}}_{i}^{^{\prime}}\mathbf{\bar{W}}_{i}\right)  ^{-}%
\mathbf{\bar{W}}_{i}^{^{\prime}}$, $\mathbf{\bar{W}}_{i}=(\mathbf{\tau}%
_{T},\mathbf{\bar{y}}_{\circ},\mathbf{\bar{y}}_{\circ,-1},...,\mathbf{\bar{y}%
}_{\circ,-q_{T}},\Delta\mathbf{y}_{i,-1},...,\Delta\mathbf{y}_{i,-p+1})$, and
$\left(  \mathbf{\bar{W}}_{i}^{^{\prime}}\mathbf{\bar{W}}_{i}\right)  ^{-}$
denotes the generalized inverse of $\left(  \mathbf{\bar{W}}_{i}^{^{\prime}%
}\mathbf{\bar{W}}_{i}\right)  $. Note that the estimates, $\hat{\phi}%
_{i,DCCE}$, are invariant to the choice of the generalized inverse. The use of
generalized inverse is particularly important if $f_{t}$ is trended, since in
that case $\mathbf{\bar{y}}_{\circ},\mathbf{\bar{y}}_{\circ,-1}%
,...,\mathbf{\bar{y}}_{\circ,-q_{T}}$ become highly multicollinear with their
pairwise correlations tending to unity as $n\rightarrow\infty$. This result is
also relevant to the choice of $q_{T}$, suggesting that small values of
$q_{T}$ should be sufficient for cross section averages to be a good proxy for
$f_{t}$ when $y_{it}$ are trended

A consistent estimator of $E\left(  \phi_{i}\right)  $ is now given by
\begin{equation}
\hat{\phi}_{DCCEMG}=n^{-1}\sum_{i=1}^{n}\hat{\phi}_{i,DCCE}, \label{phiDCCEMG}%
\end{equation}
and
\begin{equation}
\widehat{Var\left(  \hat{\phi}_{DCCEMG}\right)  }=\frac{1}{n(n-1)}\sum
_{i=1}^{n}\left(  \hat{\phi}_{i,DCCE}-\hat{\phi}_{DCCEMG}\right)  ^{2}.
\label{varphiDCCEMG}%
\end{equation}

\section{Allowing for time-varying covariates\label{TVTI}}

Up to now the focus has been on output convergence in presence of latent
(technology) factors. We now consider how the introduction of observed
country-specific covariates interact with dynamics and common factors. One can
also include observed common factors such as global climate variables, but it
is important to bear in mind that the effects such common factors cannot be
separately identified from latent common factors. Therefore, in what follows
we focus on observed covariates that vary across both time and countries.
Covariates that are time-invariant like geography will be considered in the
next section.

We augment the panel data model (\ref{parwite}) with the $k_{x}\times1$ vector
of covariates, $\mathbf{x}_{it}$, and consider the general panel data model
with current and $p_{T}$ lagged values of $\mathbf{x}_{it}$, and write the
resultant specification more compactly as:
\begin{equation}
\Delta y_{it}=\alpha_{i}+\boldsymbol{\gamma}_{i}^{\prime}\boldsymbol{f}%
_{t}+\boldsymbol{\psi}_{i}^{\prime}\mathbf{q}_{it}+u_{it}, \label{qit}%
\end{equation}
where $\boldsymbol{f}_{t}$ is a $m\times1$ vector, $\mathbf{q}_{it}%
=(y_{i,t-1},\Delta y_{i,t-1},...,\Delta y_{i,t-p_{T}-1},$ $\mathbf{x}%
_{it}^{\prime},\Delta\mathbf{x}_{i,t-1}^{\prime},...,\Delta\mathbf{x}%
_{i,t-p_{T}-1}^{\prime})^{\prime}.$ As before, the coefficients
$\boldsymbol{\psi}_{i}$ can be estimated consistently by augmenting the panel
regression with the cross-sectional averages $\mathbf{\bar{w}}_{\circ t}%
=(\bar{y}_{\circ t},\mathbf{\bar{x}}_{\circ t}^{\prime})^{\prime}$, and their
lagged values, where $\mathbf{\bar{x}}_{\circ t}=n^{-1}\sum_{i=1}%
^{n}\mathbf{x}_{it}$. These act as proxies for the latent factors. For DCCEP
and DCCEMG to be applicable in this more general set up, it is required that
$m\leq k_{x}+1$. We also consider a random coefficient model and assume
$\boldsymbol{\psi}_{i}=\boldsymbol{\psi}_{0}$ $\boldsymbol{+\eta}_{i}$, where
$\boldsymbol{\eta}_{i}$ are distributed independently over $i$ with mean zero
and a finite variance, such that conditional on $\mathbf{x}_{it}$ and
$\mathbf{f}_{t}$, output processes are stationary. The focus is on estimation
of the mean effects, $E\left(  \boldsymbol{\psi}_{i}\right)  =\boldsymbol{\psi
}_{0}$.

As discussed already, the required number of lagged values, $p_{T},$ depends
on whether $\rho_{i}$ is homogenous or not. In the homogeneous case (with
$\boldsymbol{\psi}_{i}=\boldsymbol{\psi}$) consistent estimation of
$\boldsymbol{\psi}_{0}$ can be obtained by pooled regression of $\Delta
y_{it}$ on $(1,\mathbf{q}_{it}^{\prime})^{\prime}$ augmented with
$\mathbf{\bar{z}}_{\circ t}$ and its lagged values. This estimator is known as
pooled dynamic CCE estimator and denoted by DCCEP. Specifically \ (See also
(\ref{rhoDCCEP}))
\begin{equation}
\left(
\begin{array}
[c]{c}%
\hat{\rho}_{DCCEP}\\
\boldsymbol{\hat{\beta}}_{DCCEP}%
\end{array}
\right)  =\boldsymbol{\hat{\psi}}_{DCCEP}=\left(  \sum_{i=1}^{n}\mathbf{Q}%
_{i}^{\prime}\mathbf{M}_{\bar{h}}\mathbf{Q}_{i}\right)  ^{-1}\sum_{i=1}%
^{n}\mathbf{Q}_{i}^{\prime}\mathbf{M}_{\bar{h}}\mathbf{y}_{i\circ},
\label{epsiDCCEP}%
\end{equation}
where $\mathbf{Q}_{i}$ is the $T\times k_{q}$ matrix of observations on
$\mathbf{q}_{it}$, $k_{q}=\dim(\mathbf{q}_{it})$, $\mathbf{M}_{\bar{h}%
}=\mathbf{I}_{T}-\mathbf{\bar{H}}_{s}\left(  \mathbf{\bar{H}}_{s}^{\prime
}\mathbf{\bar{H}}_{s}\right)  ^{-}\mathbf{\bar{H}}_{s}^{\prime}$,
$\mathbf{\bar{H}}_{s}\mathbf{=}\left(  \mathbf{\tau}_{T},\mathbf{\bar{W}%
,\bar{W}}_{-1},...,\mathbf{\bar{W}}_{-s_{T}}\right)  $, and $\mathbf{\bar{W}%
}=n^{-1}\sum_{i=1}^{n}\mathbf{W}_{i},\mathbf{W}_{i}$ is the $T\times(k_{x}+1)$
observations on $\mathbf{w}_{it}=(y_{it},\mathbf{x}_{it}^{\prime})^{\prime}$.
Assuming $\mathbf{x}_{it}$ are weakly exogenous, following Chudik and Pesaran
(2015), it is easily established that $\boldsymbol{\hat{\psi}}_{DCCEP}%
\rightarrow_{p}\boldsymbol{\psi}_{0}=(\rho_{0},\boldsymbol{\beta}_{0}^{\prime
})^{\prime}$ for sufficiently large $n$ and $T$. Large $T$ is required to
remove the Nickel type bias which continues to apply. $\,$

In the fully heterogeneous case where $\boldsymbol{\psi}_{i}$ are allowed to
vary across $i$, the mean effects, $\boldsymbol{\psi}_{0}$, can be estimated
using the mean group approach discussed above but with the important
difference that the country-specific regressions must now be augmented with
the cross section averages of $\mathbf{x}_{it}$ and $y_{it}$ as well as
$p_{T}$ lagged values of $y_{it}$. Namely,%
\begin{equation}
\boldsymbol{\hat{\psi}}_{DCCEMG}=n^{-1}\sum_{i=1}^{n}\boldsymbol{\hat{\psi}%
}_{i,DCCE},\text{ }\boldsymbol{\hat{\psi}}_{i,DCCE}=\left(  \mathbf{Q}%
_{i}^{\prime}\mathbf{M}_{\bar{h}}\mathbf{Q}_{i}\right)  ^{-1}\mathbf{Q}%
_{i}^{\prime}\mathbf{M}_{\bar{h}}\mathbf{y}_{i\circ}. \label{epsiDCCE}%
\end{equation}
The asymptotic variance of $\boldsymbol{\hat{\psi}}_{DCCEMG}$ can be
consistently estimated by
\begin{equation}
\widehat{Var\left(  \boldsymbol{\hat{\psi}}_{DCCEMG}\right)  }=\frac
{1}{n(n-1)}\sum_{i=1}^{n}\left(  \boldsymbol{\hat{\psi}}_{i,DCCE}%
-\boldsymbol{\hat{\psi}}_{DCCEMG}\right)  \left(  \boldsymbol{\hat{\psi}%
}_{i,DCCE}-\boldsymbol{\hat{\psi}}_{DCCEMG}\right)  ^{\prime}.
\label{VarPsiDCCEMG}%
\end{equation}
The theoretical analysis suggests setting $s_{T}$ (lag order of cross section
averages) to the integer part of $T^{1/3}$. However, as already discussed, in
the case of trended series only a few lags are required to ensure the factors
$\boldsymbol{f}_{t}$ are adequately proxied by the cross country averages.

The DCCE approach can also accommodate models with a mix of homogeneous and
heterogeneous coefficients. This arises in our empirical application where we
consider the effects of democracy on output with no time variations in the
democracy indicator for half of the countries in the panel.

\section{Time-invariant or slowly moving covariates\label{TINV}}

One important advantage of DCCE estimators of $\boldsymbol{\psi}_{0}$ are
their robustness to the processes that generate fixed effects, $\alpha_{i}$,
and factor loadings, $\gamma_{i}$. However, when it comes to estimating the
possible determinants of $\alpha_{i}$ and $\gamma_{i}$, this feature seems to
be a disadvantage, since (like TWFE) the DCCE approach eliminates the
time-invariant or slowly moving determinants and seemingly throws the baby out
with the bathwater, so to speak. It is for this reason that many researchers
do not favour FE estimation and prefer pooled least squares with $\alpha_{i}$
and $\gamma_{i}$ replaced by their determinants. Barro (2015) argues against
including fixed effects, saying "there is insufficient within country
variation in the measured institutional quality to isolate a statistically
significant effect on economic growth." Barro regressions, discussed in
Section \ref{BB}, impose the parallel trends assumption and in effect assume
that $\alpha_{i}$ is a \textit{deterministic} function of the time-invariant
or slowly moving regressors. But, as already argued, pooling can lead to
inconsistent estimators unless very strong assumption are made about
determinants of $\alpha_{i}$ and $\gamma_{i}$.

Here we follow an alternative approach, whereby we first estimate the
coefficients of the time-varying regressors, $\boldsymbol{\psi}_{i}$, using
DCCEP or DCCEMG type procedures that are invariant with respect to the values
of $\alpha_{i}$ and $\gamma_{i}$, and then follow Pesaran and Zhou (2018) and
estimate the effects of time-invariant regressors using a cross-country
regression after filtering out the effects of time-varying regressors. For
slowly moving variables, country-specific time averages can be used. This
approach originates in the pioneering contributions of Hausman and Taylor
(1981) in static panel data models with random coefficients. The present
application is further complicated due to the dynamic nature of the output
convergence process and heterogeneity, particularly when the parallel trends
assumption is relaxed. Pesaran and Zhou (2018) focus on static short T\ panels
and addresses the implications of estimation uncertainty of fixed
effects\ estimation of coefficients of time-varying variables for inference on
the time-invariant effects. Here we assume both $n$ and $T\,$are large which
considerably simplifies inference regarding the time-invariant effects.

Starting with (\ref{aizi}) and (\ref{gizi}), and denoting the vector of
time-invariant variables by $\mathbf{z}_{i}$, after filtering out the dynamics
and the effects of time-varying covariates, $\mathbf{x}_{it}$, we obtain%
\begin{equation}
a_{iT}=\bar{y}_{i\circ}-\mathbf{\bar{q}}_{i\circ}^{\prime}\boldsymbol{\psi
}=\alpha_{T}+\boldsymbol{\theta}^{\prime}\mathbf{z}_{i}+\eta_{iT}+\bar
{u}_{i\circ}, \label{FCS}%
\end{equation}
where $\mathbf{\bar{q}}_{i\circ}=(\bar{y}_{i,-1},\mathbf{\bar{x}}_{i\circ
}^{\prime})^{\prime}$, $\boldsymbol{\psi}=(\rho,\boldsymbol{\beta}^{\prime
})^{\prime}$, $\mathbf{z}_{i}=$ $\mathbf{z}_{i\alpha}\cup\mathbf{z}_{i\gamma}%
$, $\boldsymbol{\theta}$ $\ $is the $k_{z}\times1$ vector of the
time-invariant coefficients that can be identified, $\alpha_{T}=\alpha
_{\alpha}+\alpha_{\gamma}\bar{f}_{T}$, and $\eta_{iT}=\eta_{i\alpha}%
+\eta_{i\gamma}\bar{f}_{T}$. Here a pooled estimator of $\boldsymbol{\psi,}$
either TWFE or DCCEP, is used. In this paper we do not explore the use of
country-specific estimates of $\mathbf{\psi}_{i}$ for filtering out the
effects of the dynamics and the time-varying covariates, since the condition
for the validity of this approach, namely $\sqrt{n}\sup_{i}\left\Vert
\boldsymbol{\hat{\psi}}_{i}-\boldsymbol{\psi}_{i}\right\Vert \rightarrow_{p}%
0$, is unlikely to be met in practice.

Conditional on $\boldsymbol{\psi}$, the time-invariant effects are identified
if $\mathbf{z}_{i}$ is distributed independently of the composite error term,
$\xi_{iT}=\eta_{iT}+\bar{u}_{i\circ}$, or if there are instrumental variables
that are uncorrelated with the composite errors, and are sufficiently
correlated with $\mathbf{z}_{i}$. To simplify the exposition here we assume
$\mathbf{z}_{i}$ is distributed independently of $\xi_{iT}$, and assume that
$\mathbf{S}_{zz,n}=n^{-1}\sum_{i=1}^{n}\left(  \mathbf{z}_{i}-\mathbf{\bar{z}%
}\right)  \left(  \mathbf{z}_{i}-\mathbf{\mathbf{\bar{z}}_{\circ}}\right)
^{\prime}\rightarrow_{p}\mathbf{S}_{zz}$, a non-stochastic positive definite
matrix. Under these assumptions, and conditional on $\boldsymbol{\psi}$,
$\boldsymbol{\theta}$ can be estimated consistently by the least squares
regression of $a_{iT}$ on an intercept and $\mathbf{z}_{i}$, namely%
\begin{equation}
\boldsymbol{\hat{\theta}}\left(  \boldsymbol{\psi}\right)  =\left[  n^{-1}%
\sum_{i=1}^{n}\left(  \mathbf{z}_{i}-\mathbf{\bar{z}}\right)  \left(
\mathbf{z}_{i}-\mathbf{\mathbf{\bar{z}}_{\circ}}\right)  ^{\prime}\right]
^{-1}n^{-1}\sum_{i=1}^{n}\left(  \mathbf{z}_{i}-\mathbf{\mathbf{\bar{z}%
}_{\circ}}\right)  a_{iT}. \label{thetahat1}%
\end{equation}
Denoting the true value of $\boldsymbol{\theta}$ by $\boldsymbol{\theta}_{0}$,
and using $a_{iT}$ from (\ref{FCS}) we have
\[
\boldsymbol{\hat{\theta}}\left(  \boldsymbol{\psi}\right)  -\boldsymbol{\theta
}_{0}=\left[  n^{-1}\sum_{i=1}^{n}\left(  \mathbf{z}_{i}-\mathbf{\mathbf{\bar
{z}}_{\circ}}\right)  \left(  \mathbf{z}_{i}-\mathbf{\mathbf{\bar{z}}_{\circ}%
}\right)  ^{\prime}\right]  ^{-1}n^{-1}\sum_{i=1}^{n}\left(  \mathbf{z}%
_{i}-\mathbf{\mathbf{\bar{z}}_{\circ}}\right)  \left(  \eta_{iT}+\bar
{u}_{i\circ}\right)  ,
\]
and it also follows that $\sqrt{n}\left[  \boldsymbol{\hat{\theta}}\left(
\boldsymbol{\psi}\right)  -\boldsymbol{\theta}_{0}\right]  \rightarrow
_{d}N(0,\mathbf{V}_{\theta}),$ where $\mathbf{V}_{\theta}=\mathbf{S}_{zz}%
^{-1}\mathbf{\Omega}_{zz}\mathbf{S}_{zz}^{-1}$, $\mathbf{\Omega}_{zz}%
=p\lim_{n}n^{-1}\sum_{i=1}^{n}\omega_{iT}^{2}\left(  \mathbf{z}_{i}%
-\mathbf{\mathbf{\bar{z}}_{\circ}}\right)  \left(  \mathbf{z}_{i}%
-\mathbf{\mathbf{\bar{z}}_{\circ}}\right)  ^{\prime}$, and $\omega_{iT}%
^{2}=\sigma_{\eta}^{2}+T^{-1}\sigma_{iu}^{2}$.

In practice, the unknown coefficients $\boldsymbol{\psi}$ in (\ref{thetahat1})
can be replaced by TWFE, DCCEP estimators. Under slope homogeneity and
parallel trends the TWFE estimator, $\boldsymbol{\hat{\psi}}_{TWFE}$, can be
used to obtain the following feasible estimator $\boldsymbol{\theta}$
\begin{equation}
\boldsymbol{\hat{\theta}}_{TWFE}=\left[  n^{-1}\sum_{i=1}^{n}\left(
\mathbf{z}_{i}-\mathbf{\mathbf{\bar{z}}_{\circ}}\right)  \left(
\mathbf{z}_{i}-\mathbf{\mathbf{\bar{z}}_{\circ}}\right)  ^{\prime}\right]
^{-1}n^{-1}\sum_{i=1}^{n}\left(  \mathbf{z}_{i}-\mathbf{\mathbf{\bar{z}%
}_{\circ}}\right)  \left(  \bar{y}_{i\circ}-\mathbf{\bar{q}}_{i\circ}^{\prime
}\boldsymbol{\hat{\psi}}_{TWFE}\right)  \mathbf{.} \label{thetaDCCE}%
\end{equation}
Substituting for $\bar{y}_{i\circ}$ using (\ref{FCS}), we now have%
\begin{equation}
\boldsymbol{\hat{\theta}}_{TWFE}-\boldsymbol{\theta}_{0}=\boldsymbol{\hat
{\theta}}\left(  \boldsymbol{\psi}\right)  -\boldsymbol{\theta}_{0}%
-\mathbf{S}_{zz,n}^{-1}\left[  n^{-1}\sum_{i=1}^{n}\left(  \mathbf{z}%
_{i}-\mathbf{\mathbf{\bar{z}}_{\circ}}\right)  \mathbf{\bar{q}}_{i\circ
}^{\prime}\right]  \left(  \boldsymbol{\hat{\psi}}_{TWFE}-\boldsymbol{\psi
}_{0}\right)  . \label{thetaFETE}%
\end{equation}
It is easily seen that $n^{-1}\sum_{i=1}^{n}\left(  \mathbf{z}_{i}%
-\mathbf{\mathbf{\bar{z}}_{\circ}}\right)  \mathbf{\bar{q}}_{i\circ}^{\prime}$
tends to a fixed matrix, and since under parallel trends and homogeneous
slopes assumption $\boldsymbol{\hat{\psi}}_{TWFE}-\boldsymbol{\psi}%
_{0}\rightarrow_{p}\mathbf{0}$, then it also follows that $\boldsymbol{\hat
{\theta}}_{TWFE}-\boldsymbol{\theta}_{0}\rightarrow_{p}\mathbf{0}$, since
$\boldsymbol{\hat{\theta}}\left(  \boldsymbol{\psi}\right)
-\boldsymbol{\theta}_{0}\rightarrow_{p}\mathbf{0}$, as already established. In
short, $\boldsymbol{\hat{\theta}}_{TWFE}$ is a consistent estimator of
$\boldsymbol{\theta}_{0}$ when $\boldsymbol{\hat{\psi}}_{TWFE}$ is a
consistent estimator of $\boldsymbol{\psi}_{0}$.

\section{Empirical results\label{empirics}}

To illustrate the methodological issues discussed above, we use the Penn World
Tables, PWT, version 10.01, which provides data for up to 70 years over the
period 1950-2019, together with the replication data of Acemoglu et al. (2019)
and Kremer et al. (2022). The theory suggests that ignoring intercept
heterogeneity, as is done in Barro regressions, biases the estimates of the
speed of convergence towards zero. Similarly, wrongly imposing parallel trends
or homogeneity on either the speeds of convergence or the other coefficients
of the panel data model will also cause the resultant estimates to have biases
that do not vanish with more data. It is further shown that when the parallel
trends assumption does not hold then the popular TWFE estimator of the speed
of convergence tends to zero if the steady state output growth in the global
economy is non-zero. See the discussion before equation (\ref{RWD}). Thus it
is of interest to see how robust are the Barro and TWFE estimates to
nonparallel trends and dynamic heterogeneity assumptions.

For output we use the national accounts measure of real GDP at constant 2017
national prices, expressed in million 2017US\$, labelled \textit{rgdpna} in
PWT, which we denote by $Y_{it}$. \ Countries with fewer than 30 observations
are removed, giving a baseline sample of 157 countries. It is not clear what
scale measure should be used for cross-country comparisons of outputs. We
consider population ($POP_{it}$) and employment ($EMP_{it}$), and work with
log output per capita $y_{it}=\log(Y_{it}/POP_{it})$, as well as log output
per employee, $\tilde{y}_{it}=\log(Y_{it}/EMP_{it})$. There are slightly
shorter time series for employment in some countries, but always at least 30
years of data. Since we use up to 4 lags, the effective minimum number of time
series observations used in panel regressions is $26$. As we include
covariates such as capital and democracy, the number of countries and the
number of time periods can vary across different panel regressions we report.
Nevertheless, in all these regressions the average time dimension of countries
within a given panel is always over 50 years, and as a result the small $T$
bias is likely to be of second order importance in our empirical applications.
Also, for brevity we sometimes use output to refer to either of the two output variables.

For a visual picture of the time profiles of outputs across countries in
Figure 1 we plot $y_{it}$ for 99 countries covering all countries with data
from 1960 onward except for a few countries with particularly erratic output
profiles.\footnote{The excluded countries are, namely Burundi, C\^{o}te d
Ivoire , Gabon, Ghana, Guinea, Equatorial Guinea, Madagascar, Uganda Tanzania,
Venezuela, Zambia and Zimbabwe} Despite the variability it is clear that a
common (global) growth factor is driving output in this sample of countries,
but it is unclear from this figure if the parallel trends assumption is likely
to be met. But due to the upward trend in the global factor, we would expect
that any violation of the parallel trends assumption would cause a substantial
bias in Barro and TWFE estimates of the speed of convergence.%

%TCIMACRO{\FRAME{ftbpFUO}{6.3408in}{2.6775in}{0pt}{\Qct{Log per capita output
%for selected countries over the period 1960-2019}\Qcb{}}{}{tkudga00.bmp}%
%{\special{ language "Scientific Word";  type "GRAPHIC";  display "USEDEF";
%valid_file "F";  width 6.3408in;  height 2.6775in;  depth 0pt;
%original-width 23.1346in;  original-height 9.2319in;  cropleft "0";
%croptop "1";  cropright "1";  cropbottom "0";
%filename 'TKUDGA00.bmp';file-properties "XNPEU";}} }%
%BeginExpansion
\begin{figure}[ptb]%
\centering
\caption{Log per capita output for selected countries over the period
1960-2019}%
\includegraphics[
natheight=9.231900in,
natwidth=23.134600in,
height=2.6775in,
width=6.3408in
]%
{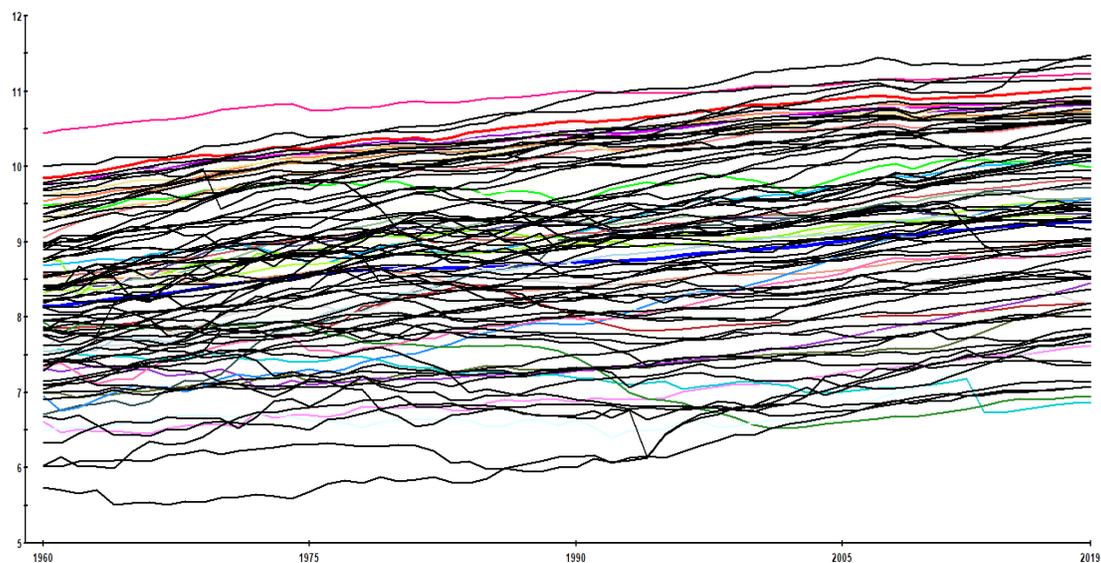}%
\end{figure}
%EndExpansion

For a more formal analysis of the trend in the global factor, we estimated
global output as a simple average of log output per capita of Asian, European,
and North American. The output indices for these regions are computed as
PPP-weighted averages, and their respective weights in the world economy are
close to 1/3. Such a measure is more satisfactory than using simple averages
of log per capita output based on a balanced panel of countries as depicted in
Figure 1. U.S. log per capita output, in dark red, appears at the top of
Figure 1, while our estimate of global output (GLO) is shown in dark blue in
the middle of Figure 1. The GLO variable has a mean growth rate of $2.4$
percent per annum. Equations (\ref{yig}) and (\ref{gtvit}) show that the
properties of $f_{t}$ are transmitted to $g_{t\text{ }}$and hence to $y_{it}%
,$unless $\gamma_{i}$ happens to be zero. It was shown in Section \ref{DCCEP}
that the DCCEP estimator of $\rho_{0}$\ is consistent for $n\,$and $T$ large,
irrespective of the nature of the process for $f_{t}$. In contrast, in Section
\ref{HetCon} it was shown that when the parallel trends did not hold and
$f_{t}$ was trended, then $\hat{\rho}_{TWFE}\rightarrow1,$ irrespective of the
value of $\rho_{0}$.

Table \ref{TableSUM} gives some summary statistics for the variables we
consider: the number of observations, overall mean, minimum, median, and
maximum. Since we are interested in both time-varying and time-invariant
variables, we give the summary statistics for the country means and the
country standard deviations. For time-invariant variables we report their
cross-country dispersions, noting that their time series dispersions is zero
by construction. The summary statistics for output and capital per employee,
not reported, are very similar to those reported for output and capital per capita.

Our empirical investigations closely follow the theoretical account set out
above. We begin with Barro cross-country regressions in Subsection
\ref{BarroSS}; go on to panel autoregressions in Subsection \ref{PanelNoCovSS}%
. We then consider panel regressions with time-varying covariates: first
including capital in Subsection \ref{PanelTVCovSS} and then reconsidering the
role of democracy in Subsection \ref{democ}. We finish with cross-country
regressions using time-invariant covariates in Subsection \ref{TimInvSS}. In
the case of panel regressions we consider: TWFE, homogeneous slope models with
parallel trends; DCCEP, homogeneous slope models with nonparallel trends; and
DCCEMG heterogeneous slope models with nonparallel trends. The DCCE estimates
are obtained using the Stata xtdcce2 package, Ditzen (2021).%

%TCIMACRO{\TeXButton{begin}{\begin{table}[H]}}%
%BeginExpansion
\begin{table}[H]%
%EndExpansion
\caption
{Summary statistics reporting dispersions of means and standard deviations of output and its possible drivers for the number of countries (n) for which there is data.}\label{TableSUM}%

\begin{tabular}
[c]{lllllll}\hline\hline
&  &  &  &  &  & \\
&  & \multicolumn{5}{c}{Dispersion of country means,{\footnotesize \ }$\bar
{y}_{i\circ}$}\\\cline{3-7}
&  &  &  &  &  & \\
{\footnotesize Selected variables} &  & Mean & S.D. & Min & Median & Max\\
& $n$ & $\bar{y}_{\circ\circ}$ & $sd_{\bar{y}}$ & {\footnotesize \ }$\bar
{y}_{\min}$ & $\bar{y}_{med}$ & $\bar{y}_{\max}$\\
& \multicolumn{1}{r}{} &  &  &  &  & \\
{\footnotesize per capita output growth} & {\footnotesize 157} &
{\footnotesize 0.018} & {\footnotesize 0.015} & {\footnotesize -0.023} &
{\footnotesize 0.019} & {\footnotesize 0.067}\\
{\footnotesize per capita capital growth} & {\footnotesize 157} &
{\footnotesize 0.026} & {\footnotesize 0.019} & {\footnotesize -0.037} &
{\footnotesize 0.026} & {\footnotesize 0.095}\\
&  &  &  &  &  & \\
{\footnotesize absolute latitude} & {\footnotesize 144} &
{\footnotesize 0.300} & {\footnotesize 0.195} & {\footnotesize 0} &
{\footnotesize 0.261} & {\footnotesize 0.722}\\
{\footnotesize ethno-linguistic fragmentation} & {\footnotesize 121} &
{\footnotesize 0.364} & {\footnotesize 0.308} & {\footnotesize 0} &
{\footnotesize 0.294} & {\footnotesize 1}\\
{\footnotesize protection against expropriation} & {\footnotesize 107} &
{\footnotesize 7.203} & {\footnotesize 1.729} & {\footnotesize 3.5} &
{\footnotesize 7.045} & {\footnotesize 10.00}\\
& \multicolumn{1}{r}{} & \multicolumn{1}{r}{} & \multicolumn{1}{r}{} &
\multicolumn{1}{r}{} & \multicolumn{1}{r}{} & \\
& \multicolumn{6}{c}{Dispersion of country standard
deviations,{\footnotesize \ }$\overline{sd}_{i\circ}$}\\\cline{2-7}
& \multicolumn{1}{r}{} & \multicolumn{1}{r}{} & \multicolumn{1}{r}{} &
\multicolumn{1}{r}{} & \multicolumn{1}{r}{} & \\
&  & Mean & S.D. & Min & Median & Max\\
& $n$ & $\overline{sd}_{\circ\circ}$ & $sd_{\overline{sd}}$ & $\overline
{sd}_{\min}$ & $\overline{sd}_{med}$ & $\overline{sd}_{\max}$\\
&  &  &  &  &  & \\
{\footnotesize per capita output growth} &
\multicolumn{1}{r}{{\footnotesize 157}} & {\footnotesize 0.056} &
{\footnotesize 0.034} & {\footnotesize 0.018} & {\footnotesize 0.047} &
{\footnotesize 0.220}\\
{\footnotesize per capita capital growth} &
\multicolumn{1}{r}{{\footnotesize 157}} & {\footnotesize 0.027} &
{\footnotesize 0.013} & {\footnotesize 0.004} & {\footnotesize 0.023} &
{\footnotesize 0.068}\\
& \multicolumn{1}{r}{} &  &  &  &  & \\
{\footnotesize absolute latitude} & \multicolumn{1}{r}{{\footnotesize 144}} &
{\footnotesize 0} & {\footnotesize 0} & {\footnotesize 0} & {\footnotesize 0}
& {\footnotesize 0}\\
{\footnotesize ethno-linguistic fragmentation} &
\multicolumn{1}{r}{{\footnotesize 121}} & {\footnotesize 0} &
{\footnotesize 0} & {\footnotesize 0} & {\footnotesize 0} & {\footnotesize 0}%
\\
{\footnotesize protection against expropriation} &
\multicolumn{1}{r}{{\footnotesize 107}} & {\footnotesize 0} &
{\footnotesize 0} & {\footnotesize 0} & {\footnotesize 0} & {\footnotesize 0}%
\\
& \multicolumn{1}{r}{} & \multicolumn{1}{r}{} & \multicolumn{1}{r}{} &
\multicolumn{1}{r}{} & \multicolumn{1}{r}{} & \multicolumn{1}{r}{}%
\\\hline\hline
\end{tabular}

{\footnotesize Notes. The number of countries is }$n.$ {\footnotesize The
first panel reports summary statistics for the }$n${\footnotesize \ country
means, }$\bar{y}_{i\circ}=T_{i}^{-1}%
%TCIMACRO{\tsum _{t=1}^{T_{i}}}%
%BeginExpansion
{\textstyle\sum_{t=1}^{T_{i}}}
%EndExpansion
y_{it}${\footnotesize , \ }$i=1,2,...,n${\footnotesize , with different time
series spans, }$T_{i}${\footnotesize , for }$i=1,2,...,n,$%
{\footnotesize \ covering the period 1950-2019, with means }$\bar{y}%
_{\circ\circ}=%
%TCIMACRO{\tsum _{i=1}^{n}}%
%BeginExpansion
{\textstyle\sum_{i=1}^{n}}
%EndExpansion
\bar{y}_{i\circ}/n,${\footnotesize \ and standard deviations, }$sd_{\bar{y}%
}=\sqrt{%
%TCIMACRO{\tsum _{i=1}^{n}}%
%BeginExpansion
{\textstyle\sum_{i=1}^{n}}
%EndExpansion
(\bar{y}_{i\circ}-\bar{y}_{\circ\circ})^{2}/(n-1)}$.{\footnotesize \ The
second panel is based on }$n${\footnotesize \ country standard deviations,
}$sd_{i\circ}=\sqrt{%
%TCIMACRO{\tsum _{t=1}^{T_{i}}}%
%BeginExpansion
{\textstyle\sum_{t=1}^{T_{i}}}
%EndExpansion
(y_{it}-\bar{y}_{i\circ})^{2}/(T_{i}-1)},$ {\footnotesize their mean
}$\overline{sd}_{\circ\circ}=n^{-1}%
%TCIMACRO{\tsum _{i=1}^{n}}%
%BeginExpansion
{\textstyle\sum_{i=1}^{n}}
%EndExpansion
\overline{sd}_{i\circ}${\footnotesize , and their standard deviations
}$sd_{sd}=\sqrt{%
%TCIMACRO{\tsum _{i=1}^{n}}%
%BeginExpansion
{\textstyle\sum_{i=1}^{n}}
%EndExpansion
(\overline{sd}_{i\circ}-\overline{sd}_{\circ\circ})^{2}/(n-1)}$%
{\footnotesize . Both panels also report minimum, median and maximum of }%
$\bar{y}_{i\circ}${\footnotesize \ and }$sd_{i\circ}${\footnotesize ,
respectively.}%
%TCIMACRO{\TeXButton{end}{\end{table}}}%
%BeginExpansion
\end{table}%
%EndExpansion

\subsection{Estimates of the speed of convergence using Barro
regressions\label{BarroSS}}

Consider the unconditional versions of the Barro regression given by setting
$\boldsymbol{\theta}=\mathbf{0}$, in (\ref{BB1}) and (\ref{BBder0}). namely
\begin{equation}
y_{iT}-y_{i0}=a_{T}+b_{T}y_{i0}+\zeta_{iT}, \label{BB1a}%
\end{equation}
where $a_{T}$ and $\zeta_{iT}$ are defined by (\ref{cT}) and (\ref{BBerror}),
respectively and $b_{T}=-(1-\rho^{T})$. The parameter of interest is the speed
of convergence, $\phi=1-\rho,$ and the mean number of years to convergence,
given by $(1-\phi)/\phi$.

While for equations that include a country specific intercept the base year
does not matter, for Barro regressions which assume intercept homogeneity,
$\alpha_{i}=\alpha,$ the base year does matter. In estimating Barro
regressions, Kremer et al. (2022, footnote 5) say "Specifically, for growth
rates we use the variable \textquotedblleft rgdpna,\textquotedblright\ real
GDP at constant 2017 national prices (2017 USD), and for growth levels we use
\textquotedblleft rgdpo,\textquotedblright\ output-side real GDP at chained
PPPs (2017 USD), as recommended by the PWT user guide." While we agree that
output series based on national accounts are appropriate for constructing the
dependent variable, $y_{iT}-y_{i0}$, it is difficult to rationalise having a
different measure of output (for example the PPP measure) on the right hand
side. Equation (\ref{BB0}) is obtained by iterating (\ref{FETE}) forward so
that the same series must be used on the right hand side as the left hand
side. Using the base year PPP measure, say $y_{i0}^{PPP},$ could be
rationalised by assuming $\alpha_{i}$ is a function of the particular basket
of goods produced in a country. Accordingly, we include both $y_{i0}$ and
$y_{i0}^{PPP}$ as initial values in (\ref{BB1a}), or equivalently by adding
$(y_{i0}-y_{i0}^{PPP})$ as an additional covariate with coefficient $c_{T}$.
Namely, we estimate the following the cross-country regression%
\begin{equation}
y_{iT}-y_{i0}=a_{T}+b_{T}y_{i0}+c_{T}\left(  y_{i0}-y_{i0}^{PPP}\right)
+\zeta_{iT}. \label{BB1b}%
\end{equation}
This equation reduces to the one estimated by Kremer et al. (2022), only if
$c_{T}=-b_{T}$, which can be tested.

While regressions of this sort can be estimated on unbalanced samples, using
$b_{T_{i}}=-(1-\rho^{T_{i}}),$ in (\ref{BB1a}), we follow the literature and
use a balanced sample. The choice of years $T=2019$ and $0=1990$ gave a large
value for $n$ and the inclusion of the post-communist countries provide a wide
range of initial outputs. Table \ref{TableBarro} reports both the linear and
non-linear estimates of equation (\ref{BB0}) (with heteroskedasticity robust
standard errors) using as dependent variable the change in the logarithm of
output per capita between 1990 and 2019: $y_{i,2019}-y_{i,1990}$. In the first
column the regressors are initial output per capita, $y_{i,1990}$ and the
difference between the national accounts and PPP measures $\left(
y_{i,1990}-y_{i,1990}^{PPP}\right)  .$ The second column drops the PPP
adjustment. It also provides comparable estimates using the logarithm of
output per employee in 2019 and 1990: $\tilde{y}_{i,2019},$ $\tilde
{y}_{i,1990}$.

In both 1990 and 2019 the two measures, output per capita and output per
employee, are very similar with a cross-country correlation of 0.98. Thus it
is not surprising that the two sets of Barro regressions give quite similar
results. The PPP adjustment term $y_{i0}-y_{i0}^{PPP}$, has\ t ratios around
two and including it has little effect on the estimated speed of convergence,
which range from 0.32 percent per annum to 0.58 percent. The restriction that
$c_{T}=-b_{T}$ is strongly rejected, with t statistics of $-3.10,$ for the per
capita equation, and $-3.96$ for the per employee one. The table also gives
the mean lag, estimated by $(1-\hat{\phi})/\hat{\phi},$ which ranges from 316
to 171 years, suggesting a very slow speed of convergence indeed. As we shall
see, the very low estimates of $\phi$ obtained using Barro regressions is
reflective of the bias towards zero of such estimates established in
Subsection \ref{BBbias}.%

%TCIMACRO{\TeXButton{begin}{\begin{table}[H]}}%
%BeginExpansion
\begin{table}[H]%
%EndExpansion
\caption
{Estimates of Barro regressions and the implied speed of convergence using log per capita and per employee output using a balanced panel covering 157 countries over 1990-2019}\label{TableBarro}%

\begin{flushleft}
{\normalsize {\small
\begin{tabular}
[c]{rccccc}\hline\hline
& \multicolumn{2}{c}{{\footnotesize Output per capita}} &  &
\multicolumn{2}{c}{{\footnotesize Output per employee}}\\\cline{2-3}%
\cline{5-6}
& {\footnotesize (1)} & {\footnotesize (2)} &  & {\footnotesize (1)} &
{\footnotesize (2)}\\\hline
\multicolumn{1}{l}{Regressors (estimates)} &  &  &  &  & \\
&  &  &  &  & \\
\multicolumn{1}{l}{$y_{i,1990}${\footnotesize \ }$\left(  {\footnotesize \hat
{b}}_{T}\right)  $} & $\underset{(0.0363)}{-0.0876}$ & $\underset{\left(
0.0346\right)  }{-0.1025}$ &  & $\underset{\left(  0.0408\right)  }{-0.1330}$
& $\underset{\left(  0.0365\right)  }{-0.1559}$\\
&  &  &  &  & \\
\multicolumn{1}{l}{${\footnotesize y}_{i,1990}-y_{i,1990}^{PPP}$ $(\hat{c}%
_{T})$} & $\underset{\left(  0.1062\right)  }{-0.2108}$ & $\underset{}{....}$
&  & $\underset{\left(  0.1117\right)  }{-0.2420}$ & $\underset{}{....}$\\
&  &  &  &  & \\
\multicolumn{1}{l}{${\footnotesize \hat{\phi}}=1-(1+\hat{b}_{T})^{1/T}$} &
$\underset{\left(  0.0014\right)  }{0.0032}$ & $\underset{\left(
0.0013\right)  }{0.0037}$ &  & $\underset{\left(  0.0016\right)  }{0.0049}$ &
$\underset{\left(  0.0015\right)  }{0.0058}$\\
&  &  &  &  & \\
\multicolumn{1}{l}{$\text{{\footnotesize Mean lag in years}}$} &
$\underset{\left(  137\right)  }{316}$ & $\underset{\left(  95\right)  }{268}$
&  & $\underset{\left(  67\right)  }{203}$ & $\underset{\left(  44\right)
}{171}$\\
\multicolumn{1}{l}{} &  &  &  &  & \\
\multicolumn{2}{l}{} &  &  &  & \\
\multicolumn{1}{l}{{\footnotesize Sample sizes}} &  &  &  &  & \\
\multicolumn{1}{l}{{\footnotesize Number of countries }$(n)$} &
${\footnotesize 157}$ & ${\footnotesize 157}$ &  & ${\footnotesize 157}$ &
${\footnotesize 157}$\\
\multicolumn{1}{l}{{\footnotesize Years }$(T)$} & ${\footnotesize 29}$ &
${\footnotesize 29}$ &  & ${\footnotesize 29}$ & ${\footnotesize 29}$\\
&  &  &  &  & \\\hline\hline
\end{tabular}
}}
\end{flushleft}

{\footnotesize Notes. The dependent variable is }${\footnotesize y}%
_{i,2019}{\footnotesize -y}_{i,1990}${\footnotesize , where }%
${\footnotesize y}_{it}${\footnotesize \ is the log output variable at time t
= 1990,2019. Estimates based on output per capita and output per employee use
logarithms of "rgdpna/pop" and "rgdpna/emp" from \textit{PWT 10.01},
respectively. }${\footnotesize b}_{T}${\footnotesize \ and }${\footnotesize c}%
_{T}${\footnotesize \ are given by the coefficients of the initial output
variable, }${\footnotesize y}_{i,1990}${\footnotesize \ and the coefficient of
the PPP adjustment variable, }${\footnotesize y}_{i,1990}{\footnotesize -y}%
_{i,1990}^{PPP}${\footnotesize \ where }${\footnotesize y}_{i,1990}^{PPP}%
${\footnotesize \ is the log per capita output in 1990 using the "rgdpo"
purchasing power parity measure from \textit{PWT 10.01}. Heteroskedasticity
robust standard errors in parentheses. The estimates are computed by running
non-linear cross-country regressions, the mean lag is estimated as
}${\footnotesize (1-\hat{\phi})/\hat{\phi}}${\footnotesize . }
%TCIMACRO{\TeXButton{end}{\end{table}}}%
%BeginExpansion
\end{table}%
%EndExpansion

\subsection{Estimates of speed of convergence based on dynamic panel data
models\label{PanelNoCovSS}}

The solutions to stochastic growth models like that in Binder and Pesaran
(1999) generate dynamic equations in output, which we represent by panel
autoregressions. Unlike the Barro regressions, these allow for intercept
heterogeneity, and are likely to be less biased. Table \ref{TabPanAR} reports
estimates of the average speed of convergence, $\phi_{0}=E\left(  \phi
_{i}\right)  $ based on panel AR(3) regressions in log output per capita
($y_{it}$, on the left panel) and log output per employee ($\tilde{y}_{it}$,
on the right panel). We report three estimates, TWFE estimates that assume
parallel trends and homogeneous dynamics, (as in (\ref{FETE})), DCCEP
estimates that allow for nonparallel trends but assume homogeneous slopes (as
in (\ref{IFE})), and DCCEMG estimates that allow for both nonparallel trends
and heterogeneous slopes, corresponding to equation (\ref{HetroInter}). The
DCCEMG estimates are computed using the error correction regressions in
$y_{it}$ (or $\tilde{y}_{it}$) augmented with cross section averages, $\bar
{y}_{\circ t}$, and their lagged values:
\begin{equation}
\Delta y_{it}=\alpha_{i}-\phi_{i}y_{i,t-1}+%
%TCIMACRO{\tsum _{s=1}^{p_{T}-1}}%
%BeginExpansion
{\textstyle\sum_{s=1}^{p_{T}-1}}
%EndExpansion
\delta_{is}\Delta y_{i,t-s}+\psi_{i0}\bar{y}_{\circ t}+%
%TCIMACRO{\tsum _{s=0}^{q_{T}-1}}%
%BeginExpansion
{\textstyle\sum_{s=0}^{q_{T}-1}}
%EndExpansion
\psi_{is}\Delta\bar{y}_{\circ,t-s}+u_{it}, \label{ardccemg}%
\end{equation}
The panel regressions are estimated using unbalanced panels, with $\bar
{y}_{\circ t}$ computed as a simple average of output of countries with data
in year $t$. The TWFE and DCCEP regressions are restricted version of the
above specification.

The lag orders, $p_{T}$ and $q_{T}$ are set close to $T_{ave}^{1/3},$ as
recommended by the theory with $q_{T}=p_{T}=3$ under homogeneous dynamics, and
$q_{T}=4>p_{T}=3$, in the case of heterogeneous dynamics.\footnote{$T_{ave}$
is the average time dimension of the panel computed as $n^{-1}\sum_{i=1}%
^{n}T_{i}$, where $T_{i}$ is the number of years of available data for country
$i$ and $n$ is the total number of countries in the panel.} The use of TWFE
estimators in the case of panels with heterogeneous dynamics results in terms
like $(\phi_{i}-\phi_{0})y_{i,t-1}$ to be included in the error term, which in
turn could result in error serial correlation leading to more lagged terms
being erroneously statistically significant. One might expect heterogeneous
models to require fewer lags. This is in fact the case, with the statistical
significance of $\Delta y_{i,t-2}$ falling as one moves from TWFE to the CCE
type estimators. Furthermore, since $f_{t}$ is likely to be trended in the
present application, and as explained earlier, fewer lags of cross section
averages are needed to adequately proxy the latent factor.

The estimates of the speed of convergence reported in Table \ref{TabPanAR} are
in accordance with the theory which shows that relaxing the homogeneity
restrictions reduces the downward bias in the estimated speed of convergence.
For the case of output per capita, the estimates of the speed of convergence
increases from 3 percent per annum when TWFE is used, to 5 percent if
nonparallel trends are allowed by using DCCEP, and to 11 percent when both the
parallel trends and slope homogeneity restrictions are relaxed if estimation
is carried out using the DCCEMG procedure. For output per employee the speeds
are slightly higher. The mean number of years to convergence falls from 24-27
estimated using TWFE, to 14-18 using DCCEP, and to 5-6 years when using DCCEMG
estimates (Table \ref{TabPanAR}). Only the DCCEMG\ estimates correspond to the
business cycle frequencies.%

%TCIMACRO{\TeXButton{begin}{\begin{table}[H]}}%
%BeginExpansion
\begin{table}[H]%
%EndExpansion
\caption
{Alternative estimates of speed of convergence based on unbalanced panel regressions of order three using PWT data covering 157 countries over the period 1950-2019}\label{TabPanAR}%

{\normalsize {\small
\begin{tabular}
[c]{rccccccc}\hline\hline
& \multicolumn{3}{c}{{\footnotesize Output per capita}} &  &
\multicolumn{3}{c}{{\footnotesize Output per employee}}\\\cline{2-4}%
\cline{6-8}
& {\footnotesize TWFE} & {\footnotesize DCCEP} & {\footnotesize DCCEMG} &  &
{\footnotesize TWFE} & {\footnotesize DCCEP} & D{\footnotesize CCEMG}\\\hline
\multicolumn{1}{l}{Regressors (estimates{\footnotesize )}} &  &  &  &  &  &  &
\\
&  &  &  &  &  &  & \\
\multicolumn{1}{l}{$y_{i,t-1}${\footnotesize \ }$\left(  {\footnotesize -}%
\hat{\phi}\right)  $} & $\underset{(0.004)}{{\footnotesize -0.027}}$ &
$\underset{(0.012)}{{\footnotesize -0.046}}$ &
$\underset{(0.012)}{{\footnotesize -0.113}}$ &  &
$\underset{(0.005)}{{\footnotesize -0.031}}$ &
$\underset{(0.014)}{{\footnotesize -0.058}}$ &
$\underset{(0.015)}{{\footnotesize -0.144}}$\\
&  &  &  &  &  &  & \\
\multicolumn{1}{l}{${\footnotesize \Delta y}_{i,t-1}$ $(\hat{\psi}_{1})$} &
$\underset{(0.030)}{{\footnotesize 0.206}}$ & $\underset{\left(  0.029\right)
}{{\footnotesize 0.141}}$ & $\underset{\left(  0.018\right)
}{{\footnotesize 0.210}}$ &  & $\underset{(0.034)}{{\footnotesize 0.185}}$ &
$\underset{(0.033)}{{\footnotesize 0.122}}$ &
$\underset{(0.016)}{{\footnotesize 0.130}}$\\
&  &  &  &  &  &  & \\
\multicolumn{1}{l}{${\footnotesize \Delta y}_{i,t-2}$ $(\hat{\psi}_{2})$} &
$\underset{(0.027)}{{\footnotesize 0.034}}$ & $\underset{\left(  0.030\right)
}{{\footnotesize -0.015}}$ & $\underset{\left(  0.014\right)
}{{\footnotesize -0.011}}$ &  & $\underset{(0.029)}{{\footnotesize 0.035}}$ &
$\underset{(0.031)}{{\footnotesize -0.019}}$ &
$\underset{(0.014)}{{\footnotesize 0.001}}$\\
&  &  &  &  &  &  & \\
\multicolumn{1}{l}{Mean lag in years} & $\underset{\left(  3.82\right)
}{{\footnotesize 27.1}}$ & $\underset{(4.87)}{{\footnotesize 18.2}}$ &
$\underset{(0.83)}{{\footnotesize 6.1}}$ &  & $\underset{\left(  3.98\right)
}{{\footnotesize 23.8}}$ & $\underset{\left(  3.70\right)
}{{\footnotesize 14.4}}$ & $\underset{\left(  0.69\right)
}{{\footnotesize 5.1}}$\\
\multicolumn{2}{l}{} &  &  &  &  &  & \\
\multicolumn{1}{l}{} &  &  &  &  &  &  & \\
\multicolumn{1}{l}{{\footnotesize Number of countries }$(n)$} &
${\footnotesize 157}$ & ${\footnotesize 157}$ & ${\footnotesize 157}$ &  &
${\footnotesize 157}$ & ${\footnotesize 157}$ & ${\footnotesize 157}$\\
\multicolumn{1}{l}{{\footnotesize T}$_{\min}\,$(years)} & ${\footnotesize 27}$
& ${\footnotesize 27}$ & ${\footnotesize 26}$ &  & ${\footnotesize 27}$ &
${\footnotesize 27}$ & ${\footnotesize 26}$\\
\multicolumn{1}{l}{{\footnotesize T}$_{ave}$} & ${\footnotesize 55.2}$ &
$55.2$ & ${\footnotesize 54.2}$ &  & ${\footnotesize 52}$ &
${\footnotesize 52}$ & ${\footnotesize 51}$\\
\multicolumn{1}{l}{{\footnotesize T}$_{\max}$} & ${\footnotesize 67}$ &
${\footnotesize 67}$ & $66$ &  & ${\footnotesize 67}$ & ${\footnotesize 67}$ &
${\footnotesize 66}$\\
&  &  &  &  &  &  & \\\hline\hline
\end{tabular}
}}

{\footnotesize Notes 1. The dependent variable is }$\Delta y_{it}%
${\footnotesize , the change in log output per capita and log output per
employee measured respectively using logarithms of "rgdpna/pop" and
"rgdpna/emp" from \textit{PWT 10.01}. Average speed of convergence, measured
by }${\footnotesize \phi=1-\rho}_{1}{\footnotesize -\rho}_{2}%
{\footnotesize -\rho}_{3}${\footnotesize , is estimated as the coefficient of
the lagged output variable. The mean lag is computed as }%
${\footnotesize (1-\hat{\phi}-\hat{\psi}}_{1}{\footnotesize -\hat{\psi}}%
_{2}{\footnotesize )/\hat{\phi}}${\footnotesize , where }${\footnotesize \psi
}_{1}{\footnotesize =-(\rho}_{2}{\footnotesize +\rho}_{3}{\footnotesize )}%
${\footnotesize , }${\footnotesize \psi}_{2}{\footnotesize =-\rho}_{3}%
${\footnotesize \ and }${\footnotesize \rho}_{s}${\footnotesize ,
}${\footnotesize s=1,2,3}${\footnotesize \ are the (average) autoregressive
coefficients. \ See also equation (\ref{ardccemg}).}

{\footnotesize Notes 2.} {\footnotesize TWFE is the two way fixed effect
estimator, DCCEP is the dynamic correlated common effect pooled estimator, and
DCCEMG is the dynamic correlated common effect mean group estimator. See
Sections (\ref{DCCEP}) and (\ref{HetConFac}) of the paper for further details.
Heteroskedasticity robust standard errors are in parentheses, Huber-White for
TWFE, the DCCE standard errors are heteroskedasticity robust. To filter out
the nonparallel trends, DCCEP augments the panel regressions with the set of
cross country averages }${\footnotesize \{\bar{y}}_{\circ,t}%
{\footnotesize ,\bar{y}}_{\circ,t-1}{\footnotesize ,\bar{y}}_{\circ
,t-2}{\footnotesize ,\bar{y}}_{\circ,t-3}{\footnotesize \}}$%
,{\footnotesize \ DCCEMG adds }${\footnotesize \bar{y}}_{\circ,t-4}%
${\footnotesize  to the set}$;$ {\footnotesize the additional lagged cross
section average is added to account for dynamic heterogeneity.\ Stata routine
xtdcce2 was used for estimation. }%
%TCIMACRO{\TeXButton{end}{\end{table}}}%
%BeginExpansion
\end{table}%
%EndExpansion

\subsection{Panel growth regressions with capital\label{PanelTVCovSS}}

We now consider panel regressions that include time-varying covariates which
we denote by the $k_{x}\times1$ vector $\mathbf{x}_{it}$. Augmenting
(\ref{ardccemg}) with time-varying covariates and their lagged values we have%
\begin{align}
\Delta y_{it}  &  =\alpha_{i}-\phi_{i}y_{i,t-1}+%
%TCIMACRO{\tsum _{s=1}^{p_{T}-1}}%
%BeginExpansion
{\textstyle\sum_{s=1}^{p_{T}-1}}
%EndExpansion
\delta_{is}\Delta y_{i,t-s}+\boldsymbol{\beta}_{i}^{\prime}\mathbf{x}_{it}+%
%TCIMACRO{\tsum _{s=0}^{p_{T}-1}}%
%BeginExpansion
{\textstyle\sum_{s=0}^{p_{T}-1}}
%EndExpansion
\boldsymbol{\lambda}_{is}^{^{\prime}}\Delta\mathbf{x}_{i,t-s}\label{DCCEMGK}\\
&  +\mathbf{\psi}_{i0}^{\prime}\mathbf{\bar{w}}_{\circ t}+%
%TCIMACRO{\tsum _{s=1}^{s_{T}-1}}%
%BeginExpansion
{\textstyle\sum_{s=1}^{s_{T}-1}}
%EndExpansion
\mathbf{\psi}_{is}^{\prime}\Delta\mathbf{\bar{w}}_{\circ,t-s}+u_{it},\nonumber
\end{align}
where $\mathbf{\bar{w}}_{\circ t}=(\bar{y}_{\circ t},\mathbf{\bar{x}}_{\circ
t}^{\prime})^{\prime}$, $\mathbf{\bar{x}}_{\circ t}$ is the cross section
average of $\mathbf{x}_{it}$, and $\boldsymbol{\pi}_{i}=-\boldsymbol{\beta
}_{i}/\phi_{i}$ is the long run elasticity of output with respect to
$\mathbf{x}_{it}$. A large number of covariates have been considered in the
literature as possible elements of $\mathbf{x}_{it}$. Here we focus on two
such variables, namely physical capital, and, a prominent institutional
measure, democracy, considered in Sub-section \ref{democ}. The problem of how
best to select a sub-set of relevant variables from a large number of
potential covariates poses new methodological challenges and involves
application of penalized regression techniques to dynamic panels with latent
factors. This is beyond the scope of this empirical application, which is
primarily intended to illustrate the methodological issues discussed in this paper.

Our choice of capital as a covariate is based on production function
approaches to growth theories that postulate output in the long run to be a
function of inputs of capital and labour as in a constant returns Cobb-Douglas
production function, with the technology regarded as a latent
factor.\footnote{Constant returns was not rejected using the DCCE estimators
but was using TWFE.} For the capital variable we use log capital per capita,
$k_{it}=\log(K_{it}/POP_{it})$ to explain $y_{it}=\log(Y_{it}/POP_{it})$, and
use log capital per employee, $\tilde{k}_{it}=\log(K_{it}/EMP_{it})$ to
explain $\tilde{y}_{it}=\log(Y_{it}/EMP_{it})$.\footnote{The capital measure
$K_{it}$ labelled rnna in PWT is capital stock at constant 2017 national
prices (in millions 2017US\$).}\ The impact effect of the capital input on
growth is likely to be subject to simultaneity bias since cyclical demand and
supply factors will complicate the identification of the short-run effects of
capital accumulation. However, the long run effects (that we focus on) are
more likely to be consistently estimated. CCE type estimators that allow
capital and output to be jointly determined by the same set of latent factors
representing tastes, technology, and demography are likely to reduce omitted
variable bias.\footnote{Here we use a regression explaining output to estimate
the long run coefficients. An alternative approach which does not involve
specifying the dynamics or a direction of causation is described in Chudik,
Pesaran and Smith (2025).}

The panel data used for estimation is unbalanced and includes countries with
at least $30$ years of data, rendering a panel with $157$ countries and a
minimum of $26$ data points, noting that the maximum lag order of the panel is
set to $s_{T}=4$. The estimation results are reported in Table
\ref{TabYPC_KPC1} for the same set of specifications as in Table
\ref{TabPanAR}. As before, allowing for heterogeneity and nonparallel trends
results in larger estimates for the speed of convergence, while at the same
time reducing the required lag orders. We also see further increases in the
estimated speed of convergence (irrespective of the estimation method or the
choice of output and capital variables) once we condition on the capital
variable. This is in line with the literature that finds conditional
convergence to be faster than unconditional convergence.

\textbf{ }%

%TCIMACRO{\TeXButton{begin}{\begin{table}[H]}}%
%BeginExpansion
\begin{table}[H]%
%EndExpansion
\caption
{Alternative estimates  of unbalanced dynamic panel regressions  of order three explaining the change in log output  by log capital  using PWT data covering 157 countries over the period 1950-2019.}\label{TabYPC_KPC1}%

\begin{flushleft}
{\normalsize {\small
\begin{tabular}
[c]{rccccccc}\hline\hline
& \multicolumn{3}{c}{{\footnotesize Output per capita}} &  &
\multicolumn{3}{c}{{\footnotesize Output per employee}}\\\cline{2-4}%
\cline{6-8}
& {\footnotesize TWFE} & {\footnotesize DCCEP} & {\footnotesize DCCEMG} &  &
{\footnotesize TWFE} & {\footnotesize DCCEP} & {\footnotesize DCCEMG}\\\hline
\multicolumn{1}{l}{{\footnotesize AR Coefficients}} &  &  &  &  &  &  & \\
&  &  &  &  &  &  & \\
$y_{i,t-1}${\footnotesize \ (output)} & $\underset{\left(  0.007\right)
}{-{\footnotesize 0.051}}$ & $\underset{\left(  0.031\right)
}{-{\footnotesize 0.131}}$ & $\underset{\left(  0.025\right)
}{-{\footnotesize 0.349}}$ &  & $\underset{\left(  0.009\right)
}{-{\footnotesize 0.057}}$ & $\underset{\left(  0.030\right)
}{-{\footnotesize 0.183}}$ & $\underset{\left(  0.029\right)
}{-{\footnotesize 0.375}}$\\
&  &  &  &  &  &  & \\
${\footnotesize \Delta y}_{i,t-1}$ & $\underset{\left(  0.030\right)
}{{\footnotesize 0.163}}$ & $\underset{\left(  0.023\right)
}{{\footnotesize 0.061}}$ & $\underset{\left(  0.020\right)
}{{\footnotesize 0.118}}$ &  & $\underset{\left(  0.031\right)
}{{\footnotesize 0.178}}$ & $\underset{\left(  0.029\right)
}{{\footnotesize 0.118}}$ & $\underset{\left(  0.021\right)
}{{\footnotesize 0.132}}$\\
&  &  &  &  &  &  & \\
${\footnotesize \Delta y}_{i,t-2}$ & $\underset{\left(  0.028\right)
}{{\footnotesize 0.028}}$ & $\underset{\left(  0.028\right)
}{-{\footnotesize 0.050}}$ & $\underset{\left(  0.016\right)
}{{\footnotesize 0.005}}$ &  & $\underset{\left(  {\footnotesize 0.030}%
\right)  }{{\footnotesize 0.028}}$ & $\underset{\left(  0.044\right)
}{{\footnotesize 0.013}}$ & $\underset{\left(  0.018\right)
}{{\footnotesize 0.027}}$\\
&  &  &  &  &  &  & \\
\multicolumn{1}{l}{{\footnotesize DL Coefficients}} &  &  &  &  &  &  & \\
&  &  &  &  &  &  & \\
$k_{i,t-1}${\footnotesize \ (Capital)} & $\underset{\left(  0.005\right)
}{{\footnotesize 0.027}}$ & $\underset{\left(  0.026\right)
}{{\footnotesize 0.084}}$ & $\underset{\left(  0.045\right)
}{{\footnotesize 0.169}}$ &  & $\underset{\left(  0.005\right)
}{{\footnotesize 0.027}}$ & $\underset{\left(  0.025\right)
}{{\footnotesize 0.115}}$ & $\underset{\left(  0.034\right)
}{{\footnotesize 0.180}}$\\
&  &  &  &  &  &  & \\
${\footnotesize \Delta k}_{it}$ & $\underset{\left(  0.111\right)
}{{\footnotesize 1.173}}$ & $\underset{\left(  0.194\right)
}{{\footnotesize 1.225}}$ & $\underset{\left(  0.159\right)
}{{\footnotesize 1.926}}$ &  & $\underset{\left(  0.044\right)
}{{\footnotesize 0.821}}$ & $\underset{\left(  0.058\right)
}{{\footnotesize 0.884}}$ & $\underset{\left(  0.065\right)
}{{\footnotesize 0.904}}$\\
&  &  &  &  &  &  & \\
${\footnotesize \Delta k}_{i,t-1}$ & $\underset{\left(  0.112\right)
}{{\footnotesize -0.705}}$ & $\underset{\left(  0.165\right)
}{{\footnotesize -0.516}}$ & $\underset{\left(  0.137\right)
}{{\footnotesize -0.815}}$ &  & $\underset{\left(  0.050\right)
}{{\footnotesize -0.352}}$ & $\underset{\left(  0.044\right)
}{{\footnotesize -0.279}}$ & $\underset{\left(  0.044\right)
}{{\footnotesize -0.285}}$\\
&  &  &  &  &  &  & \\
${\footnotesize \Delta k}_{i,t-2}$ & $\underset{\left(  0.066\right)
}{-{\footnotesize 0.069}}$ & $\underset{\left(  0.085\right)
}{-{\footnotesize 0.074}}$ & $-\underset{\left(  0.121\right)
}{{\footnotesize 0.043}}$ &  & $\underset{\left(  0.038\right)
}{-{\footnotesize 0.010}}$ & $\underset{\left(  0.050\right)
}{-{\footnotesize 0.050}}$ & $\underset{\left(  0.043\right)
}{-{\footnotesize 0.029}}$\\
&  &  &  &  &  &  & \\
\multicolumn{1}{l}{} &  &  &  &  &  &  & \\
{\footnotesize Long run effect of capital} & $\underset{\left(  0.043\right)
}{{\footnotesize 0.524}}$ & $\underset{\left(  0.105\right)
}{{\footnotesize 0.640}}$ & $\underset{\left(  0.122\right)
}{{\footnotesize 0.484}}$ &  & $\underset{\left(  0.044\right)
}{{\footnotesize 0.470}}$ & $\underset{\left(  0.068\right)
}{{\footnotesize 0.627}}$ & $\underset{(0.084)}{{\footnotesize 0.481}}$\\
\multicolumn{2}{l}{} &  &  &  &  &  & \\
&  &  &  &  &  &  & \\
{\footnotesize Number of Countries }$(n)$ & ${\footnotesize 157}$ &
${\footnotesize 157}$ & ${\footnotesize 157}$ &  & ${\footnotesize 157}$ &
${\footnotesize 157}$ & ${\footnotesize 157}$\\
{\footnotesize Year \ \ \ \ \ \ \ \ \ \ \ \ \ \ \ \ \ \ \ \ T}$_{\min}$ &
${\footnotesize 27}$ & ${\footnotesize 27}$ & ${\footnotesize 26}$ &  &
${\footnotesize 27}$ & ${\footnotesize 27}$ & ${\footnotesize 26}$\\
{\footnotesize T}$_{ave}$ & ${\footnotesize 55.2}$ & ${\footnotesize 55.2}$ &
${\footnotesize 54.2}$ &  & ${\footnotesize 52.0}$ & ${\footnotesize 52.0}$ &
${\footnotesize 51.0}$\\
{\footnotesize T}$_{\max}$ & ${\footnotesize 67}$ & ${\footnotesize 67}$ &
${\footnotesize 66}$ &  & ${\footnotesize 67}$ & ${\footnotesize 67}$ &
${\footnotesize 66}$\\
&  &  &  &  &  &  & \\\hline\hline
\end{tabular}
}}
\end{flushleft}

{\footnotesize Notes. }$k_{it}$ {\footnotesize is either the logarithm of
capital per capita "rnna/pop" or per employee "rnna/emp" from\textit{ PWT
10.01}, matching the way the output variable is scaled.} {\footnotesize To
filter out the nonparallel trends, DCCEP augments the panel regressions with
the cross country averages }${\footnotesize \{\bar{y}}_{\circ,t}%
{\footnotesize ,\bar{k}}_{\circ,t}{\footnotesize ,\bar{y}}_{\circ
,{\footnotesize t-1}}{\footnotesize ,\bar{k}}_{\circ,t-1}{\footnotesize ,\bar
{y}}_{\circ,{\footnotesize t-2}}{\footnotesize ,\bar{k}}_{\circ
,{\footnotesize t-2},}{\footnotesize \bar{y}}_{\circ,t-3}{\footnotesize ,\bar
{k}}_{\circ,t-3}{\footnotesize \}}$, {\footnotesize and the DCCEMG augments
the panel regressions with this set plus }${\footnotesize \bar{y}}_{\circ
,t-4}{\footnotesize ,\bar{k}}_{\circ,t-4}$ {\footnotesize to account for
dynamic heterogeneity. Speed of convergence, }${\footnotesize \phi=1-\rho}%
_{1}{\footnotesize -\rho}_{2}{\footnotesize -\rho}_{3}${\footnotesize , is
estimated as the coefficient of the lagged output variable, }%
${\footnotesize y}_{i,t-1}${\footnotesize . See also Note 2 to Table
\ref{TabPanAR}}%
%TCIMACRO{\TeXButton{end}{\end{table}}}%
%BeginExpansion
\end{table}%
%EndExpansion

We also report estimates of the average long run elasticity of output with
respect to capital, across the three estimation methods and the two measures
of output and capital variables. Under some assumptions the long-run
elasticity should also equal the share of capital in income. Pesaran and Yang
(2021) discuss the various estimates of the US share of capital used in the
literature, which tend to be in the range $0.3-0.4$. The estimates we obtain
for the long-run elasticity are somewhat higher, ranging from $0.47$ to
$0.64$, with the DCCEMG estimates at $0.48$ $(0.12)$ (using output per capita)
and $0.48$ $(0.08)$ (using output per employee) falling at the lower end.
However, our estimates refer to an average across a wide range of countries.
Over the $126$ countries with available data, the average labour share in the
PWT data is $0.54$, that would suggest a coefficient on capital of around
$0.46$. It is interesting that the DCCEMG estimates, that fully allow for
heterogeneity across countries, are close to this. As discussed above allowing
for a latent factor that simultaneously drives capital accumulation and the
output process reduces endogeneity bias (if any) for estimation of long run
coefficients.\footnote{The working paper version of this study, Pesaran and
Smith (2026), reports results using the Barro-Lee measure of years of
education. Including education did not change the estimates of the effect of
capital very much, and education is only statistically significant when the
capital stock variable is excluded and the panel is estimated by DCCEP using
per capita measures. The poor performance of the education measure may be a
consequence of there being severe problems in measuring human capital.}

\subsection{Acemoglu et al. (2019) panel growth regressions with
democracy\label{democ}}

There has been extensive research on the economic impact of democratization,
most prominently by Acemoglu et al. (2019), and it is of interest to
investigate the extent to which Acemoglu et al.'s estimates are robust to
relaxing the assumptions of dynamic homogeneity and parallel trends. To focus
on this problem we use their replication data set which is publicly available.
Acemoglu et al. (2019, p.50) argue that existing measures of democracy suffer
from measurement error and they develop a dichotomous democracy indicator,
zero for non-democracy one for democracy, which combines several indices to
purge spurious changes in each. They estimate a dynamic TWFE panel data model
using four lags of log output per capita. This corresponds to our error
correction specification with three lagged output changes. Their estimates are
based on $n=175$ countries for the years 1960-2010, but some countries only
have six years of data, which is too few to relax the parallel trends
assumption using CCE estimators. Accordingly, we consider a subset of $n=148$
countries, that had at least 24 years of data, leaving a minimum of 20 time
series observations per country for estimation, after allowing for lags.

Their democracy indicator shows time series variation in $73$ of the $148$
countries; in $75$ countries it does not change. This lack of time series
variation does not cause any problems when using pooled estimators such as
TWFE and DCCEP, since these estimation methods assume the democracy variable
to have the same effect across all countries, irrespective of whether their
democracy indicator varies over time or not. This is a strong assumption which
is difficult to test. We report estimates for two versions of DCCEMG that
differ depending on whether homogeneity of the coefficient of the democracy
variable is imposed or not. The first version, denoted by DCCEMG*, imposes the
same coefficient on the democracy variable across all countries, thus
including all the $148$ countries, but differs from TWFE or DCCEP since it
allows for dynamic heterogeneity. The second version, denoted by DCCEMG,
allows for full parameter heterogeneity and uses the smaller sample of $73$
countries whose democracy indicator varies over time.

The estimation results are reported in Table \ref{TabDemocracy}. The first
column gives the Acemoglu et al.'s TWFE estimates based on their full sample
of $175$ countries and match their estimate of the long run effect of
democracy, namely $21.240$ $(7.215)$, given in column 3 of their Table
2.\footnote{The bracketed figures give standard errors clustered by country.}
The subsequent columns give our TWFE estimates using the $n=148$ sample, which
is also used to produce the estimates in the columns under DCCEP and DCCEMG*
that impose a homogeneous democracy coefficient. The final column gives the
DCCEMG estimates using the $n=73$ sample of countries with a time-varying
democracy indicator, but using $\bar{y}_{\circ t}$ and $\bar{d}_{\circ t}$
computed as cross section averages of $y_{it}$ and $d_{it}$ corresponding to
the $n=148$ sample. While the point estimates of the long run effects for the
three DCCE estimators are similar, the $n=73$ estimate in the final column is
not comparable to the other two DCCE estimates. It is an average treatment
effect on the treated, where the treatment is a change in the democracy indicator.

A comparison of the first two columns of Table \ref{TabDemocracy} shows that
dropping countries with less than $24$ years of data makes virtually no
difference to the TWFE estimates. Looking across the first row of the table
shows that, as in the preceding tables, relaxing parallel trends and dynamic
homogeneity result in increased estimates of the average speed of convergence
across countries, as predicted by the theory. Also, whereas all three lagged
output changes are statistically significant when using TWFE, the lagged
changes are less significant in the case of the other estimates that relax the
homogeneity restrictions. Turning to the impact effect of democracy, the point
estimates are very similar to those obtained by Acemoglu et al. (2019), though
their statistical significance falls as we allow for a greater degree of
heterogeneity, and become statistically insignificant if we condition on
countries where the democracy indicator varies. A similar pattern can be
observed if we consider the long run estimates. The estimates of the long run
coefficient of the democracy variable fall as we relax the parallel trends and
dynamic homogeneity assumptions. The statistically significant estimate
reported by Acemoglu et al. (2019) become either insignificant or only
marginally significant, although all the estimates have the expected positive
sign. This loss of statistical significance is not due to multicollinearity
between the country-specific democracy indicator, $d_{it}$, and the cross
section average, $\bar{d}_{\circ t}$.\footnote{We are grateful to Daron
Acemoglu for suggesting this possibility.} When the latter is dropped,
democracy is less significant. For instance the DCCEP estimate falls from
$0.947$ with a $t$-ratio of $1.65$ when $\bar{d}_{\circ t}$ is included, to
$0.748$ with a$\ t$-ratio of $1.27$\ when $\bar{d}_{\circ t}$ is excluded from
the panel regression.%

%TCIMACRO{\TeXButton{begin}{\begin{table}[H]}}%
%BeginExpansion
\begin{table}[H]%
%EndExpansion
\caption
{Alternative estimates of equations explaining the change in log output per capita using the dichotomous democracy indicator in unbalanced panel regressions using Acemoglu et al. (2019) data 1960-2010.}\label{TabDemocracy}%

\begin{flushleft}
{\normalsize {\small
\begin{tabular}
[c]{rccccccc}\hline\hline
&  &  & \multicolumn{4}{c}{{\footnotesize Output per capita}} & \\\cline{3-8}%
\cline{6-8}
&  & \multicolumn{2}{c}{TWFE} &  & \multicolumn{3}{c}{Dynamic CCE}%
\\\cline{3-4}\cline{3-4}\cline{6-8}
&  & Acemoglu et al. & This paper &  & {\footnotesize DCCEP} &
{\footnotesize DCCEMG}$^{\ast}$ & DCCEMG\\
$y_{i,t-1}${\footnotesize \ (Lagged log output)} &  & $\underset{\left(
0.005\right)  }{-{\footnotesize 0.037}}$ &
$\underset{(0.005)}{{\footnotesize -0.036}}$ &  &
$\underset{(0.019)}{{\footnotesize -0.164}}$ & $\underset{\left(
0.022\right)  }{{\footnotesize -0.218}}$ & $\underset{\left(  0.035\right)
}{{\footnotesize -0.348}}$\\
&  &  &  &  &  &  & \\
${\footnotesize \Delta y}_{i,t-1}$ &  & $\underset{\left(  0.036\right)
}{{\footnotesize 0.275}}$ & $\underset{(0.035)}{{\footnotesize 0.287}}$ &  &
$\underset{\left(  0.043\right)  }{{\footnotesize 0.210}}$ &
$\underset{\left(  0.025\right)  }{{\footnotesize 0.186}}$ &
$\underset{\left(  0.034\right)  }{{\footnotesize 0.207}}$\\
&  &  &  &  &  &  & \\
${\footnotesize \Delta y}_{i,t-2}$ &  & $\underset{\left(  0.023\right)
}{{\footnotesize 0.069}}$ & $\underset{(0.025)}{{\footnotesize 0.058}}$ &  &
$\underset{\left(  0.034\right)  }{{\footnotesize 0.018}}$ &
$\underset{\left(  0.018\right)  }{{\footnotesize -0.054}}$ &
$\underset{\left(  0.027\right)  }{{\footnotesize -0.016}}$\\
&  &  &  &  &  &  & \\
${\footnotesize \Delta y}_{i,t-3}$ &  & $\underset{\left(  0.017\right)
}{{\footnotesize 0.043}}$ & $\underset{(0.019)}{{\footnotesize 0.047}}$ &  &
$\underset{\left(  0.028\right)  }{{\footnotesize 0.057}}$ &
$\underset{\left(  0.015\right)  }{{\footnotesize 0.030}}$ &
$\underset{\left(  0.021\right)  }{{\footnotesize 0.086}}$\\
&  &  &  &  &  &  & \\
$d_{it}${\footnotesize \ (Democracy indicator)} &  & $\underset{\left(
0.226\right)  }{{\footnotesize 0.787}}$ &
$\underset{(0.226)}{{\footnotesize 0.769}}$ &  & $\underset{\left(
0.574\right)  }{{\footnotesize 0.947}}$ & $\underset{\left(  0.654\right)
}{{\footnotesize 1.285}}$ & $\underset{\left(  1.034\right)
}{{\footnotesize 1.499}}$\\
&  &  &  &  &  &  & \\
\multicolumn{1}{l}{{\footnotesize Long run effect}} &  &  &  &  &  &  & \\
\multicolumn{1}{l}{{\footnotesize of Democracy}} &  & $\underset{\left(
7.21\right)  }{{\footnotesize 21.24}}$ &
$\underset{(7.47)}{{\footnotesize 21.53}}$ &  &
$\underset{(3.59)}{{\footnotesize 5.77}}$ &
$\underset{(3.06)}{{\footnotesize 5.90}}$ & $\underset{\left(  3.01\right)
}{{\footnotesize 4.30}}$\\
&  &  &  &  &  &  & \\
\multicolumn{1}{l}{{\footnotesize Sample Sizes}} &  &  &  &  &  &  & \\
&  &  &  &  &  &  & \\
\multicolumn{1}{l}{{\footnotesize Number of countries} $n$} &  &
${\footnotesize 175}$ & ${\footnotesize 148}$ &  & ${\footnotesize 148}$ &
${\footnotesize 148}$ & ${\footnotesize 73}$\\
\multicolumn{1}{l}{{\footnotesize Years}} &  &  &  &  &  &  & \\
$T_{\min}$ &  & ${\footnotesize 6}$ & ${\footnotesize 20}$ &  &
${\footnotesize 20}$ & ${\footnotesize 20}$ & ${\footnotesize 20}$\\
$T_{ave}$ &  & ${\footnotesize 36.2}$ & ${\footnotesize 40}$ &  &
${\footnotesize 40}$ & ${\footnotesize 40}$ & ${\footnotesize 41.3}$\\
$T_{\max}$ &  & ${\footnotesize 47}$ & ${\footnotesize 47}$ &  &
${\footnotesize 47}$ & ${\footnotesize 47}$ & ${\footnotesize 47}%
$\\\hline\hline
\end{tabular}
}}
\end{flushleft}

{\footnotesize Notes. The dependent variable, }${\footnotesize \Delta y}_{it}%
${\footnotesize , the change in log output per capita, and }${\footnotesize d}%
_{it}${\footnotesize , the dichotomous democracy indicator, are both from the
replication data set provided by Acemoglu et al. (2019). The first column,
headed Acemoglu et al., replicates their results, subsequent columns exclude
countries where the minimum }${\footnotesize T}_{i}$ {\footnotesize used for
estimation was less that 20. To filter out the nonparallel trends, DCCEP and
DCCEMG use cross section averages }$\{{\footnotesize \bar{y}}_{\circ
,t}{\footnotesize \bar{y}}_{\circ,t-1}{\footnotesize ,\bar{y}}_{\circ
,t-2}{\footnotesize ,\bar{y}}_{\circ,t-3}{\footnotesize ,\bar{y}}_{\circ
,t-4},{\footnotesize \bar{d}}_{\circ,t}\}.$ {\footnotesize DCCEMG* imposes
that the coefficient of democracy is the same across all }${\footnotesize 148}%
$ {\footnotesize countries, whilst \thinspace DCCEMG allows for full
heterogeneity across the 73 countries in the panel with time-varying democracy
indicators}. {\footnotesize See also the notes to Table \ref{TabPanAR}.}%
%TCIMACRO{\TeXButton{end}{\end{table}}}%
%BeginExpansion
\end{table}%
%EndExpansion

\subsection{Regressions with time-invariant covariates\label{TimInvSS}}

Finally, we turn to the effects of time invariant covariates on $\alpha_{i}$
and $\gamma_{i},$ following the procedure set out in Section \ref{TINV}. For
the reasons set out there, we use pooled rather than the country specific
estimates to filter out the time varying components. From the large number of
time-invariant variables considered in the literature and recently documented
by Kremer et al. (2022), we consider an indicator of geography,
\textit{absolute lattitude}; an indicator that has often been associated with
conflict, \textit{ethnolinguistic fractionalization}; and an indicator of
property rights, \textit{protection from expropriation}. To these we add our
PPP indicator, $y_{i,1990}-y_{i,1990}^{PPP}$, to control for possible
systematic differences in the national accounts and PPP measures of output in
the initial year of the sample. There are also many other variables that one
could consider, such as alternative measures of climate change, technological
adaptation, governance approaches, and institutional features. However, due to
the challenging new methodological issues that surround the problem of
variable selection in the context of high dimensional heterogeneous dynamic
panel data models, we confine our empirical analysis to a few variables
highlighted as potentially important in the literature. An extension of our
empirical analysis to selection from a large set of covariates is beyond the
scope of the present paper and must be the subject of a separate investigation.

The estimation of the time-invariant effects follows two steps: in the first
step we estimate the coefficients of the time-varying effects used to estimate
$a_{iT}$ in (\ref{FCS}). To avoid the small $T$ bias of some of the individual
country estimates, we only consider the pooled estimates, namely TWFE and
DCCEP, which we denote by $\boldsymbol{\hat{\psi}}_{TWFE}$\textbf{ }and
$\boldsymbol{\hat{\psi}}_{DCCEP}$, respectively. We obtain these estimates
using panel ARDL(2,2) regressions in log output and log capital (either per
capita or per employee). We also increase the minimum number of observations
used for these first stage panel regressions, so that the average number of
time periods used is over $60$ years. In the second stage, we run
cross-country regressions of $\hat{a}_{iT}=\bar{y}_{i\circ}-\mathbf{\bar{q}%
}_{i\circ}^{\prime}\boldsymbol{\hat{\psi}}$ , with $\boldsymbol{\hat{\psi
}=\hat{\psi}}_{TWFE}$\textbf{ }or $\boldsymbol{\hat{\psi}}_{DCCEP}$, for
$i=1,2,...,100$, on an intercept and four time-invariant regressors, namely
\textit{absolute latitude}, \textit{ethnolinguistic fractionalization},
\textit{protection from expropriation, and }$y_{i,1990}-y_{i,1990}^{PPP}$.

The estimation results are summarized in Table \ref{TabPanINV}. The estimated
coefficient of \textit{absolute latitude }is not statistically significant. In
contrast, the variables \textit{ethnolinguistic fractionalization }and
\textit{protection from expropriation} are both statistically significant with
the expected signs: a higher probability of conflict or weaker protection of
property rights both have negative output effects. The PPP correction
coefficient is positive and significant or on the margin of significance.
These outcomes are robust to the choice of the estimates of $\boldsymbol{\psi
}$, namely $\boldsymbol{\hat{\psi}}_{TWFE}$\textbf{ }and $\boldsymbol{\hat
{\psi}}_{DCCEP}$.%

%TCIMACRO{\TeXButton{begin}{\begin{table}[H]}}%
%BeginExpansion
\begin{table}[H]%
%EndExpansion
\caption
{Cross-country  regressions estimating the coefficients (standard errors) of time-invariant regressors using different procedures to filter the time-varying effects.}\label{TabPanINV}%

\begin{flushleft}
{\normalsize {\small
\begin{tabular}
[c]{rccccc}\hline\hline
& \multicolumn{2}{c}{{\footnotesize Output per capita}} &  &
\multicolumn{2}{c}{{\footnotesize Output per employee}}\\\cline{2-3}%
\cline{5-6}
& {\footnotesize TWFE} & {\footnotesize DCCEP} &  & {\footnotesize TWFE} &
{\footnotesize DCCEP}\\\hline
\multicolumn{1}{l}{} &  &  &  &  & \\
\multicolumn{1}{l}{\textbf{Dependent variable, }$\hat{a}_{i}$} &  &  &  &  &
\\
&  &  &  &  & \\
\multicolumn{1}{l}{\textit{absolute latitude (}$z_{i1}$)} &
$\underset{(1.289)}{0.383}$ & $\underset{(2.568)}{0.832}$ &  &
$\underset{\left(  1.283\right)  }{0.747}$ & $\underset{\left(  3.390\right)
}{1.353}$\\
&  &  &  &  & \\
\multicolumn{1}{l}{ethnolinguistic fractionalization \textit{(}$z_{i2}$)} &
$\underset{(0.608)}{-1.708}$ & $\underset{(1.341)}{-3.062}$ &  &
$\underset{\left(  0.658\right)  }{-1.643}$ & $\underset{\left(  1.831\right)
}{-4.554}$\\
&  &  &  &  & \\
\multicolumn{1}{l}{\textit{protection from expropriation (}$z_{i3}$)} &
$\underset{(0.122)}{0.699}$ & $\underset{(0.236)}{0.952}$ &  &
$\underset{\left(  0.135\right)  }{0.627}$ & $\underset{\left(  0.336\right)
}{1.219}$\\
&  &  &  &  & \\
\multicolumn{1}{l}{$y_{i,1990}-y_{i,1990}^{PPP}$ \textit{(}$z_{i4}$)} &
$\underset{(0.272)}{0.704}$ & $\underset{(0.709)}{1.374}$ &  &
$\underset{\left(  0.328\right)  }{0.883}$ & $\underset{\left(  0.984\right)
}{2.675}$\\\hline\hline
\end{tabular}
}}
\end{flushleft}

{\footnotesize Notes. The reported estimates are computed by least squares
regressions of }$\hat{a}_{i}=\bar{y}_{i\circ}-\mathbf{\bar{q}}_{i\circ
}^{\prime}\boldsymbol{\hat{\psi}}${\footnotesize , on an intercept and
}$\mathbf{z}_{i}=(z_{i1},z_{i2},z_{i3},z_{i4})^{\prime}${\footnotesize , for
}$i=1,2,...,{\footnotesize n=100}$, {\footnotesize where }$\mathbf{\bar{q}%
}_{i\circ}${\footnotesize is the time averages of lagged output and capital
variables for country }$i${\footnotesize , }$\boldsymbol{\hat{\psi}%
}=\boldsymbol{\hat{\psi}}_{TWFE}$ {\footnotesize or }$\boldsymbol{\hat{\psi}%
}_{DCCEP}$ {\footnotesize are }$\,${\footnotesize TWFE and DCCEP estimators
obtained} {\footnotesize using ARDL(2,2) panel regressions in log output and
log capital (either per capita or per employee). Data on }$\mathbf{z}_{i}$
{\footnotesize are from Kremer et al. (2022) and are all divided by 100 to
scale up the estimates provided. Standard errors are heteroskedasticity
robust. The maximum time period used for estimation of }$\boldsymbol{\hat
{\psi}}\ ${\footnotesize \ was 68 years. }${\footnotesize T}_{\min
}{\footnotesize =48}$ {\footnotesize and }${\footnotesize T}_{ave}%
{\footnotesize =63}$ {\footnotesize years for regressions based per capita
measures, and }${\footnotesize T}_{\min}{\footnotesize =38}$
{\footnotesize and }${\footnotesize T}_{ave}{\footnotesize =61}$
{\footnotesize years for regressions based on per employee measures. }%
%TCIMACRO{\TeXButton{end}{\end{table}}}%
%BeginExpansion
\end{table}%
%EndExpansion

\section{Concluding remarks\label{conc}}

The empirical results presented in the previous section are in line with the
theory and show that large biases can result from wrongly assuming parallel
trends and/or homogeneous dynamics. Barro or pooled regression estimates of
the speed of convergence are biased towards zero unless all slope coefficients
in the panel are homogeneous and the heterogeneity of the fixed effects can be
modelled perfectly: an impossible undertaking. TWFE estimates allow for
intercept heterogeneity, but continue to be substantially biased if its two
main assumptions of parallel trends and dynamic homogeneity are invalidated.
Most crucially, when the latent factor is trended (deterministic or
stochastic) and its effects differ across countries, the TWFE\ estimate of the
speed of convergence tends to zero, no matter what is its true value.

To deal with nonparallel trends and dynamic heterogeneity we consider DCCEP
and DCCEMG estimators that are the dynamic versions of the CCE\ estimator
originally proposed by Pesaran (2006). DCCEP relaxes the parallel trends
assumption but maintains dynamic homogeneity, whilst DCCEMG allows for both.
We show that DCCEP produces consistent estimates even if the latent factors
driving the growth process are trended, with or without unit roots.

In the empirical applications and in line we our theoretical results, we find
that the mean convergence time of output to its steady state trend falls from
hundreds of years if one uses Barro estimates, to 27 years if one uses TWFE,
and to 6 years if one uses DCCEMG estimates, which closely match the business
cycle frequency in OECD countries. We have relatively quick convergence to a
country specific steady state trend, but there is no evidence of global
cross-country convergence.

When capital is added it has a sensible long run coefficient. Relaxing the
parallel trends assumption reduces the endogeneity problem, since the
introduction of a latent global factor with a country specific effect reduces
the omitted variable problem arising from common factors driving both output
and capital.

We also investigated the effects of time-invariant variables on average
deviations of output from its steady state trend, and found statistically
significant positive effects from higher levels of \textit{protection from
expropriation} and statistically significant negative effects from high levels
of \textit{ethnolinguistic fractionalization}.

There remain a number of areas for further research both in the econometric
methods and in the determinants of the differences in output across countries.
These include

\begin{itemize}
\item The high dimensional selection problem in panels for a more
comprehensive empirical analysis of the determinants of output from the large
number of fast and slow moving variables suggested in the literature.

\item Inference on the role of time-invariant effects and how to separate
their effects on the level of output ($\alpha_{i}$) and the trend output
growth ($\gamma_{i}$).

\item How to distinguish between time-varying and time-invariant effects of
climate. The time-varying effects have been extensively investigated in the
literature. See, for example, Kahn, Mohaddes, Ng, Pesaran, Raissi, and Yang
(2021), and references cited therein.

\item Convergence between countries, perhaps within clubs, can be investigated
by looking at pairwise convergence, as in Pesaran (2007).
\end{itemize}

Although this paper has focussed on economic growth similar issues come up in
other research areas. For example, Westerlund, Karabiyik, Narayan and Narayan
(2022) consider similar econometric issues in the context of applications to
corporate finance.

\pagebreak\bigskip

\begin{center}
{\LARGE References}
\end{center}

Acemoglu, D., S. Johnson, J.A. Robinson (2001) The colonial origins of
comparative development: an empirical investigation, \textit{The American
Economic Review}, 91: 1369-1401.

Acemoglu, D., S. Naidu, P. Restrepo, J.A. Robinson (2019). Democracy does
cause growth. \textit{Journal of Political Economy}, 127: 47-100.

Aghion, P., P. Howitt (1992) A model of growth through creative destruction,
\textit{Econometrica,} 60:323-351.

Barro, R. J. (1991). Economic growth in a cross section of countries.
\textit{Quarterly Journal of Economics}, 106: 407-44.

Barro, R.J., X. Sala-i-Martin (1992) Convergence, \textit{Journal of Political
Economy}, 100: 223-251.

Barro, R.J. (2015) Convergence and modernisation, \textit{The Economic
Journal} 125: 911-942.

Baumol, W.J. (1986) Productivity growth, convergence, and welfare: what the
long-run data show, \textit{American Economic Review}, 76: 1072-1085.

Binder, M., M.H. Pesaran (1999) Stochastic growth models and their econometric
implications, \textit{Journal of Economic Growth}, 4: 139-183.

Brock, W.A., and L.J. Mirman (1972) Optimal economic growth and uncertainty:
the discounted case, \textit{Journal of Economic Theory}, 4, 479-513.

Chauvet, M., C. Yu (2006) International business cycles: G7 and OECD
countries, \textit{Economic Review}, Federal Reserve Bank of Atlanta, First
Quarter: 43-54.

Chudik, A., M.H. Pesaran (2015) Common correlated effects estimation of
heterogeneous dynamic panel data models with weakly exogenous regressors,
\textit{Journal of Econometrics}, 188: 393-420.

Chudik, A., M.H. Pesaran, R.P.Smith (2025) Analysis of multiple long run
relations in panel data models, https://arxiv.org/abs/2506.02135v3

De Vos, I., G. Everaert (2021) Bias-corrected common correlated effects pooled
estimation in dynamic panels, \textit{Journal of Business and Economic
Statistics}, 39: 294-306.

Ditzen, J. (2021) Estimating long-run effects and the exponent of
cross-sectional dependence: an update to xtdcce2, \textit{The Stata Journal},
21: 687-707.

Donaldson, J.B., and R. Mehra (1983): Stochastic Growth with Correlated
Production Shocks, \textit{Journal of Economic Theory}, 29, 282-312.

Durlauf, S. (2009) The rise and fall of cross-country growth regressions,
\textit{History of Political Economy}, 41 (annual supplement): 315-333.

Everaert, G., T. De Groote (2016) Common correlated effects estimation of
dynamic panels with cross-sectional dependence, \textit{Econometric Reviews},
35: 428-463.

Friedman M. (1992) Do old fallacies ever die? \textit{Journal of Economic
Literature}, 30: 2129-32.

Granger, C.W.J. (1980) Long memory relationships and the aggregation of
dynamic models, \textit{Journal of Econometrics}, 14: 227--238.

Hausman, J.A., W.E. Taylor (1981) Panel data and unobservable individual
effects, \textit{Econometrica}, 49: 1377-1398

Hayakawa, K., M.H. Pesaran, L. V. Smith (2023) Short T dynamic panel data
models with individual, time and interactive effects, \textit{Journal of
Applied Econometrics, }38: 940-967.

Islam, N. (1995) Growth empirics: A panel data approach. \textit{Quarterly
Journal of Economics} 110: 1127--1170.

Johnson, P., C. Papageorgiou (2020) What remains of cross-country convergence,
\textit{Journal of Economic Literature,} 58: 129-175.

Juodis, A,. H. Karabiyik, J. Westerlund (2021) On the robustness of the pooled
CCE estimator, \textit{Journal of Econometrics}, 220: 325-348.

Kahn, M.E., K. Mohaddes, R.N.C. Ng, M.H. Pesaran, M. Raissi, J-C. Yang (2021)
Long-term macroeconomic effects of climate change: a cross-country analysis,
\textit{Energy Economics, 104: 105624.}

Kapetanios, G., M.H. Pesaran, T. Yamagata (2011) Panels with non-stationary
multifactor error structures, \textit{Journal of Econometrics}, 160: 326-384.

Kremer, M., J. Willis, Y. You (2022) Converging to convergence, \textit{NBER
Macroeconomics Annual}, 36: 337-412.

Lee, K., M.H. Pesaran, R.P. Smith (1997) Growth and convergence in a
multi-country empirical stochastic Solow model, \textit{Journal of Applied
Econometrics}, 12: 357-392.

Lee, K., M.H. Pesaran, R.P. Smith (1998), Growth empirics: a panel data
approach - a comment, \textit{Quarterly Journal of Economics}, 113: 319-323.

Mankiw, N.G., D. Romer, D.N. Weil (1992). A contribution to the empirics of
economic growth, \textit{Quarterly Journal of Economics}, 107: 407--37.

Maseland, R. (2021) Contingent determinants, \textit{Journal of Development
Economics}, 151: 102654.

Merton, R.C. (1975) An asymptotic theory of growth under uncertainty,
\textit{Review of Economic Studies}, 42, 375-393.

Moon, H.R., M. Weidner (2017) Dynamic linear panel regression models with
interactive fixed effects, \textit{Econometric Theory}, 33: 158-195.

Nerlove, M. (1998) Growth rate convergence, fact or artifact? An essay on the
use and misuse of panel data econometrics, WP 98-21, Department of
Agricultural and Resource Economics, The University of Maryland.

Nerlove, M. (1999) Properties of alternative estimators of dynamic panel
models: an empirical analysis of cross-country data for the study of economic
growth, in Chapter 6 of \textit{Analysis of Panels and Limited Dependent
Variable Models: In honour of G.S. Maddala}, by C. Hsiao, K. Lahiri, L-F Lee,
and M. H. Pesaran (Editors), Cambridge University Press, Cambridge.

Nickell, S. (1981) Biases in dynamic models with fixed effects,
\textit{Econometrica}, 49: 1417-1426.

Ong, K. (2024) Do countries converge to their steady states at different
rates? \textit{Open Economies Review} 35: 723--749

Patel, D., J. Sandefur, A. Subramanian (2021) The new era of unconditional
convergence, \textit{Journal of Development Economics}, 152: 102687.

Pesaran, M.H. (2006) Estimation and inference in large heterogeneous panels
with a multifactor error structure, \textit{Econometrica} 74: 967-1012.

Pesaran, M.H. (2007) A pair-wise approach to testing for output and growth
convergence, \textit{Journal of Econometrics}, 138: 312-355.

Pesaran, M.H., A. Chudik (2014) Aggregation in large dynamic panels,
\textit{Journal of Econometrics}, 178: 273-285.

Pesaran, M.H., R.P. Smith (1995) Estimating long-run relationships from
dynamic heterogenous panels, \textit{Journal of Econometrics, 68: 79-113.}

Pesaran, M.H., R.P. Smith (2026) The output convergence debate revisited:
lessons from recent developments in the analysis of panel data models,
\textit{Cambridge Working Papers in Economics} 2607.

Pesaran M.H., Y. Xie (2026) How to detect network dependence in latent factor
models? A bias-corrected CD test, \textit{Econometric Theory}, published online.

Pesaran, M.H., L.Yang (2021) Estimation and inference in spatial models with
dominant units, \textit{Journal of Econometrics}, 221: 591-615.

Pesaran, M.H., L.Yang (2024) Heterogeneous autoregressions in short T panel
data models, \textit{Journal of Applied Econometrics}, 39: 1359-1378.

Pesaran, M.H., Q. Zhou (2018) Estimation of time-invariant effects in static
panel data models, \textit{Econometrics Reviews}, 37: 1137-1171.

Robinson, P. M. (1978) Statistical inference for a random coe cient
autoregressive model, \textit{Scandinavian Journal of Statistics,} 5: 163-168.

Sarafidis, V., T. Wansbeek (2021) Celebrating 40 years of panel data analysis:
Past, present and future, \textit{Journal of Econometrics}, 220: 215-226.

Solow, R.W. (1956) A contribution to the theory of economic growth,
\textit{Quarterly Journal of Economics}, 70, 65-94.\smallskip\ 

Swan, T.W. (1956) Economic growth and capital accumulation, \textit{Economic
Record}, 32, 334-361.

Smith, R.P. (2024) Econometric aspects of convergence: a survey, \textit{Open
Economies Review}, 35: 701--721.

Temple, J. (1999) The new growth evidence, \textit{Journal of Economic
Literature}, 37: 112-156.

Wansbeek, T., A. Kapteyn (1989) Estimation of the error-components model with
incomplete panels, \textit{Journal of Econometrics}, 41: 341-361.

Westerlund, J. (2018) CCE in panels with general unknown factors, \textit{The
Econometrics Journal}, 21: 264-276.

Westerlund, J., H. Karabiyik, P.K. Narayan and S. Narayan (2022) Estimating
the speed of adjustment of leverage in the presence of interactive effects,
\textit{Journal of Financial Econometrics}, 20: 942-960.

\end{document}